\documentclass[11pt,preprint]{aastex}
\usepackage{natbib}
\usepackage{apjfonts}
\usepackage{epsfig}

\newcommand{\Msun}{\hbox{$\rm\thinspace M_{\odot}$}}
\newcommand{\msun}{\rm M_{\odot}}
\newcommand{\Zsun}{\hbox{$\rm\thinspace Z_{\odot}$}}
\newcommand{\zsun}{\rm Z_{\odot}}

\shorttitle{Molecular gas, CO, and star formation in galaxies}
\shortauthors{Pelupessy \& Papadopoulos}
\begin{document}

\title{Molecular gas, CO,  and star formation in galaxies: emergent
empirical relations, feedback, and the evolution of very gas-rich systems}
\author{Federico I. Pelupessy\altaffilmark{1}, Padelis P.
Papadopoulos\altaffilmark{2}}
\altaffiltext{1} {Leiden Observatory, Leiden University, P.O. Box 9513, 
2300 RA Leiden, The Netherlands}
\altaffiltext{2} {Argelander-Institut f\"ur Astronomie,  Auf dem H\"ugel 71, 
  D-53121 Bonn, Germany}
  
\begin{abstract}

 We use time-varying models of the coupled evolution of the HI, $\rm H_2$
 gas phases and stars in galaxy-sized numerical simulations to: a) test for 
 the emergence of the Kennicutt-Schmidt (K-S) and the $\rm H_2$-pressure 
 relation, b) explore a realistic $\rm  H_2$-regulated star formation recipe
 which brings forth a neglected and  potentially significant SF-regulating
 factor, and c) go beyond typical  galactic environments (for which these
 galactic empirical relations are  deduced) to explore the early evolution
 of very gas-rich galaxies.  In this work we model low mass galaxies
 ($M_{\rm baryon} \le 10^9 \msun$), while incorporating an independent
 treatment of CO formation and destruction, the most important
  tracer molecule of H2 in galaxies, along with that for the
  H2 gas itself. We find that both the K-S and  the $\rm
 H_2$-pressure  empirical relations  can robustly  emerge in galaxies after
 a dynamic equilibrium sets in between the various ISM states, the  stellar
 component and its feedback  ($\rm T\ga 1\,Gyr$). The only  significant
 dependence of  these relations seems to  be for the  CO-derived (and  thus
 directly observable) ones,  which  show a strong dependance  on the  ISM
 metallicity.  The  $\rm H_2$-regulated star formation recipe successfully
 reproduces the morphological and quantitative aspects of previous 
 numerical models while doing away with the star formation efficiency
 parameter.  Most of the $\rm HI\rightarrow H_2$ mass  exchange is found
 taking place under highly    non-equilibrium    conditions     
 necessitating    a time-dependent treatment even in  typical ISM
 environments. Our dynamic models indicate that the CO
 molecule can be a poor, non-linear, $\rm H_2$  gas tracer.

 Finally, for early evolutionary  stages ($\rm T\la 0.4\,Gyr$) we find
 significant and systematic deviations of the true star formation from  that
 expected  from  the K-S  relation,  which are  especially pronounced and 
 prolonged for  metal-poor systems.  The  largest such deviations  occur for
 the  very gas-rich galaxies, where deviations of  a factor $\sim  3-4$ in
 global star formation rate  can take place with  respect to those  expected
 from the CO-derived  K-S relation.  This is  particularly important since
 gas rich systems at high redshifts  could  appear as having unusually high 
 star-formation rates with respect  to their CO-bright $\rm H_2$ gas
 reservoirs. This points to a possibly serious deficiency of  K-S 
 relations  as  elements  of the  sub-grid  physics  of  star formation  in 
 simulations  of   structure  formation  in  the  Early Universe.

\end{abstract}
 

\section{Introduction\label{sec:intro}}

In spite of the fact that the general character  of the cycle through
which galaxies  convert their ISM to  stars has been known  for a long
time,  it  has  proven  to  be remarkably  difficult  to  formulate  a
predictive theoretical framework  for this process. Indeed, we know 
that throughout most of the Universe star formation takes place in 
molecular gas complexes \citep[e.g.][]{Solomon2005, Omont2007}.
The $\rm  HI\rightarrow H_2$  phase  transition is
conditioned by a combination  of sufficiently high HI column densities
and pressures, and consequently star formation tends to concentrate in
high  density regions,  e.g.  in  the  central parts  of galaxies,  in
spiral arms or high-pressure concentrations  of gas formed by bulk gas
motions or swept up by the shocks from supernovae and stellar winds of
OB associations.  The link between molecular gas and star formation is
so tight that  it has even been used to infer  the distribution of the
former by that of the  latter, when the CO-$\rm H_2$ conversion factor was
still  considered  very uncertain  \citep{Rana1986}.   In the  Galaxy,
where   this  link   is  best   studied   \citep[e.g.][and  references
therein]{Blitz1997}, the latest  results confirm star formation always
taking  place  in  CO-bright  molecular  clouds, even  at  very  large
galactocentric  distances  \citep{Kobayashi2008}.   In other  galaxies
this tight association has been verified in all cases where sufficient
angular resolution is  available \citep[e.g.][]{Wong2002}.  Thus it is
fair to say that {\it  $\rm H_2$ formation is a necessary prerequisite for
star  formation in  galaxies,}  and incorporating  it in  galaxy-sized
numerical simulations  of gas and  stars is the single  most important
step currently missing from a realistic rendering of star formation in
such models.

Following the early  and widespread observational evidence establishing
the $\rm H_2$-(star formation) link, the  inclusion of the $\rm H_2$ gas phase
in   numerical  models   of   galaxies  has   occured  only   recently
\citep{Pelupessy2006, Dobbs2006, Robertson2008}, and the refinement of
these models is an area of ongoing research. This is mainly due to
the difficulty of tracking the dynamic and thermodynamic evolution of
$\rm H_2$ and its precursor phase, the Cold Neutral Medium HI 
\citep[$\rm n \sim 5-100$  cm$^{-3}$, $\rm T_k  \sim 60-200$K,][]{Wolfire2003}
in galaxy-sized numerical models and due to the strong $\rm H_2$ 
self-shielding complicating radiative transfer models of its far-UV 
radiation-induced destruction.   The  first  problem  has prevented  most 
efforts  from properly  tracking the  $\rm  HI\rightarrow H_2$  phase 
transition  in galaxies  without  resorting  to  simplifying  
steady-state  solutions \citep{Hidaka2002,   Robertson2008}  
(suitable   only   for  quiescent galactic environments), or  to 
semi-empirical multiphase models \citep[e.g.][]{Semelin2002} with 
limited predictive value.  The second problem confounds
even  numerical simulations tracking  the $\rm HI\rightarrow H_2$ phase
transition in individual gas  clouds where local approximations of the
self-shielding HI/$\rm H_2$  volume (necessary for  numerically manageable
solutions) can make the  $\rm H_2$ gas  mass fraction  a strong
function       of      the      chosen       numerical      resolution
\citep[e.g.][]{Glover2007a}.   Finally   a  secondary,  yet  important
problem of such single gas  cloud simulations is posed by the constant
boundary  conditions  assumed during  their  evolution,  which are  an
unlikely setting for  real gas clouds immersed in  the ISM environment
of a galaxy.  In such environments cloud boundary conditions that
powerfully  influence the  $\rm HI\rightarrow  H_2$  phase transition,
such  as the  ambient  FUV  radiation field  and  pressure, change  on
timescales comparable  or shorter than ``internal''  cloud dynamic and
thermodynamic  timescales,  especially   in  vigorously  star  forming
environments \citep[e.g.][]{Parravano2003, Wolfire2003, Pelupessy2006}.

Despite  the   aforementioned  difficulties the incorporation of the  $\rm
H_2\leftrightarrow HI$  gas phase interplay, and its strong role
as star formation regulator, in numerical models holds the promise of
large  improvements in  their handling  of galaxy  evolution,  and the
possibility  of unveiling  new,  hitherto neglected,  aspects of  star
formation  feedback  on   the  ISM.   In  this  paper   we  apply  our
numerical  models for the  coupled evolution of  gas (HI,
$\rm H_2$)  and  stars  \citep{Pelupessy2006}  to the evolution 
of low mass galaxies ($M_{\rm baryon}<10^9 \msun$) in order to explore
two  new  key directions, the first of which is  the  emergence  of  two  
important  empirical relations   found   for   galaxies   in  the   local  
Universe:   the Kennicutt-Schmidt (K-S)  and the $\rm H_2$--pressure  relation. 
Secondly we will make a investigation of very
gas-rich systems (more typical of the Early Universe), and check 
whether the aforementioned  relations remain  valid during  their  
evolution.  The
latter  is  of  crucial  importance  given  the  prominant  role  such
empirical   relations   are   given   in   describing   the   sub-grid
star-formation/gas  interplay in  cosmological  simulations of  galaxy
evolution (where the resolution  limitations imposed by the simulation
of large  volumes preclude a detailed description  of star formation).
Finally along  with our original time-dependent treatment  of the $\rm
HI\rightarrow H_2$  phase transition we also include the  CO molecule,
allowing  direct   comparisons  to   the  {\it  observed}   $\rm H_2$  gas
distributions, and a new independent investigation of  the   CO-$\rm H_2$ 
relation  within  the  dynamical   setting  of  an evolving galaxy. 

 The  structure  of  our  work  is  as  follows:  in
section~\ref{sec:model}  we present  relevant features  of  the model,
show  semi-analytical predictions for  the $\rm H_2$-pressure  relation, and
formulate  the  extention  of  the  model  so  that  includes  CO,  in
section~\ref{sec:sims} we present  our detailed numerical simulations,
and  investigate  the  K-S,  $\rm H_2$--pressure and  CO--$\rm H_2$  empirical
relations.  In  section~\ref{sec:disc} we investigate  and discuss the
validity of  the important K-S relation during  early galaxy evolution
stages,  and  for  very   gas-rich  systems.  We  then  summarize  our
conclusions in section~\ref{sec:concl}.

\section{Incorporating the molecular gas phase: a dynamical approach}
\label{sec:model}

 The  $\rm  HI\rightarrow  H_2$   phase  transition  in  galaxies,  as
catalyzed by dust grains, have been studied extensively ever since the
strong self-shielding  nature of $\rm H_2$  in its dissociation  by far-UV
(FUV)  photons  has  been recognized  \citep{Spitzer1975,  Savage1977,
Federman1979}. These  theoretical \citep{Elmegreen1989, Elmegreen1993,
Papadopoulos2002},  and observational  \citep{Honma1995}  studies made clear
that  ISM pressure,  ambient FUV field,  as well  as metallicity play major 
roles in  the $\rm HI\rightarrow  H_2 $  phase transition. The role  of
pressure in particular  has been highlighted  over a wide range  of galaxy 
properties \citep{Blitz2004},  and quantified  in an empirical
$\rm H_2$-pressure relation derived by \cite{Blitz2006}.  Such a relation
(herafter  B-R relation) along with  the well-established K-S relation
\citep{Kennicutt1989,Kennicutt1998} are the most encompassing observational
benchmarks that galaxy models must pass before they can be  trusted in 
their predictions.  Moreover,  by linking  gas and star  formation (K-S 
relation), and  the $\rm H_2$  phase (the  true star formation fuel)  to
``macroscopic'' ISM  environmental parameters such as  pressure  (B-R
relation)  these  empirical  relations are  natural choices for  any
sub-grid  formulation that   relates gas and  star  formation  in 
simulations of  cosmological  volumes  where sub-grid recipes for star
physics at kpc scales become necessary.

 Currently there is no evidence  that the K-S  and B-R relations hold in the
extreme and very gas-rich star forming galaxies discovered at  high
redshifts \citep[e.g.][]{Walter2003},  and there  is even tentative  evidence 
that  the  K-S  relation obtained  in  the  local Universe may  not be
applicable  in UV/optically selected  galaxies at high redshifts
\citep{Tacconi2008}. Detailed galaxy-sized simulations   of gas and stars
are thus important tools for exploring the  robustness and possible
limitations of these emperical relations in a systematic fashion. Key 
features of  our  galaxy-sized TREE/SPH  numerical models of gas+stars that
make them appropriate for such purposes are:

\begin{itemize}

\item Non-equilibrium  treatment of the gas  thermodynamics, resulting
 in gas with $\rm (n, T_{kin})\sim (0.1\,cm^{-3}, 10^4\,K)$  (the WNM HI phase) 
 to $\rm (n,  T_{kin})\sim (100\,cm^{-3}, 40\,K)$,  (i.e. the  
 CNM HI  and the resulting $\rm H_2$ phase).

\item  Tracking  temporally  and  spatially varying  radiation  fields
      (profoundly  influencing  the   $\rm  HI\rightarrow  H_2$  phase
      transition  within   a  galaxy)  using   time-dependent  stellar
      evolution libraries.

\item  A time-dependant sub-grid  physical model  of the  HI/$\rm H_2$
       mass exchange that  readily incorporates the varying ambient
       conditions expected for the ISM within an evolving galaxy.

\item Star formation controlled by a Jeans-mass instability criterion.

\end{itemize}

\noindent

The  latter  is  enabled  by   the  fact  that  our  code  tracks  the
gravitational and  thermodynamical state of  the gas and can identify  
gravitationally unstable  regions down  to the  temperatures and densities
typical  of Giant  Molecular Clouds (GMCs).   It is  in such regions  
that   strong   observational   evidence   and   theoretical considerations 
\citep[e.g.][]{Elmegreen2000, Elmegreen2002}  suggests  that  star formation
occurs.  Finally, note that given the continuous mass  exchange  between 
the  WNM  and  the CNM  gas  phase,  and  the non-equilibrium  conditions 
often  found  for the  former  even  for quiescent environments  in the
Milky  Way \citep[]{Wolfire2003}, any successfull  time-dependent treatment 
of  the $\rm  HI\leftrightarrow H_2$ mass exchange  within evolving galaxies
must  track the ISM thermodynamics.

\subsection{The $\rm H_2$ model}

 The  subgrid cloud  structure  model used  by \cite{Pelupessy2006}  to
 describe the $\rm HI\leftrightarrow H_2$ mass exchange is constructed using
 widely observed ISM scaling laws \citep{Larson1981, Heyer2004}, shown to 
 hold  generally for  gas  clouds  virialized  under a  background pressure
 \citep{Elmegreen1989}.  A major development since  our first use of this
 sub-grid cloud representation in our numerical models is that these scaling
 laws have now been found to hold for extragalactic GMCs as well
 \citep{Bolatto2008}. Below we describe its main features, while more
 details can be found in \cite{Pelupessy2006}.

 For a cloud with radius $R$, mean density $\langle n \rangle$,
 and internal  density profile $\rm  n(r)\propto 1/r$, consisting  of a
 molecular  core and  an outer  HI gas  layer of  a  $HI\rightarrow H_2$
 transition column density $N_{tr}({\rm  HI})$, under irradiation by an
 external  stellar UV  field the  molecular
 fraction can be expressed as
\begin{equation}
\label{eq:fm}
f_{{\rm H}_2} \equiv \frac{M(H_2)}{M_c}=\exp \left[-3\frac{N_{\rm tr}}{\langle
n \rangle R}\right].
\end{equation}

 Here we will assume that the gaseous ISM is composed of structure 
conforming to the widely observed density-size scaling relation 
\citep{Larson1981, Pelupessy2006},
\begin{equation}
\label{eq:nmean}
\langle n \rangle R = 4.7 \times 10^{21} \left(\frac{P_e/k_B}{10^4\ {\rm K\
cm}^{-3}}\right)^{1/2}\ {\rm cm}^{-2}.
\end{equation}
 Hence, a calculation of the thickness of the neutral layer $N({\rm HI})$ 
 gives the local molecular fraction from equations \ref{eq:fm} 
 and \ref{eq:nmean}.  For this  transition column  density $N({\rm HI})$ a
 differential  equation  can  be  formulated that  describes  the  time
 evolution
\begin{equation}
\label{eq:dntrdt}
\tau_f \frac{d \sigma N_{\rm tr}}{dt}=
r_{\rm dis} e^{-\sigma N_{\rm tr}}-\sigma N_{c} 
\left(e^{\sigma N_{\rm tr}/\sigma N_{c}}-1\right),
\end{equation}
 where $\tau_f=1/ (2 n R_f)$  is the $\rm H_2$ formation timescale. The $\rm H_2$
 formation   rate  function   $R_f$  depends   on   temperature  $T_k$,
 metallicity  $Z$, and  a normalization  parameter $\mu$  (encoding the
 uncertainties inherent in its absolute value,  with 
 $\mu=\mu_0=3.5$ corresponding to the \cite{Jura1974} formation
 rate of $R_f\sim 3 \times 10^-{17}$ s$^{-1}$ at $T=100$K), as
\begin{equation}
\label{eq:rf}
R_f = 3.5\times 10^{-17} \mu \, Z\, {\left(\frac{T_k}{100\ K}\right)^{1/2}} \,
 S_H(T_k) \gamma_{H_2}\, {\rm cm}^3\  {\rm s}^{-1}.
\end{equation}

\noindent
 The function $ S_{H}(T_k)$ expresses  the HI sticking probability on a
 dust grain  and forming  an $\rm H_2$ molecule  that then  detaches itself
 from the grain  with a probability $\rm \gamma  _{H_2}$. Here we adopt
 $S_{H}(T_k)=  [1+(k_B  T_k/E_{\circ})]^{-2}$  ($\rm  E_{\circ}/k_B\sim
 100$ K)  obtained by the  study of \cite{Buch1991} and  $\rm \gamma
_{H_2}\sim 1$. 
The dimensionless parameter
\begin{equation}
r_{\rm dis} \equiv \frac{G k_{\circ}}{n_e R_f}\Phi 
\end{equation}
 measures the  relative balance  of the $\rm H_2$  dissociation 
versus  the $\rm H_2$  formation, with $k_\circ=4 \times 10^{-11}  \rm{s}^{-1}$
being the (unshielded) $\rm H_2$ dissociation rate and $G$
the far-UV radiation field in Draine  field  units 
\citep[$2 \times 10^7$ photons s$^{-1}$ cm$^2$ between 11.2 and 13.6
eV,][]{Draine1978}.   
The dimensionless  factor  $\Phi$ is  an integral  of the
self-shielding  function over  the $\rm H_2$  column, which encompasses the 
details of $\rm H_2$  self-shielding \citep{Goldshmidt1995, Pelupessy2006}.
  For
a  detailed explanation  of the solution of Equation ~\ref{eq:dntrdt} within
our dynamical model  and key dependencies  of the $\rm  HI\rightarrow H_2$ phase 
transition the  reader  is refered  to \cite{Pelupessy2006}.  The numerical 
simulations presented in the  section \ref{sec:sims} use the solution to the fully 
time-dependent  Equation~\ref{eq:dntrdt}. For the moment we will consider the 
equilibirum solutions first to gain some qualitative
insight in the B-R relation.

\subsection{Equilibrium results: The B-R relation}
\label{sec:aneq}

\begin{figure*}
 \centering
 \epsscale{0.49}
 \plotone{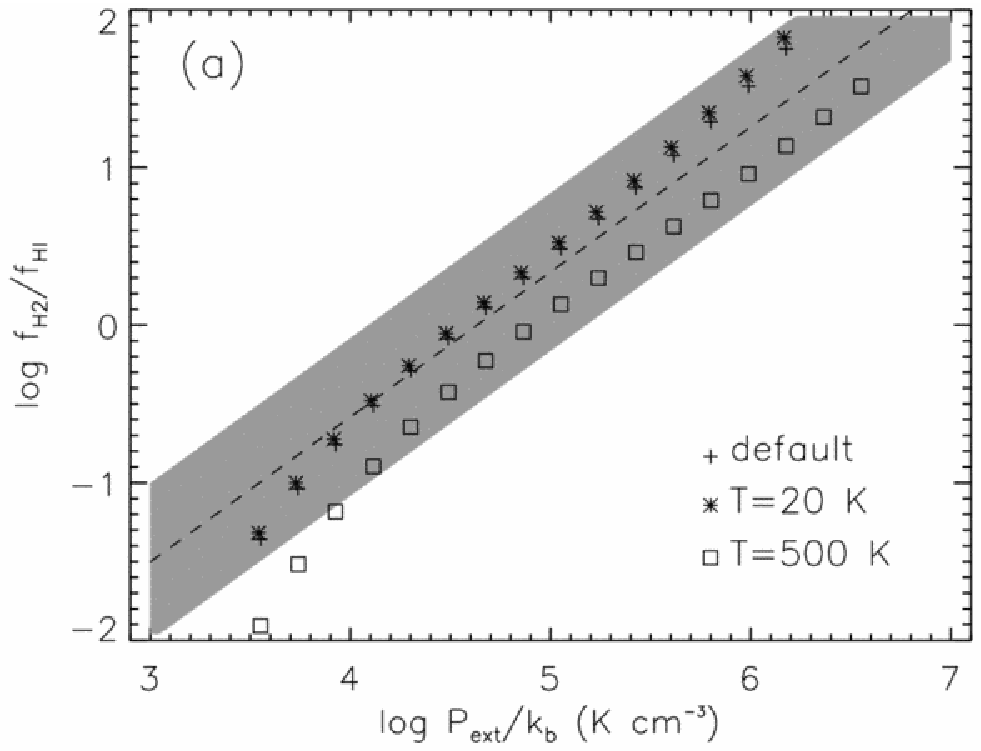}
 \plotone{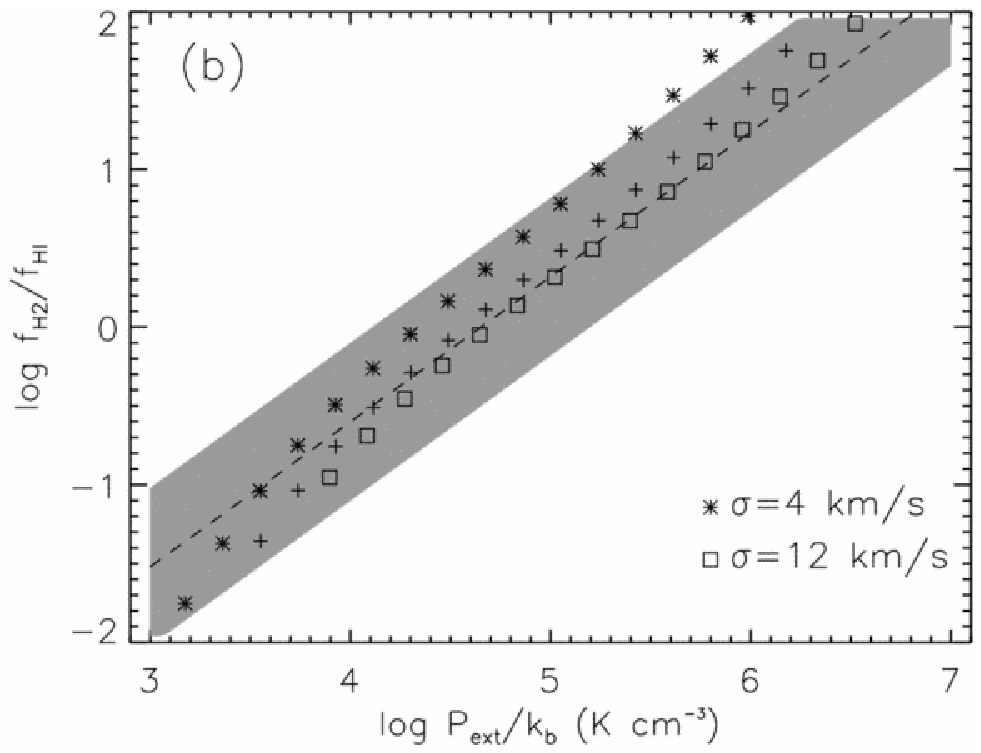}
 \plotone{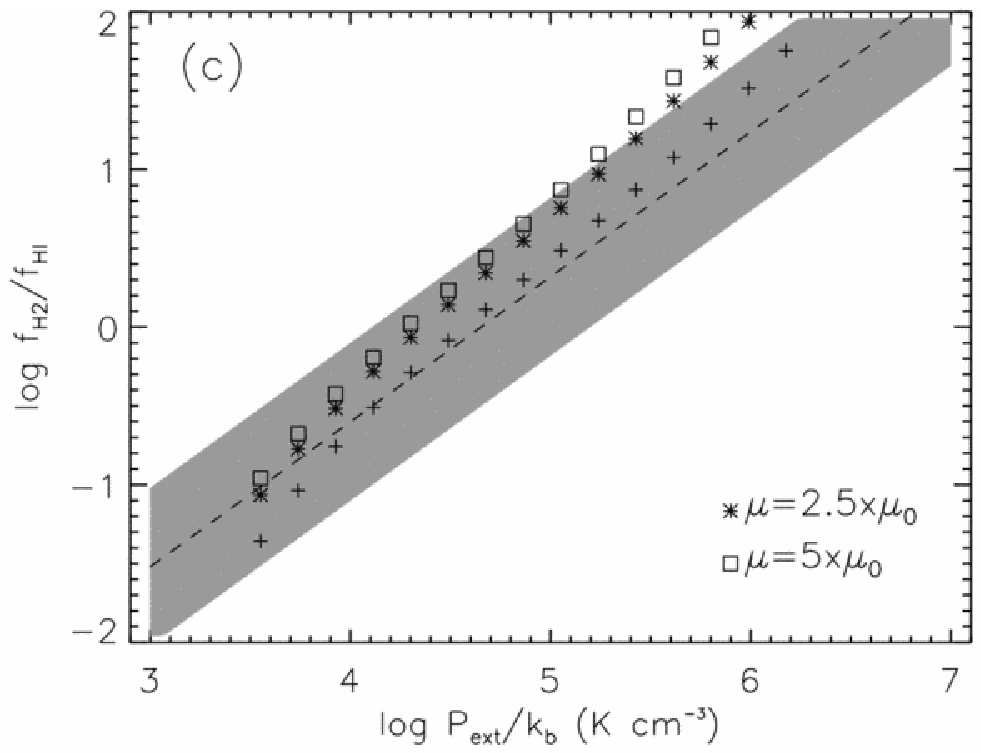}
 \plotone{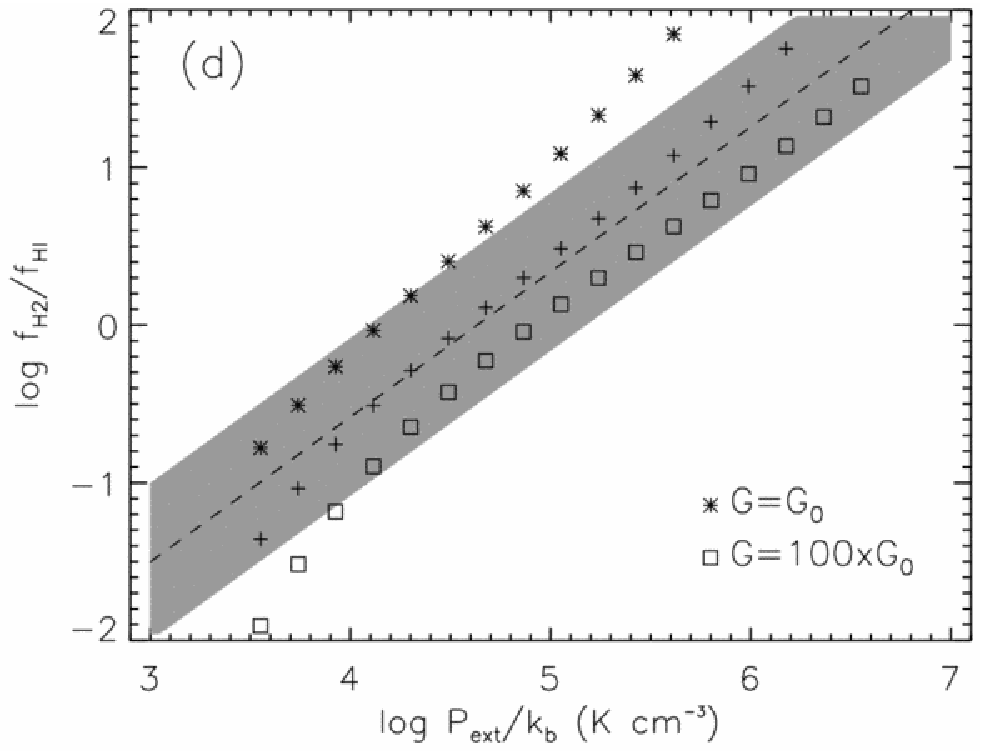}
 \plotone{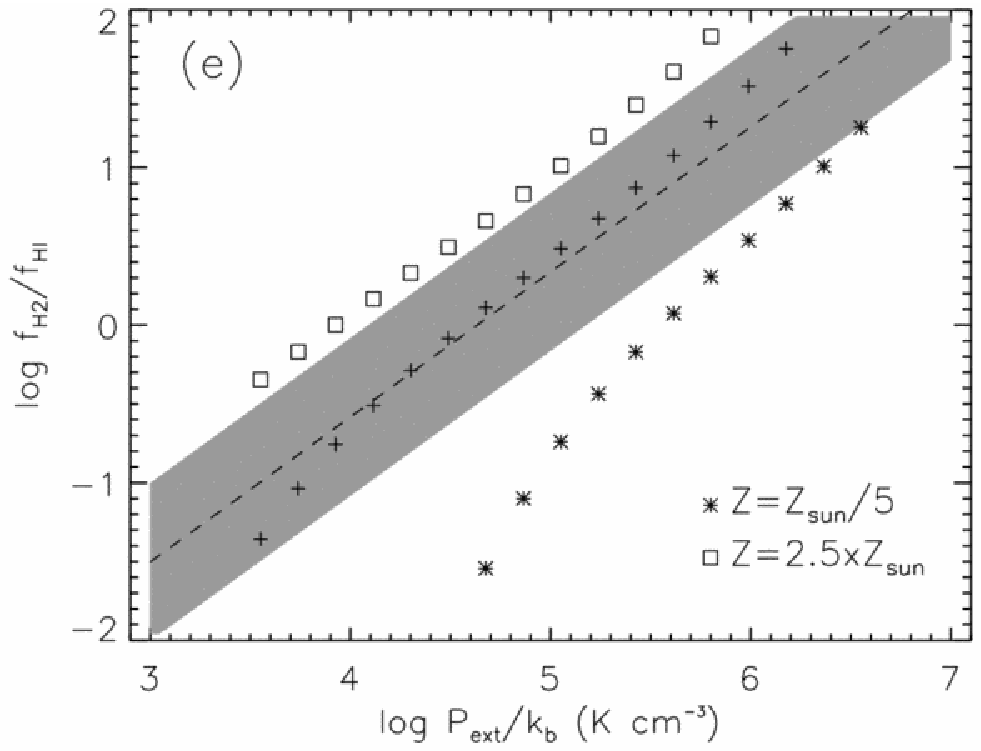}
 \plotone{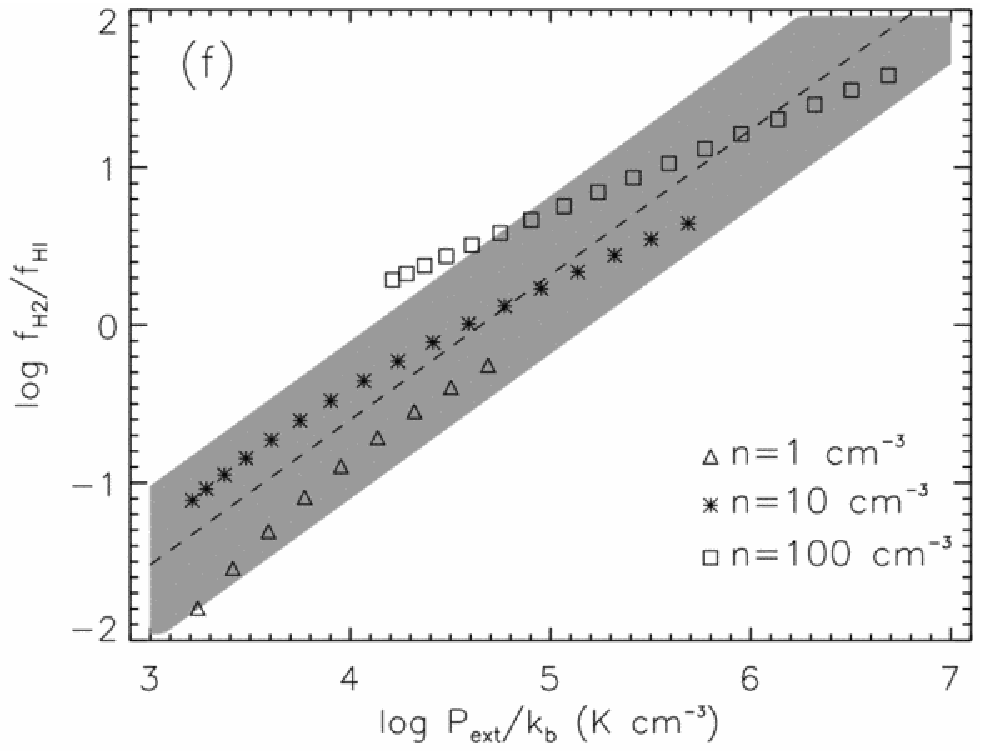}
 \caption{Equilibrium expectations for $f_{{\rm H}_2}/f_{\rm HI}$ (points) 
 for a range of physical parameters (section 2) versus the observed B-R relation
 and its scatter (dashed line and shaded area). In panels (a)-(e)
 the plus symbols are plotted with the same default set of parameters:
 $T=100$K, $\sigma=8$ km$/$s, $\mu=\mu_0$, $G=10\times G_0$ and $Z=\zsun$. }
 \label{fig:pext_an} 
\end{figure*}

The equilibrium transition column density for the fiducial case where the 
density, radiation field etc of a given patch of ISM is constant in time, 
is given by \citep{Pelupessy2006}, 
\begin{equation}
\label{eq:Ntr}
N_{tr}(HI)= \frac{\nu}{\sigma_{FUV}} 
\ln \left(1+ \frac{3}{2 \nu} \frac{G k_{0}}{R_f n} \Phi\right),
\end{equation}
where      $\nu=n       R      \sigma_{FUV}/\left(\frac{3}{2}+n      R
\sigma_{FUV}\right)$.      Together     with     Equations~\ref{eq:fm}
and~\ref{eq:nmean}  one can  calculate  the corresponding  equilibrium
molecular   to  neutral   ratio  $R_m   \equiv   f_{{\rm  H}_2}/f_{\rm
HI}=f_{{\rm H}_2}/(1-f_{{\rm  H}_2})$. The latter depends  only on the
following local  conditions of an  ISM gas parcel: the  density ($n$),
temperature  ($T_k$),  metallicity ($Z$)  and  impinging UV  radiation
field($G$) as well as the local turbulent velocity field with velocity
dispersion $\sigma$ through the total external pressure
\begin{equation}
\label{eq:pext}
P_e/k_B = n \left(T+54 \sigma^2 \right) {\rm K cm}^{-3}
\end{equation}
(needed  in Equation~\ref{eq:nmean}).  In  Figure~\ref{fig:pext_an} we
show the resulting $R_m$-$P_e$ relation and compare it to the observed
one  (Blitz \&  Rosolowsky  2006).  The  shaded  regions indicate  the
scatter  in the observational  B-R relation,  both for  the individual
measurements in a given galaxy  as well as between different galaxies.
For  interpreting the  plots  in Figure~\ref{fig:pext_an}  we need  to
consider the  following: a) the  observed B-R relation is  one between
the projected molecular-atomic  ratio and midplane pressure (estimated
from  projected quantities),  and thus  not  exactly the  same as  the
theoretical   points   in   Figure~\ref{fig:pext_an}  that   use  the
volume-averaged   local  $R_m$ ratio  and  pressure,  and b)  these
results correspond to  ISM equilibrium.  In practice the  ISM may not
be even in an  approximate equilibrium, especially for low density/low
pressure regions where the  timescales to equilibrium are the longest.
For the  moment we defer discussion of  projection and non-equilibrium
effects   to  the   investigation  of   realistic  galaxy   models  in
Section~\ref{sec:sims}.

A number of important points becomes apparent from the panels in 
Figure \ref{fig:pext_an} namely:  
\begin{itemize}

\item  for a  wide range  of parameters  a B-R  type of  relation does
emerge. Variations  in temperature (fig.~\ref{fig:pext_an}a), velocity
dispersion    (fig.~\ref{fig:pext_an}b),     and    radiation    field
(fig.~\ref{fig:pext_an}d)  as  well as  the  formation rate  parameter
$\mu$ (fig.~\ref{fig:pext_an}c)  have only a minor  effect. This shows
that a  (B-R)-type relation is  plausible from a theoretical  point of
view,  while  its  various  functional dependances  remain  within  its
 observational scatter expected within and between galaxies.

\item  Metallicity has a  more pronounced  effect on  the $\rm H_2$-pressure
relation.   Figure~\ref{fig:pext_an}e shows  that for  low metallicity
environments  the  theoretical relation  tends  to fall  significantly
below  and  outside  the  nominal  range, while  it  steepens  at  low
pressures.  Systematic studies of  the B-R relation in low metallicity
galaxies may  reveal such  deviations. Note however  that it  is still
possible  to shift  the theoretical  points  back to  the nominal  B-R
relation by assuming e.g.  a lower radiation field.

\item The  plotted relations in general have  slopes and normalization
similar  to  the observed  B-R  relation,  but  not necessarily  equal
(though remaining  mostly within the  expected observational scatter).
Introducing a  secondary dependence of  one of the other  variables on
pressure  can easily produce  exact matches  of the  B-R slope.   In a
stationary model  there is  no unique way  of doing this,  though. For
example, postulating a G dependency on pressure ($ G \propto P^{1/2}$)
or a  velocity dispersion dependency $\sigma \propto  P^{1/4}$ or some
suitable combination of  those will result in a  relation with a slope
close  to the  observed  value ($\sim  0.92$).   While such  secondary
relations are  plausible (e.g.  ISM environments  with large pressures
tend to  host more vigorous  star-formation and will thus  have higher
$G$'s), it is  uncertain whether they indeed emerge  in a more realistic
time-dependent galaxy-size model of gas and stars.

\item The pressure dependence  of $R_m$ expressed in the observational B-R 
relation  could  effectively   boil  down  solely  to  a  density dependence
given that midplane pressures observationally are estimated using   a 
constant   velocity  dispersion   (the   dominant  pressure contributor  in
the  CNM ISM).   Previous  work though  has deduced  a direct   
dependence   of    $\rm   R_m$    on   the    ISM   pressure
\citep{Elmegreen1993}.  To distinguish  between these possibilities in
Figure~\ref{fig:pext_an}f we investigate  the $\rm H_2$-pressure relation at
constant density (i.e.  varying the pressure only through the velocity
dispersion).  It  can be seen that  the B-R relation  is still present but 
tends to  flatten  (especially for  high  densities) to  $\propto P^{0.5}$.
Additional sources of pressure (magnetic fields, ram pressure, shocks) may 
be expected  to behave  similarly:  increasing $R_m$  but not  as strongly
as an increase in density would do it.

\end{itemize}

\noindent
Finally, unlike  the investigation of  B-R relation a similar  one for
the K-S relation must involve  the full dynamical treatment allowed by
our models given that in  our approach star formation is controlled by
the Jeans mass criterion (a  dynamical one) rather than any parametric
formulation.

\subsection{The CO model} 
\label{sec:CO}

Direct  detection  of $\rm H_2$  gas  is  difficult  given that  its  lowest
transition  $\rm  S(0):J_u-J_l=2-0$  at   28$\mu  $m  still  has  $\rm
E_{20}/k_B\sim 510\,K$,  much too high to be  substantially excited by
the typically  much colder $\rm H_2$ gas ($\rm  T_{k}\sim 15-60\,K$). This
fact, along  with its small Einstein  coefficient (A$_{20}=2.94 \times
10^{-11}\ s^{-1}$), diminishes its  luminosity, while at 28$\mu $m the
Earth's   atmosphere  is  mostly   opaque,  further   compounding  the
observational diffuculties of  its detection.  These difficulties made
the  next most  abundant molecule  after $\rm H_2$  itself, CO  (with $\rm
[CO/H_2]\sim 10^{-4}$ for Solar  metallicities) and its easily excited
rotational   lines   (mostly  CO   J=1--0   at   115\,GHz  with   $\rm
E_{10}/k_B\sim  5.5\,K$  and   $\rm  n_{crit}\sim  400\,cm^{-3}$)  the
molecular gas  tracer of choice via the  so-called CO-$\rm H_2$ conversion
factor \citep[e.g.]{Dickman1986,Solomon1987}
It must be noted that  {\it all} fundamental relations that involve
the $\rm H_2$ gas distribution in galaxies have been deduced for CO-bright
$\rm H_2$ gas. This may not encompass the bulk of the molecular gas
phase, especially in metal-poor and/or FUV-aggressive ISM environments
\citep{Maloney1988, Pak1998, Bolatto1999}.
 Such conditions  can be  found in  spiral disks  at  large galactocentric
distances  because  of well-known  metalicity  gradients
\citep[e.g.][and  references  therein]{Henry1998, Garnett1998}, as
well  as in  dwarf irregular  galaxies  \citep{Israel1997, Madden1997}.
Finally metal-poor  systems with  significant star  formation rates
(and thus strong FUV radiation fields) such  as Ly-break galaxies are
also known at high redshifts \citep{Steidel1999}.

 Thus in  our models it would  be instructive to  examine the specific
distribution   of  the   CO-rich,   conventionally  observable   $\rm H_2$
gas. Moreover, by tracking the  evolving ISM environment in which real
gas clouds are immersed, we can identify conditions regions and epochs
in   which  CO-bright   $\rm H_2$  gas   in  galaxies   may  significantly
underestimate its true distribution during the evolutionary track of a
galaxy. Including it in simulations allows a dynamic examination  of 
the CO-$\rm H_2$ relation,  contrasting  and  complementing   those  based  
on  static Photo-Dissociation Regions (PDRs) models 
\citep{Pak1998, Bolatto1999}.

For the CO model we adopt a similar model as for $\rm H_2$:
the size  of the CO dominated  region within a spherical  cloud can be
estimated by considering the width of the C$^+$ layer that surrounds a
mixed C$^0$, CO inner region
\citep[][and    references therein]{Papadopoulos2004}.  
 The dominant reaction channels for the
formation and destruction of $C^+$ that determine its layer are:
\begin{eqnarray}
C^{0}+\nu & \rightarrow & C^{+} + e^{-} \nonumber \\
C^{+} + e^{-} & \rightarrow & C^{0} + \nu  \\
C^{+} + H_{2} & \rightarrow  & C H_{2}^{+} + \nu \nonumber
\end{eqnarray}
Following \cite{Rollig2006},  the radius $r_+$ beyond  which the
gas  is $C^+$-dominated  in  spherical FUV-illuminated  clouds with  a
uniform density n can be estimated from
\begin{equation}
\label{eq:cco}
3\times 10^{-10} {\rm s}^{-1}\ G\ E_2[\xi _{FUV} A_v(r_+)] = 
n_{\rm H}\ \left(a_c X_c + 0.5\ k_c\right),
\end{equation}
where $X_c =  [C/H]= 1.4\times 10^{-4} Z$, the  factor $\xi _{FUV}\sim
2-4$ accounts for the absorption  at FUV wavelengths, and $E_2$ is the
second order exponential integral
\begin{equation}
E_2[\xi _{FUV} A_v(r_+)] = 
\int _1 ^{\infty} \frac{e^{-\mu \xi _{FUV} A_v(r_+)}}{\mu ^2} d\mu.
\end{equation}
The recombination and  radiative association rate coefficients  for the reactions: 
$C^+ +  e^- \rightarrow C + \nu$ and   $C^+  + H_2\rightarrow  CH^+ _2 +  \nu $  
 that destroy C$^+$ are 
$a_c=3\times 10^{-11}\, {\rm cm}^{3}\, {\rm s}^{-1}$ and 
$k_c=8\times 10^{-16}\,  {\rm cm}^3\,  {\rm s}^{-1}$, 
while the extinction $A_v$ from $r_+$ to the edge of the cloud R, given
by (again for a $n \propto r^{-1}$ density profile) 
\begin{equation}
\label{eq:cAv}
A_v(r_+)=  0.724\,  \sigma _{v}\, Z\, n\, R\, \ln\left(\frac{R}{r_+}\right).
\end{equation}
The product $n R$ can  be eliminated using the linewidth-size relation
Equation~\ref{eq:nmean}.  Equation~\ref{eq:cco}  is solved numerically
for $r_+$ by  simple root finding.  From this  we obtain the CO-bright
part  of  the $\rm H_2$  cloud  through  $f_{\rm  CO}=( r_+/R  )^2$. 
Given the fact that CO formation happens at much higher densities ($n \gtrsim
1000$), and correspondingly shorter timescales, than $\rm H_2$ formation,  
the stationary  treatment adopted  for the  CO chemistry  is appropiate.

\section{The dynamical model: gas+stars galaxy simulations}
\label{sec:sims}

\subsection{Simulation code}
\label{sec:code}

The code  calculates gravity using a TREE  code \citep{Barnes1986} and
gas dynamics using the SPH formalism \citep[see e.g.][]{Monaghan1992},
and the conservative  formulation of \cite{SpringelHernquist2002}.  An
advanced model  for the  ISM medium is  used, a star  formation recipe
based  on  a  Jeans   mass  criterion,  and  a  well-defined  feedback
prescription.   The  code  is   described  and  tested  in  detail  in
\cite{Pelupessy2005, Pelupessy2004},  below  we  will  only   give  a  brief
description of the relevant physical ingredients.

\subsubsection{ISM model}
\label{sec:ismmodel}

Our   ISM  model   is   similar,  albeit   simplified,   to  that   of
\cite{Wolfire1995, Wolfire2003}.   We solve for  the thermal evolution
of the gas including a  range of collisional cooling processes, cosmic
ray heating and ionization. From the viewpoint of our application here
the  most important  feature is  the tracking  of the  WNM and  CNM HI
phases. The latter is where  high densities and low temperatures allow
the  $\rm H_2$  molecules  to form  and  survive,  with  the $\rm H_2$  gas  phase
(Section~\ref{sec:model}) then  naturally completing the ISM treatment. 

The FUV  luminosities of  the stellar particles,  which are  needed to
calculate  the photoelectric  heating from  the local  FUV  field, are
derived  from Bruzual  \&  Charlot \citep[][and  updated]{Bruzual1993}
population  synthesis  models  for  a  Salpeter IMF  with  cutoffs  at
0.1~\Msun  and  100~\Msun.  In the present  work we do  not account for
dust extinction of UV light, other than that from the natal cloud: for
a young stellar  cluster we decrease the amount  of UV extinction from
75\% to 0\% in 4 Myr \citep[see][]{Parravano2003}.

\subsubsection{Implementation of the $\rm H_2$ model}
\label{sec:imp}

The gas particles  in the code are assigned, in  addition to the usual
density $\rho$,  internal energy $u$,  etc, a varying  local molecular
gas  fraction $f_m$.   We then  use Eq.~\ref{eq:dntrdt}  to  track the
evolution  of  $\sigma  N_{\rm  tr}$,  and  thus  $f_m$  (through  eq.
\ref{eq:fm}  and eq.   \ref{eq:nmean}), during  a  simulation timestep
$dt$.  If the  temperatures are too high for  $\rm H_2$ formation to occur
we  solve for the  evolution of  $f_{m}$ using  pure photo-destruction
while  for $\rm  T_k>3000\ K$,  we treat  the  collisional destruction
process   of   the   remnant    molecular   gas   using   rates   from
\cite{Martin1998}.

The  density that enters in those equations  is assumed  to be  the mean
density $\langle n  \rangle$ given by the SPH  density at the particle
position,  and the  temperature the  particle temperature  (both taken
constant during the timestep).  The radiation field is calculated from
the  distribution of stars,  assuming no  other extinction  apart from
that in the  natal clouds. For the macroscopic  pressure $P_e$ we need
the local velocity  dispersion $\sigma$.  For this we  take the formal
SPH estimate
\begin{equation}
\sigma_j^2 = \sum_i \frac{m_i}{\rho_j} (v_i-\langle v \rangle_j)^2 W( |r_{i j}|, h_j) 
\end{equation}
with $v_i$ and $m_i$ the particle velocities and masses, 
$\langle  v \rangle_j$  the local bulk velocity.

\subsection{Star formation: SN feedback, and $\rm H_2$ as an additional SF regulator}
\label{sec:sfcrit}

The coldest  and densest gas  in our simulations  is found in  the CNM
phase with the  $\rm H_2$ formation occuring on the formation timescales
of the Giant molecular clouds (GMC) complexes  embedded in this phase.
Following  gravitational  instabilities further  ``down''  in the  CNM
phase would require additional physics (e.g.  CO and $\rm H_2$O cooling of
dense  molecular cores,  the emergence  of the  IMF, etc)  as  well as
demanding  much higher  numerical resolution,  currently unattainable.
At this point it is appropriate to introduce a prescription to further
track the star formation process.

The first assumption we make is that star formation is governed by the
gravitational  unstability of  gas  clouds, with  a region  considered
unstable to star formation if  its local Jeans mass $\rm M_J<M_{ref}$,
where $\rm M_{ref} \approx 10^{4-5}  \msun$ is a reference cloud mass.
The    exact   value    of    $\rm   M_{ref}$    is   not    important
\citep{Gerritsen1997}, and provided that it is always well-resolved by
our simulation  this star-formation criterion  precludes the emergence
of numerical artifacts that can result from insufficient resolution of
the Jeans  mass. Moreover  an $\rm M_{ref}$  smaller than  typical GMC
masses  (as  the  chosen  values  are)  makes  the  Jeans  instability
criterion select conditions  ``deep'' into the dense parts  of the CNM
phase.  This makes  this criterion a good assumption  for the onset of
star formation, mirroring the irreversibility of the (gas)$\rightarrow
$(stars)  transition observed in  nature once  dense CNM  clouds form.
Once a region is deemed unstable it proceeds towards star formation by
converting some fraction $\epsilon_{SF}$  of the gas particle to stars
after a  delay time.  This  delay is taken  to be proportional  to the
local    free    fall     time:    $\rm    \tau_{sf}=f_{sf}    t_{ff}=
\frac{f_{sf}}{\sqrt{4 \pi G \rho}}$.  The delay factor $\rm f_{sf}$ is
uncertain, but from observations a value $\rm f_{sf} \approx 10$ seems
 necessary to account for the observed inefficiency of star formation
\citep{Zuckerman1974}.   The   actual rate  of  star   formation  is  then
determined  by balance  between  gas  cooling and  the  far-UV and  SN
heating.  We  will refer  to this star  formation model as  the Simple
Delay (SD) model.

Our ISM model allows to set  the local $\rm H_2$ gas mass fraction
as  star formation  regulator in  the dynamical
setting of  an evolving galaxy.  Irrespective  whether $\rm H_2$ formation
ahead of star formation is  incidental (e.g.  cold and dense gas forms
$\rm H_2$  on its  ``way''  to gravitational  collapse  and eventual  star
formation) or instrumental (e.g.  $\rm H_2$  {\it must} form first so that
CO and  other powerful molecular  coolants can form and  ``drive'' the
gravitational collapse further towards denser and colder ISM regimes),
this is  a very important step  towards a better,  much more realistic
rendering  of the  star  formation process  in  numerical models.   We
implement this  molecular regulated (MR) star  formation by converting
the  molecular ($\rm  f_m$) mass  fraction of  an unstable  (i.e.  $\rm
M_J<M_{ref}$) gas  particle to stars  (with a minimum value of 
$\rm  f_m=0.125$, corresponding to a particle mass of $\sim 60 \msun$, 
to avoid the creation  of very small star particles).  Unlike the
(SD) recipe  that needs an  adhoc $\epsilon_{SF}$ value, the  (MR) one
contains a physical basis for this part of the star formation modeling
and   thus   no   longer  needs   a  star   formation  efficiency
parameter $\epsilon_{SF}$.

The mechanical energy output of  stars is reasonably well known but it
has been proven difficult to include its feedback (i.e. supernovae and
stellar winds) self-consistently in galaxy-sized ISM simulations.  The
reason for this is that  the effective energy of such feedback depends
sensitively  on the  energy radiated  away in  thin shells  around the
bubbles  created, which  would need  prohibitively high  resolution to
follow.   In SPH  codes there  have  been conventionally  two ways  to
account  for feedback:  by changing  the thermal  energy input  and by
acting  on  particle  velocities.   Both are  unsatisfactory,  as  the
thermal  method  suffers  from  overcooling \citep{Katz1992}  and  the
kinetic   method   is   too    efficient   in   disturbing   the   ISM
\citep{Navarro1993}.  Here  we use a  method based on the  creation of
{\it pressure particles} that act as normal SPH particles in the limit
of the particle mass $m_{\rm p}$ going to zero \citep{Pelupessy2004,Pelupessy2005}. 
 Such a pressure particles is associated with every newly formed star 
particle and will receive its feedback energy, acting on the surrounding 
gas particles through the usual SPH particle forces in the limit that
$m_{p} \rightarrow 0$ while simultaneously keeping the product of 
particle mass and specific thermal energy, $m_{p} \times u_{p}$, 
fixed. The thermal evolution (the time dependence of $m_{p} \times u_{p}$)
is specified by adiabatic expansion  and the energy
input  from young  stars.   For  this energy  injection  rate we  take
$\dot{E}=\epsilon_{sn} n_{sn} E_{sn}/ \Delta t$, with $E_{sn}=10^{51}$
erg,  $\epsilon_{sn}=.1$, $n_{sn}=0.009$  per \Msun  ~and  $\Delta t=3
\times 10^7$ yr. The efficiency $\epsilon_{sn}$ thus assumes that 90\%
of the energy is radiated away in thin, dense shells.

\subsection{Galaxy models}
\label{sec:models}

The galaxy models we use stem  from the analytic disc galaxy models of
\cite{Mo1998}, constructed as  described in \cite{Springel2005}.  They
consist  of a disk  consisting of  a stellar  and a  gaseous component
embedded  in a dark  halo. The  stellar disk  has an  exponential disk
radial profile (with scale length $R_{\star}$),
\begin{equation}
\label{eq:sech2}
\rho_{disk}(R,z) = \frac{\Sigma_0}{2 h_z} \exp(-R/R_{\star}) {\rm sech}^2(z/h_z).
\end{equation}
The gas disk is set up in vertical hydrostatic equilibrium with a surface 
density profile consisting of an exponential component (so proportional 
to the stellar density) and a more extended component
\begin{equation}
\label{eq:siggas}
 \rm  \Sigma= \Sigma_g/(1+R/R_{g})
\end{equation}
 cutoff at a radius $8 \times R_{g}$. Note that this distribution is
 necessary to match closer the observed gas distributions that typically 
 extent well beyond the stellar ones.  Apart from the radial profiles 
 the stellar and gas disk are initialized as smooth initial conditions and the 
 gas is setup with a constant temperature (8000K).
 Finally the dark halo has an Hernquist profile 
\begin{equation}
\label{eq:hqprof}
\rho_{\rm halo}(r)=\frac{M_{\rm halo}}{2 \pi}\frac{a}{r (r+a)^3}.
\end{equation}
The Hernquist  scale parameter is  related to the more  familiar scale
parameter $r_s$ and the concentration  index $c$ of Navarro, Frenk and
White (NFW) profiles \citep[see][for details]{Springel2005}. We do not
include a bulge component here.

We take models of different size by choosing the total baryonic mass,
the disk being  a mass fraction $f_{\rm baryon}$ of the  total mass, and consider
galaxies ranging in mass  from $M_{\rm baryon}=10^8 \msun$ to $M_{\rm baryon}=10^9
\msun$,  with $f_{\rm baryon}=0.041$.   The  smaller scale  is representative  of
dwarf galaxies and the bigger of  small disk galaxies. We vary the gas
mass fraction of  the disk from $f_{\rm gas}=0.5$ (for the  low mass model) to
$f_{\rm gas}=0.2$ (for  the high mass  model), roughly mirroring  the observed
correlation  between gas  fraction and  galaxy size.   The  total halo
virial mass is fixed by $M_{\rm vir}=M_{\rm baryon}/f_d$ ($M_{\rm vir}=M_{\rm
halo}+M_{\rm baryon}$), which gives the virial
velocity and radius through the~relations
\begin{eqnarray}
M_{\rm vir} & = & \frac{v_{\rm vir}^3}{10 G H(z)},\label{eq:vir1} \\
R_{\rm vir} & = & \frac{v_{\rm vir}}{10 G H(z)},\label{eq:vir2} \\
\end{eqnarray}
assuming a virial overdensity  $\Delta=200$. Determining the $c$ index
gives the  halo scale  length. The metallicity  of each model  will be
taken to  be constant during the  run, but we  will consider different
metallicities, namely models at  solar metallicity $\zsun$ and at $0.2
\times \zsun$.  These models  are summarized in Table \ref{tab:models}
as models  A1 to C1.   Other parameters of  the models are  not varied
here:  we fix  the  spin parameter  at  $\lambda=0.05$. The  $\lambda$
implicitly determines the  scale lengths of the gas  and stellar disk,
as  the angular  momentum in  the disk  is assumed  to scale  with the
angular momentum in  the parent halo. The scale  height of the stellar
component  is taken  to  be a  fraction  of the  radial scale  length:
$z_h=0.2-0.4 \times  R_{\star}$. The  galaxy models are  realized with
mass resolutions for the gas particles of $200-10^3$ \Msun, also given
in Table~\ref{tab:models}.  Finally two models with very high gas mass
fractions  are also  run  (D1 and  E1),  representing extreme  systems
expected  at early epochs  of galaxy  evolution. For  these we  set an
initial gas  mass fraction of  $f_{\rm gas}=0.99$.

\begin{table*}
 \caption[]{Overview  of  galaxy  model  parameters. The  letters  A-E
 indicate different  structural properties and/or  metallicities while
 numbering  indicates different resolutions used on  otherwise identical
 models. The  gas distributions  of models D1  and E1 consist  of equal
 mass exponential  and extended disk, the other  models consist purely
 of the extended distribution eq.~\ref{eq:siggas}} \centering
  \label{tab:models}
  \begin{tabular}{ c | r  r  r  r  r   }
  \hline
  \hline
  Model & $M_{\rm baryon}$ & $M_{\rm vir}$          & $f_{\rm gas}$ & $m_{par}$ & $Z$  \\
  \hline
  A1    & $10^8$     & $2.3\times10^9$     & 0.5 & 200  & 0.2 \\
  A2    & $10^8$     & $2.3\times10^9$     & 0.5 & 500  & 0.2 \\
  B1    & $10^9$     & $2.3\times10^{10}$  & 0.2 & 500  & 0.2 \\
  B2    & $10^9$     & $2.3\times10^{10}$  & 0.2 & 1000 & 0.2 \\
  C1    & $10^9$     & $2.3\times10^{10}$  & 0.2 & 500  & 1. \\
  D1    & $10^{9}$   & $2.3\times10^{11}$  & .99 & 1000 & 1. \\
  E1    & $10^{9}$   & $2.3\times10^{11}$  & .99 & 1000 & 0.2 \\
  \hline
\end{tabular}
\end{table*}

\subsection{Runs}
\label{sec:runs}

In  addition to different  galaxy models  A1-E1 we  also test  our two
different   star  formation   recipes   (MR  and   SD)  described   in
Section~\ref{sec:sfcrit}.   Each model  is  run well  beyond the  time
strong  evolutionary effects take place  (investigated in  section
\ref{sec:disc}) and until a dynamic  equilibrium for the star formation  sets
in.  At this evolutionary point,  i.e.  $\sim 1$  Gyr of simulation time 
after the start of  the simulation, we analyze the  resulting gas
distributions. Given  that at  present chemical  enrichment effects 
(influencing ISM thermodynamics  and $\rm H_2$  formation) and cosmological
infall are  not included,  evolving our models for  much longer  timescales
is of  limited value, simply resulting in a  steady  depletion  of  their 
gas reservoirs.

\subsection{Results}
\label{sec:res}

Table~\ref{tab:result}  gives   an  overview  of   the  molecular  gas
fractions and star formation rates of our runs. From this table it can
be  seen that  the  low metallicity  models  ($Z=\zsun/5$, A  B and  E
models)  have a  molecular fraction  $f_m \sim  0.03-0.05$,  while the
models at  solar metallicity (C  and D models)  reach up to  $f_m \sim
0.4-0.6$.  Models  with the  same metallicity reach  similar molecular
fractions while the structural parameters of the galaxy models seem to
have only a minor influence on the global molecular fraction, at least
over the limited  range explored here.  It is  also important to point
out that the SPH particle mass has little effect on the basic physical
quantities   examined  here,   suggesting   that  adequate   numerical
resolution  has  been  reached  to describe  the  physical  mechanisms
considered.

 A comparison of SD and MR star formation recipes shows that the latter
 results in less  molecular gas and increased SFRs. A useful measure of the
 star formation efficiency is the gas consumption timescale $\tau_x=M_x  /
 SFR$, also  given in Table~\ref{tab:result}, and calculated separately for
 atomic and molecular hydrogen.  For the low metallicity models we find
 typically $\tau_{HI} \sim 5-8$ Gyr and $\tau_{H_2} \sim 0.15-0.3$ Gyr,
 while for the high metallicity models the timescales for HI and $\rm H_2$ become
 comparable,  with  $\tau_{HI}\sim  1.5-3$ Gyr and  $\tau  _{H_2}\sim 
 0.6-5$   Gyr.   These  results  seem  largely independent  of   the  SF 
 model   adopted  and  reflect   a  general characteristic of low-Z versus 
 high-Z systems, namely that the former are much more WNM-dominated than 
 the latter. This can be seen in the last column of Table~\ref{tab:result},
 where the $\rm M_{CNM}/M_{WNM}$ ratio is also tabulated (CNM is taken  to
 be all gas colder than 1000 K, WNM gas with $1000<T<10000$). 
  For systems  that are  WNM-dominated,
 hydrogen will  be overwhelmingly atomic,  and  the  large  disparity 
 between the  HI  and  $\rm H_2$  gas consumption  timescales simply  reflects
 the  one between  atomic and molecular gas reservoirs and the fact no
 Jeans-unstable regions occur in  the  WNM  phase  and  thus  star 
 formation  can  never  directly ``consume''  its  mass. For  high 
 metalicities  the mass  allocation between WNM HI and CNM HI and  $\rm H_2$
 becomes more even and so are the corresponding consumption timescales. Note
 that for the SD model alone  one might be tempted to conclude that  the
 short $H_2$ gas consumption timescale at low $Z$ (or equivalently the  low
 $f_m$) has something to do with the fact this model is formulated 
 independent of molecular gas (thus unrealistically converting gas into 
 stars before molecules can form).  This is not the case: the MR model has 
 \emph{lower} consumption timescales at low $Z$ (partly due to a general
 higher SF),  and lower molecular fractions. It seems that even in the MR
 model the  outcome of the star formation model is not constrained by the
 chemistry of $\rm H_2$ formation but by the conversion of WNM to CNM.   The
 difference between high metallicity gas and low metallicity gas is  that 
 molecular gas forms in smaller reservoir of CNM gas. Once there, evolution
 to star formation occurs faster than  the large scale processes driving gas
 down to the CNM, so   it is not necessary for a large reservoir of $\rm H_2$ to
 form.

 Compared with the SD model the MR star formation model has a 
$\sim 50\%$ lower molecular fraction and a $\sim 50\%$ higher rate of SF.
This increase in efficiency of SF in terms of its molecular mass
means that the $\rm H_2$ gas consumption timescale is a factor 
$\sim 3-4$ times shorter. The reason for the smaller amount of $\rm H_2$ 
gas is that the MR recipe selects SF sites that are on average denser 
CNM regions (where $\rm H_2$  forms), and that these regions are then 
converted to stars, resulting in a more efficient consumption of 
molecular gas. 
At  least globally,  M(HI+$\rm H_2$)/SFR  seems  relatively 
insensitive to  the  SF recipe chosen.  Given that stars  form
unequivocally only out  of the $\rm H_2$-rich  regions   of  the  CNM   phase 
this  suggests   that  any self-regulating  mechanism responsible  for 
distributing the  gas between the SF  and the non-SF phase remains  broadly
similar in these two SF recipes. Of course we must reiterate that the MR
star formation recipe is the more physical of the two, doing away with the
adhoc $\rm \epsilon_{SF}$ parameter  typically used  in numerical 
models.  The emergence of a robust global efficiency M(HI+$\rm H_2$)/SFR out of
dynamic galaxy models where only cold, dense, and $\rm H_2$-rich gas is allowed
to form  stars  confirms  the  trustworthiness  of (K-S)-type 
phenomenological relations.  This  is because  the latter often relate the
total gas mass to star formation, irrespective of its thermodynamic  state 
or  molecular  gas fraction.

\begin{table*}
 \caption[]{Results: molecular masses and fractions for runs. } 
 \centering
  \label{tab:result}
  \begin{tabular}{ c c | r  r  r  r  r  r  r  r  r }
  \hline
  \hline
  Run & SF    &  $M_{\rm gas}$ &  $M_{HI}$      &  $M_{H_2}$     & $f_{H_2}$ & $R$ & $SFR$      & $\tau_{HI}$ & $\tau_{H_2}$ & $\frac{f_{CNM}}{f_{WNM}}$ \\
      & model & ($10^7 \msun$) & ($10^7 \msun$) & ($10^7 \msun$) &           &     & (\Msun/yr) & ($10^9$ yr) & ($10^9$ yr) & \\
  \hline
  A1  & SD & 4.5 & 4.0 & 0.2  & 0.045 & 0.05  & 0.005 & 5.8 & 0.3 & 0.64\\
      & MR & 4.4 & 4.1 & 0.1  & 0.023 & 0.024 & 0.006 & 4.7 & 0.11 & 0.41\\
  A2  & SD & 4.4 & 4.0 & 0.21 & 0.047 & 0.052 & 0.005 & 6.0 & 0.31 & 0.64\\
      & MR & 4.2 & 3.8 & 0.1  & 0.024 & 0.027 & 0.006 & 4.7 & 0.13 & 0.4\\
  B1  & SD & 19.0 & 17. & 0.86 & 0.045 & 0.049 & 0.01 & 11.8 & 0.58 & 0.61\\
      & MR & 19.0 & 17. & 0.4  & 0.02  & 0.022 & 0.015 & 7.8 & 0.17 & 0.39\\
  B2  & SD & 19.1 & 17. & 0.84  & 0.044 & 0.048 & 0.009 & 12.3 & 0.59 & 0.59\\
      & MR & 18.6 & 17. & 0.34 & 0.02  & 0.02 & 0.015 & 8.4 & 0.16 & 0.38\\
  C1  & SD & 18.6 & 6.7 & 11. & 0.61 & 1.7  & 0.014 & 3.2 & 5.4 & 5.28\\
      & MR & 17.1 & 11. & 5.4 & 0.3  & 0.49 & 0.031 & 2.5 & 1.2 & 1.81\\
  D1  & SD & 88. & 41. & 46. & 0.52 & 1.1 & 0.14 & 2.0 & 2.2 & 3.06\\
      & MR & 76. & 53. & 20. & 0.27 & 0.4 & 0.24 & 1.6 & 0.6 & 1.32\\
  E1  & SD & 90. & 83. & 4.  & 0.045 & 0.05 & 0.1  & 4.9 & 0.24 & 0.59\\
      & MR & 89. & 84. & 1.8 & 0.02  & 0.02 & 0.17 & 3.5 & 0.08 & 0.34\\
  \hline
\end{tabular}
\end{table*}

\begin{figure*}
 \centering
 \epsscale{0.16}
 \plotone{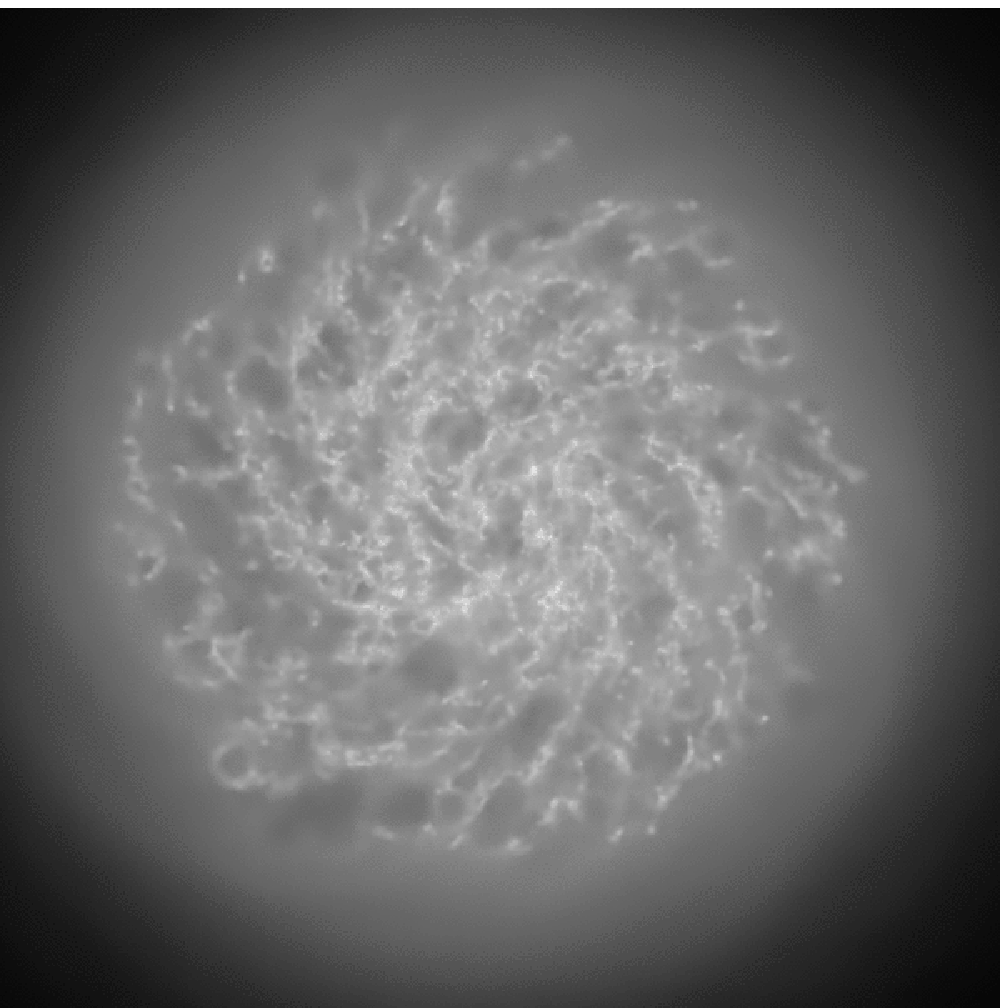}
 \plotone{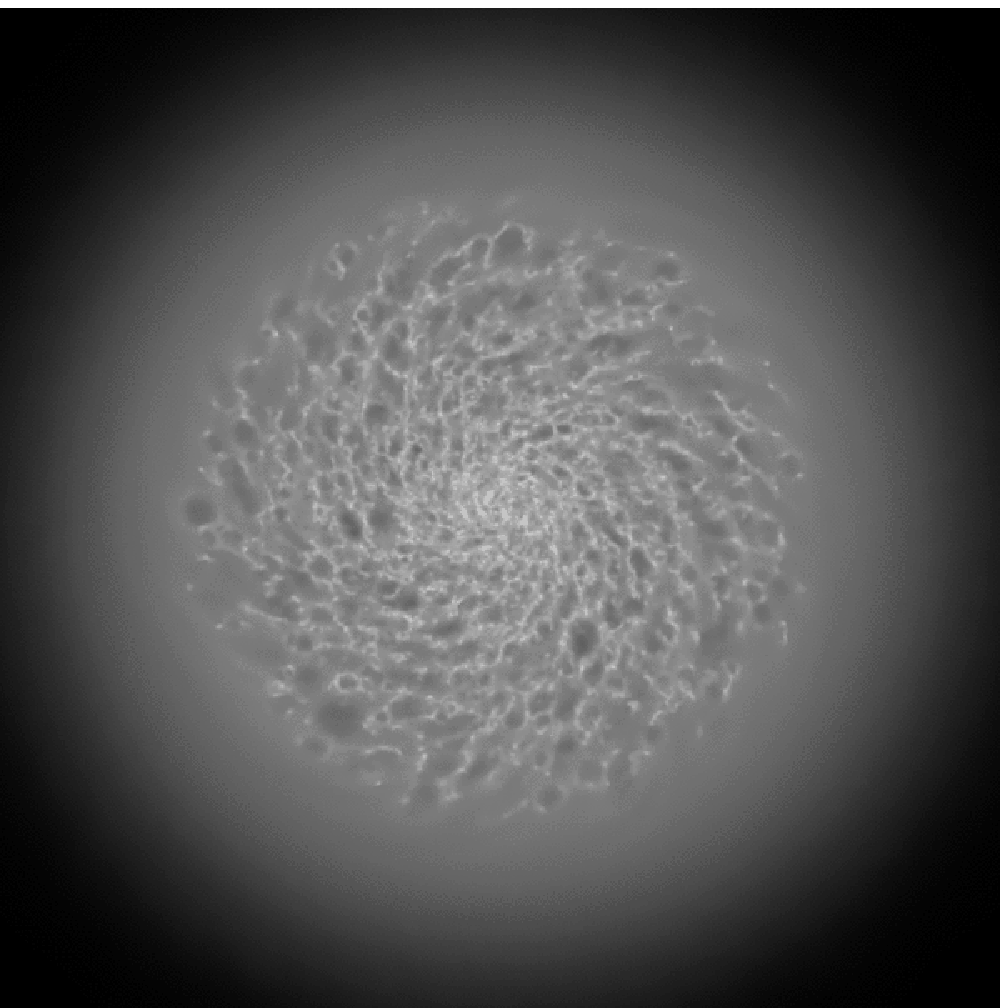}
 \plotone{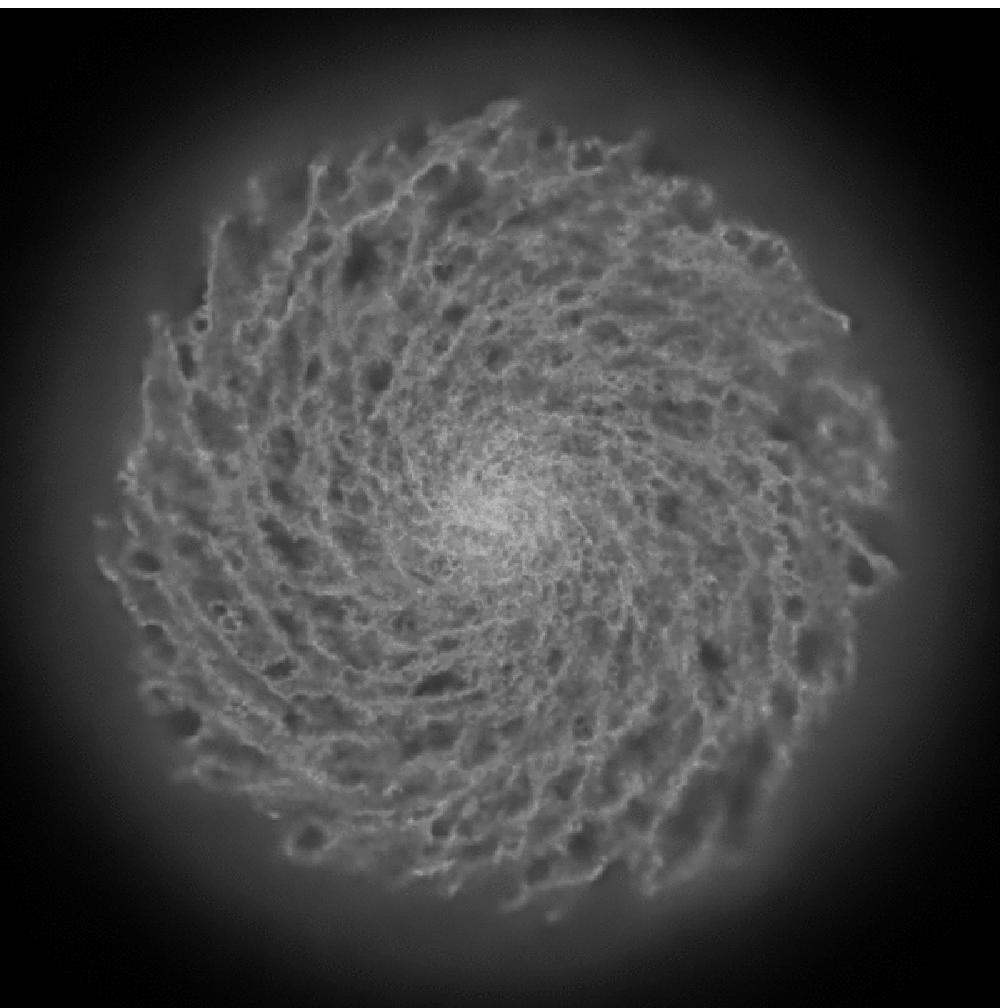}
 \plotone{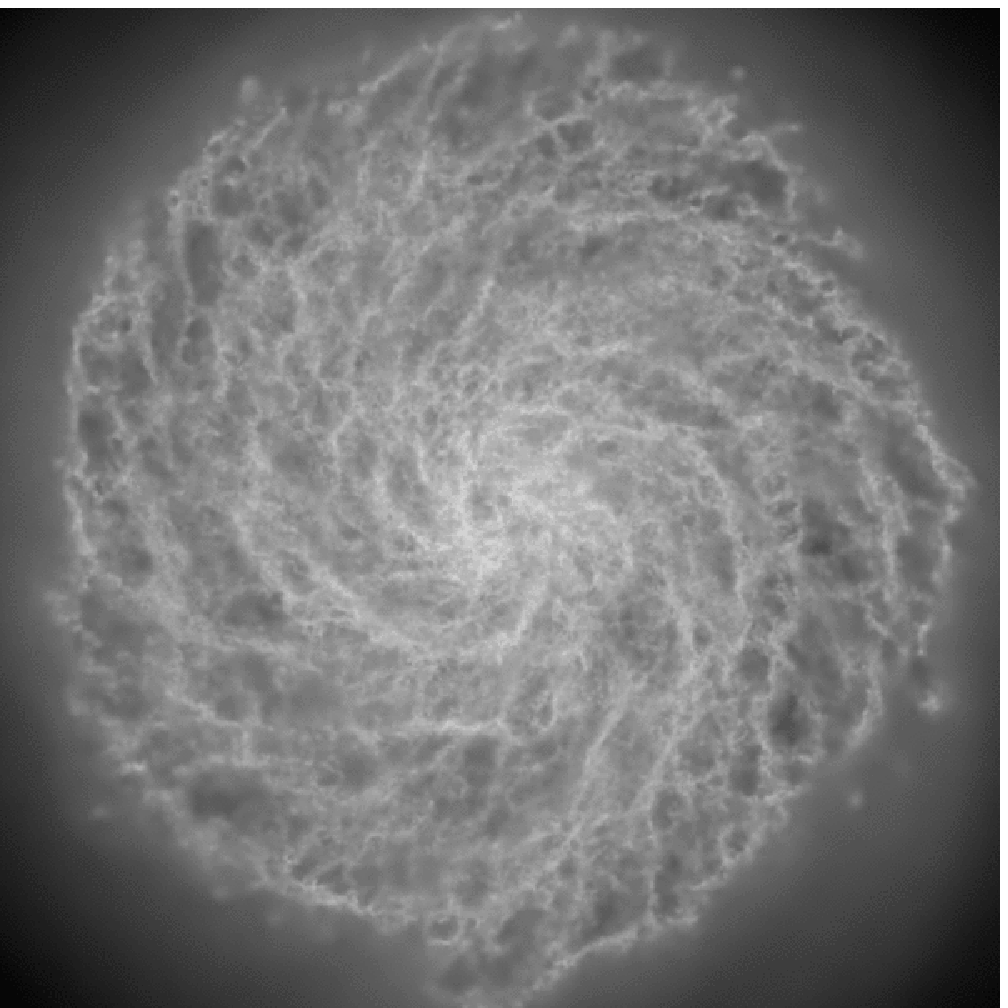}
 \plotone{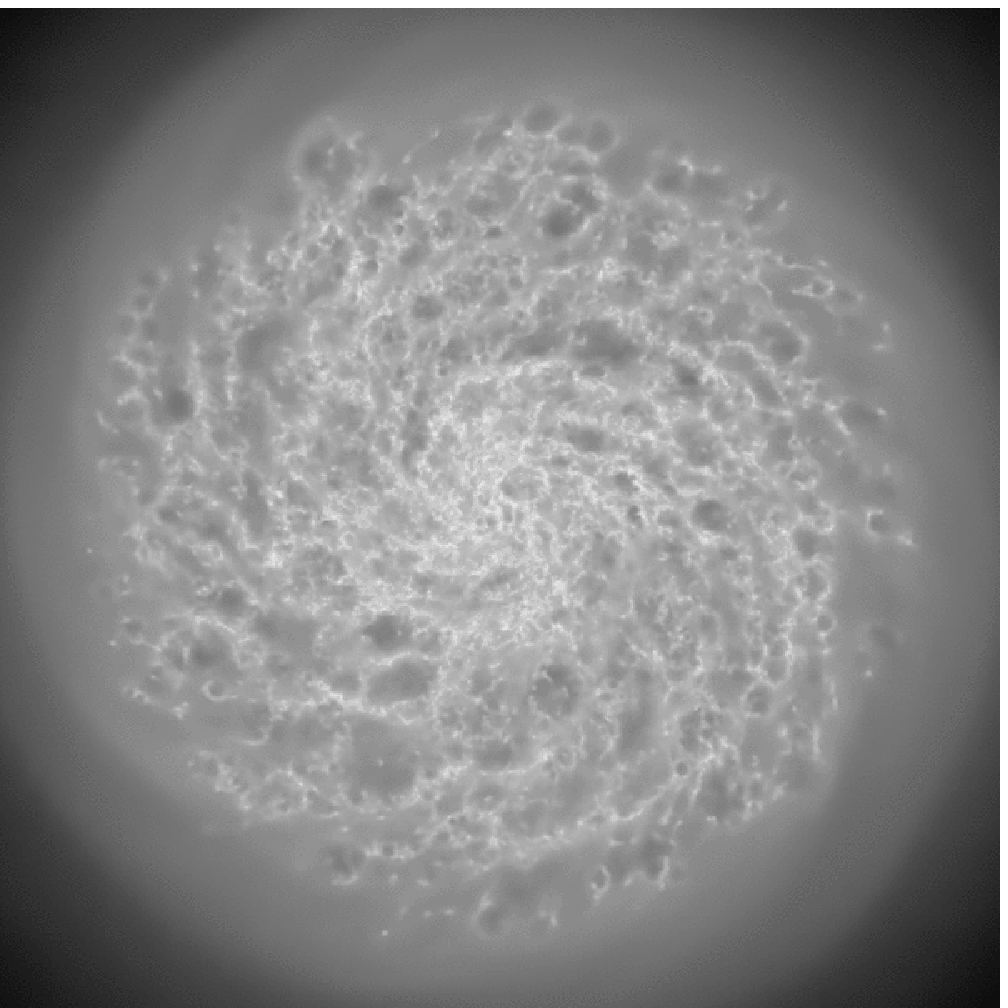}
 \epsscale{0.062}
 \plotone{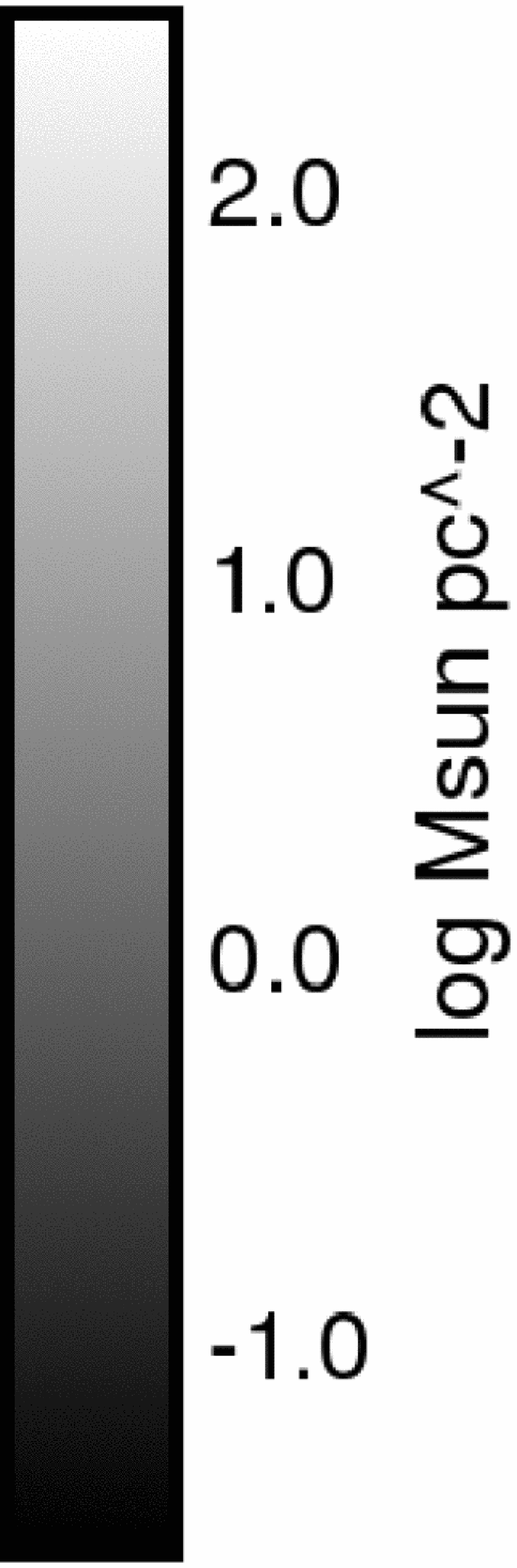}
 \epsscale{0.16}

 \plotone{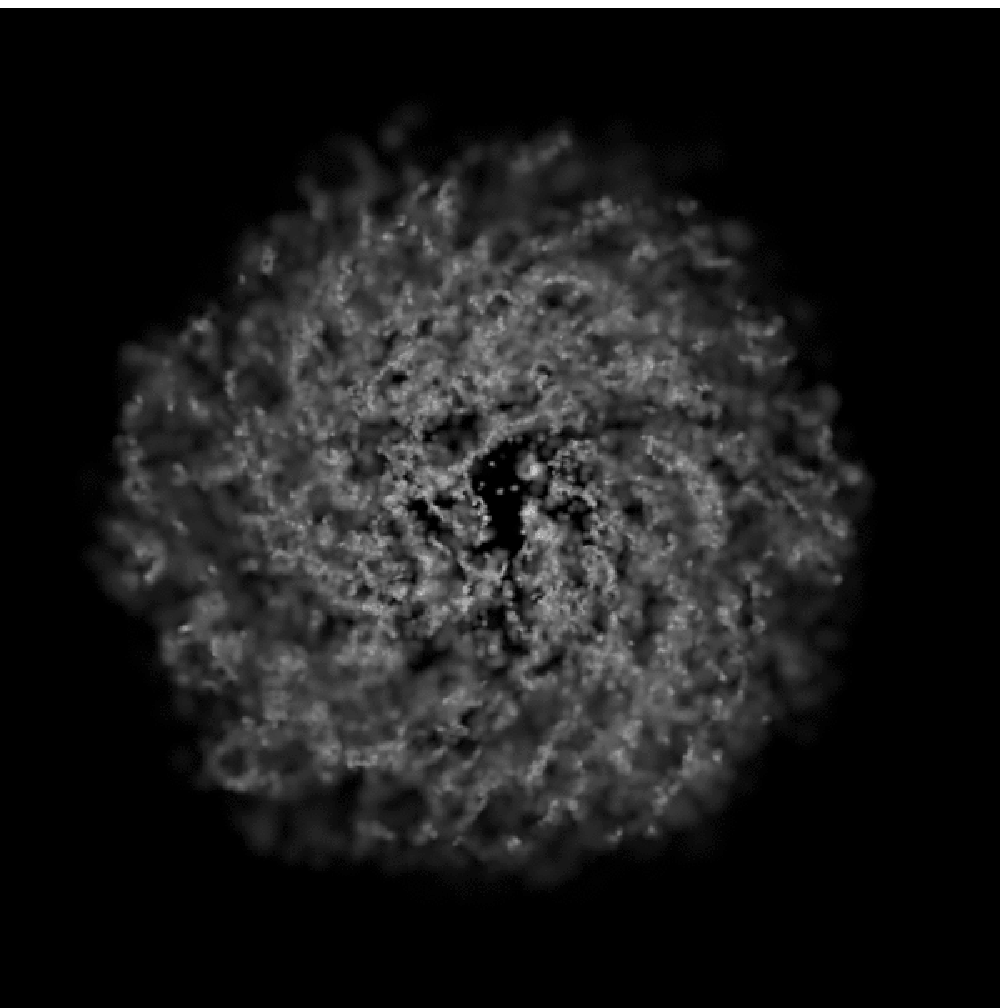}
 \plotone{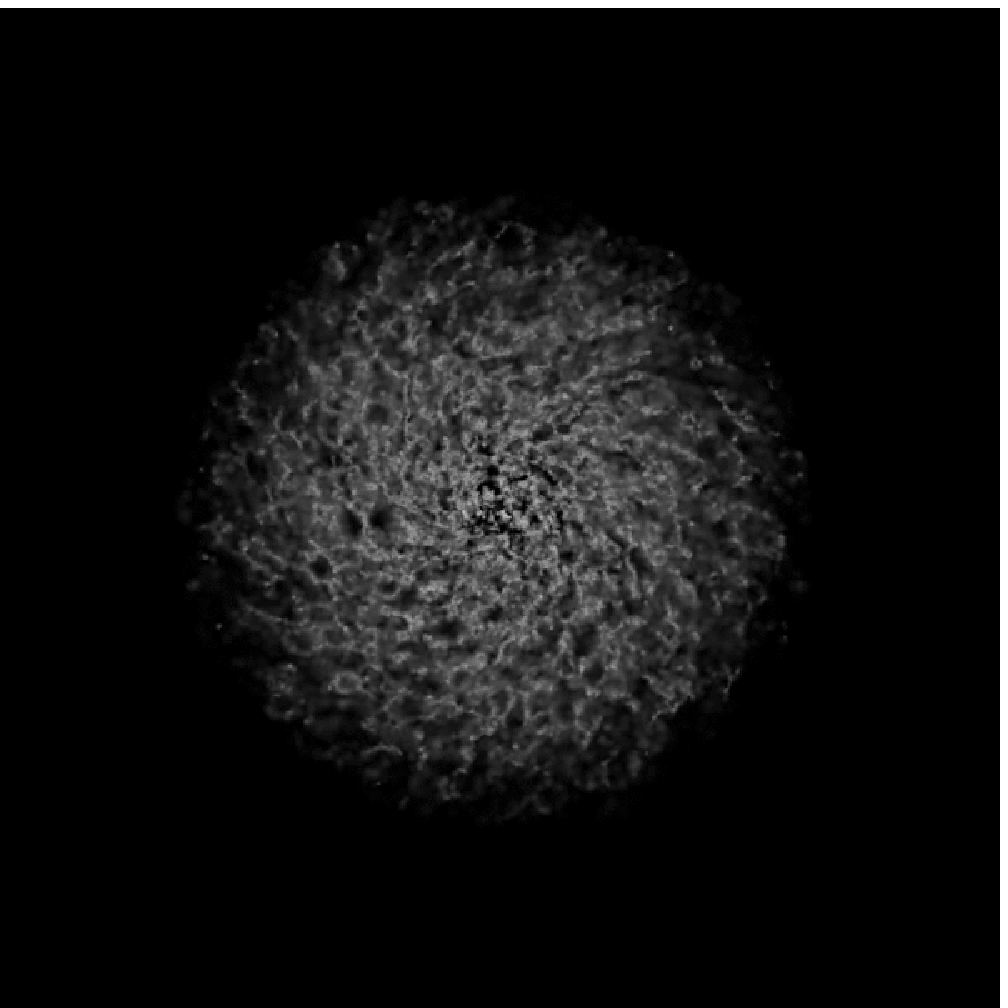}
 \plotone{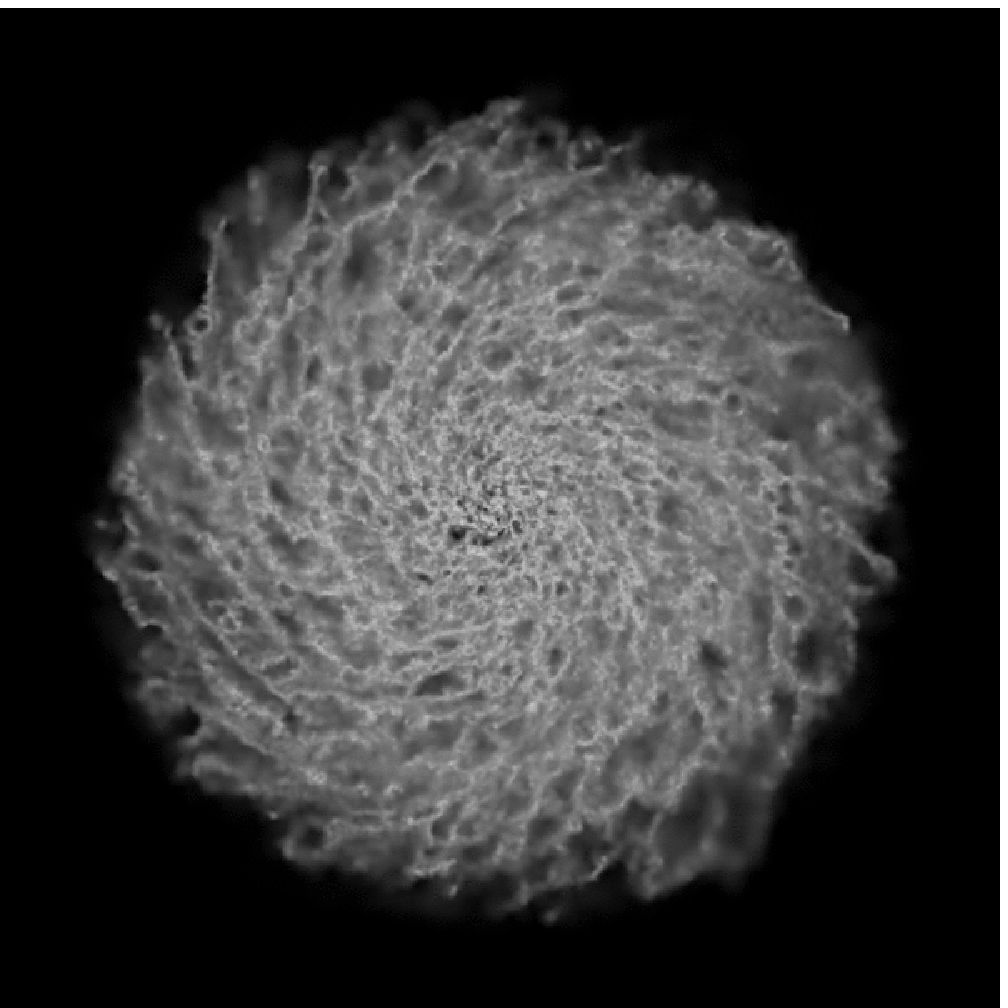}
 \plotone{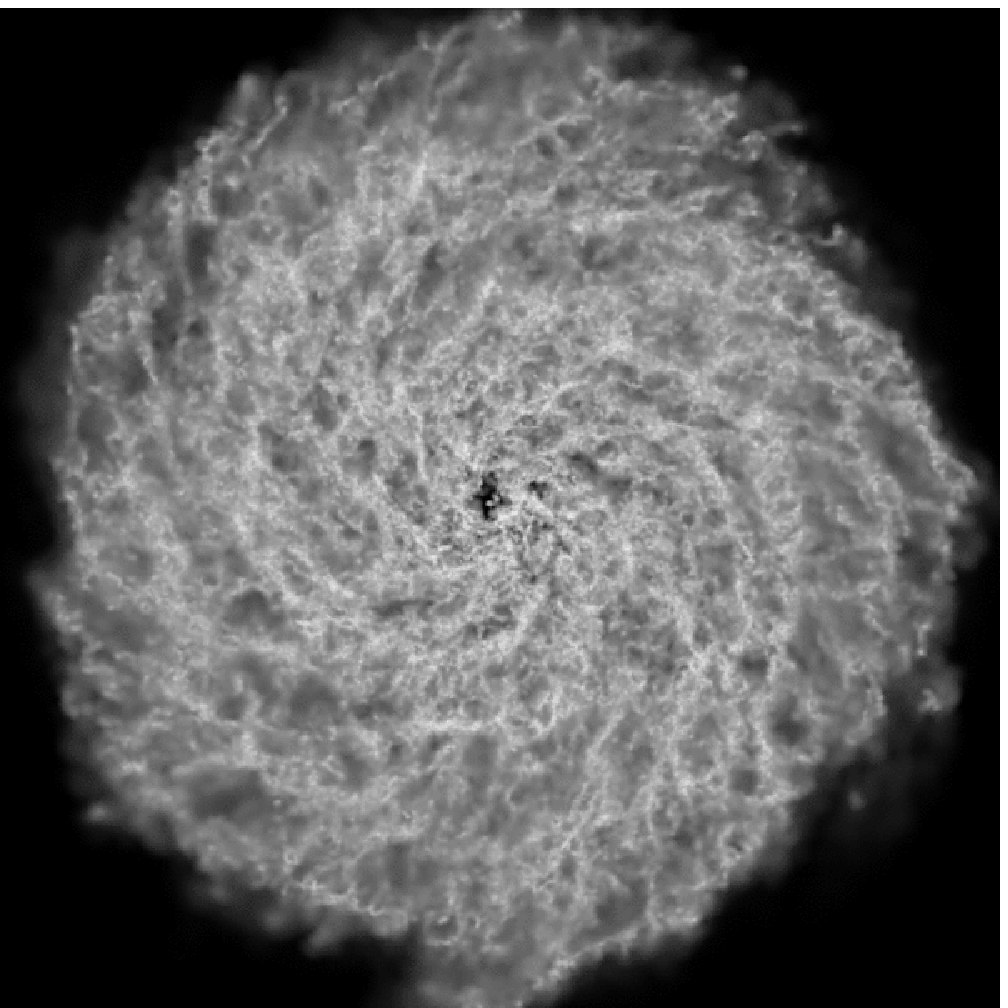}
 \plotone{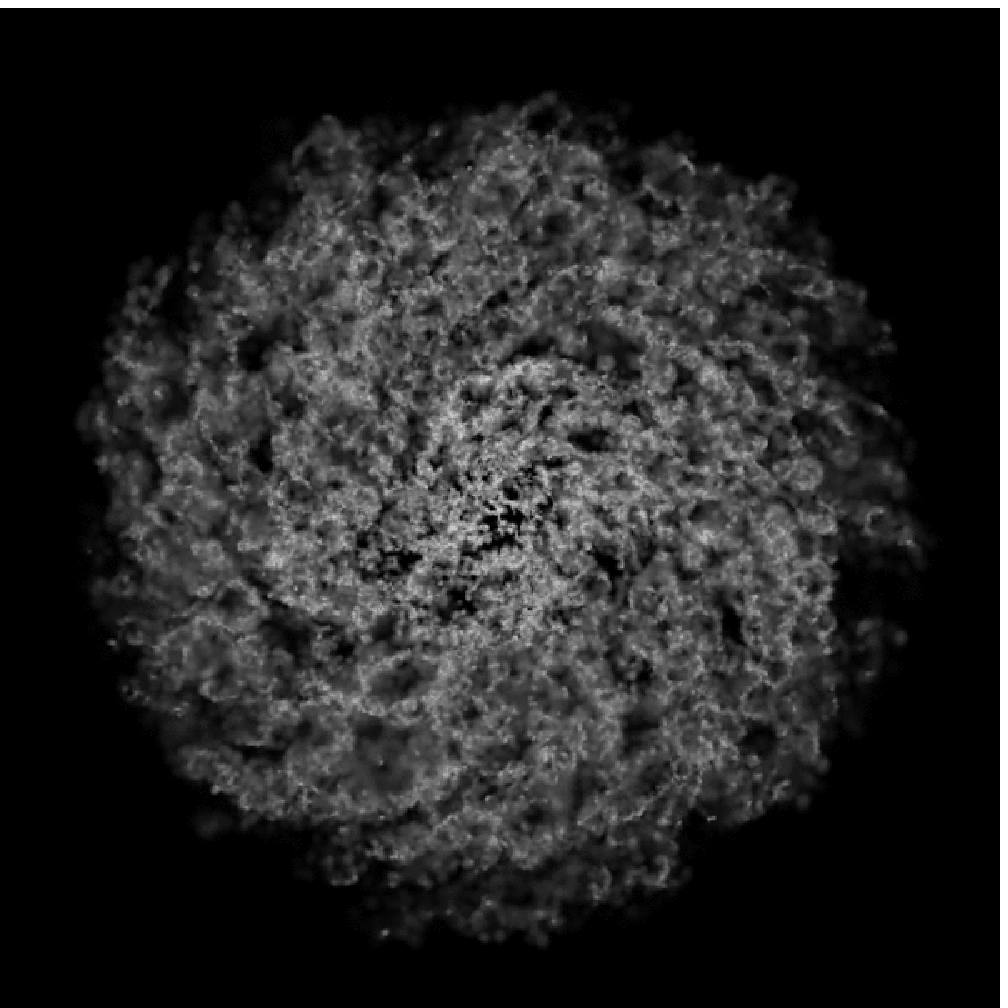}
 \epsscale{0.062}
 \plotone{figs/bar2.ps}
 \epsscale{0.16}

 \plotone{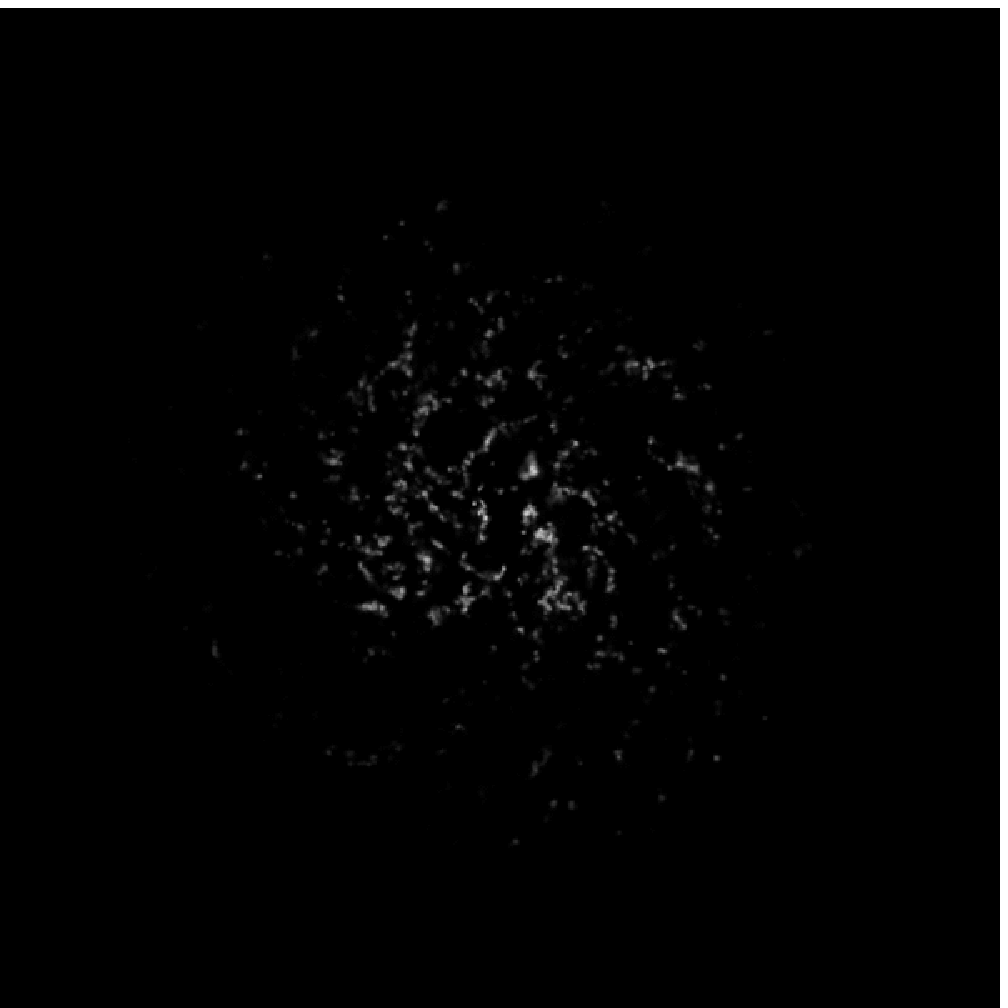}
 \plotone{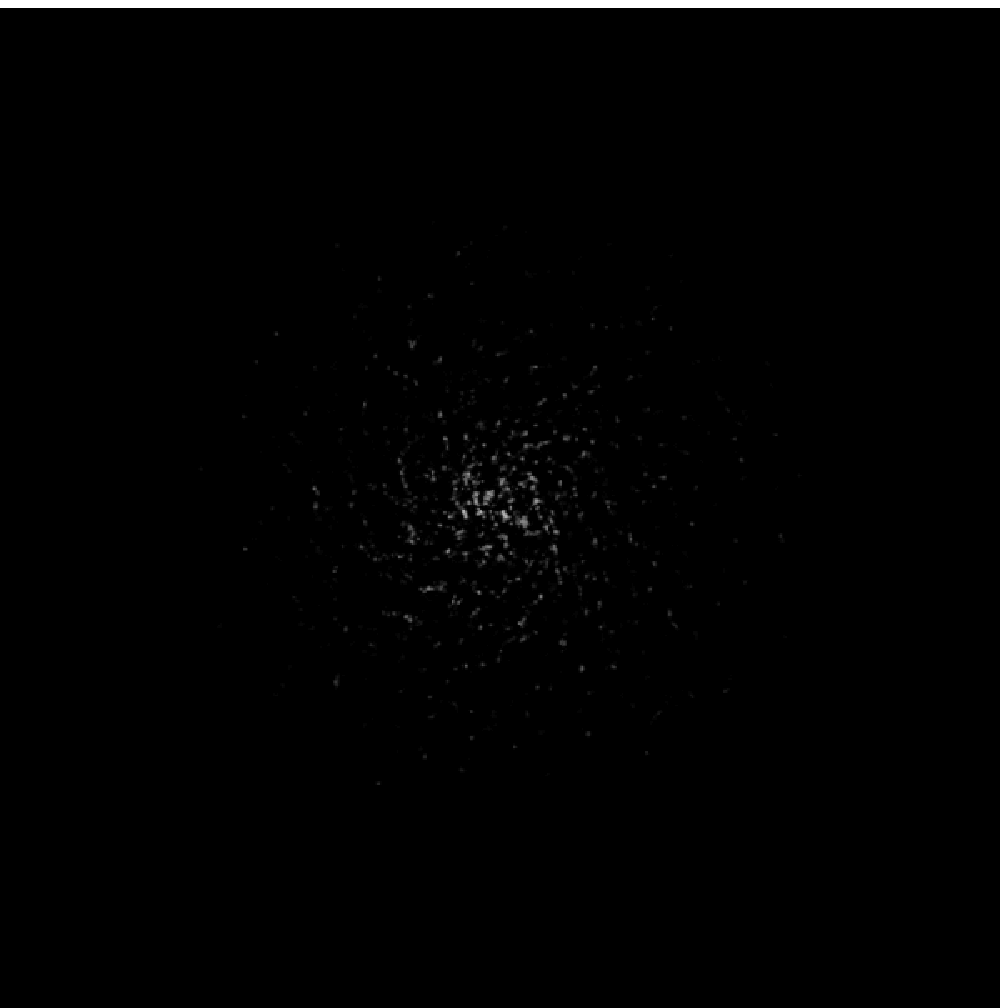}
 \plotone{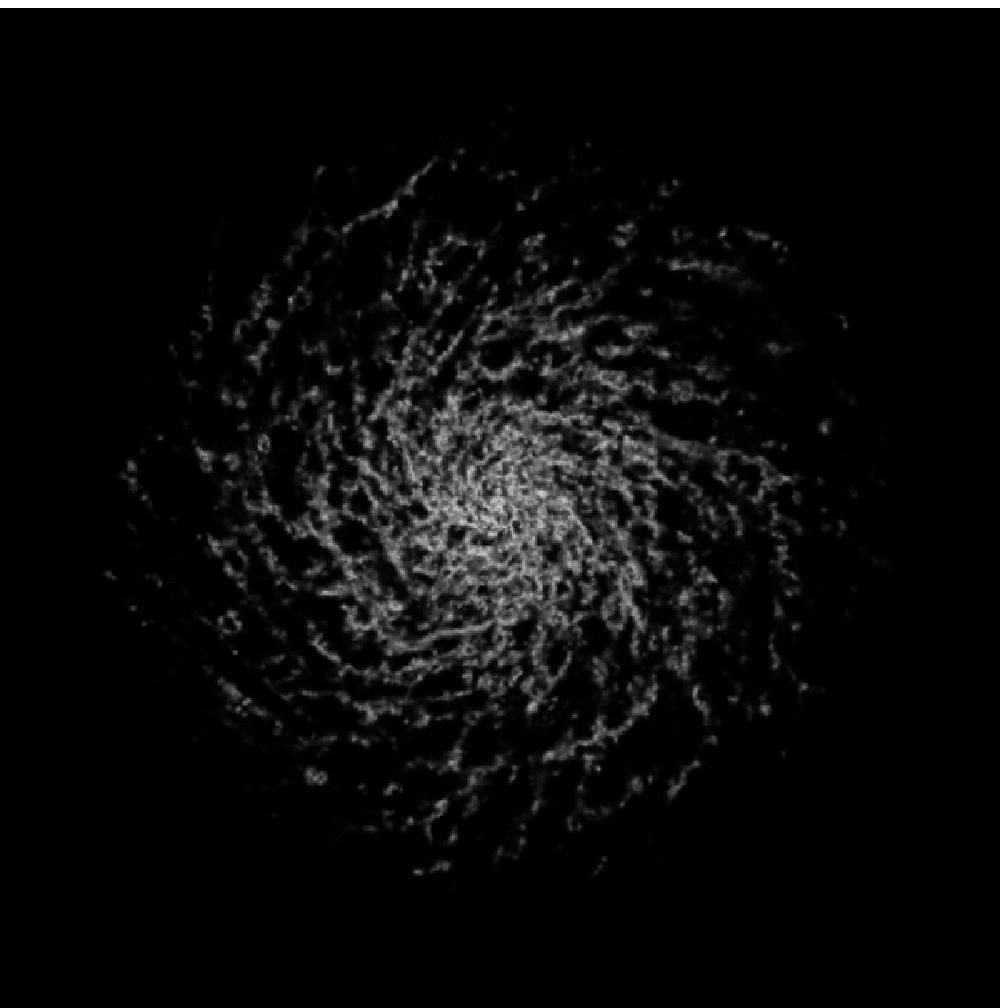}
 \plotone{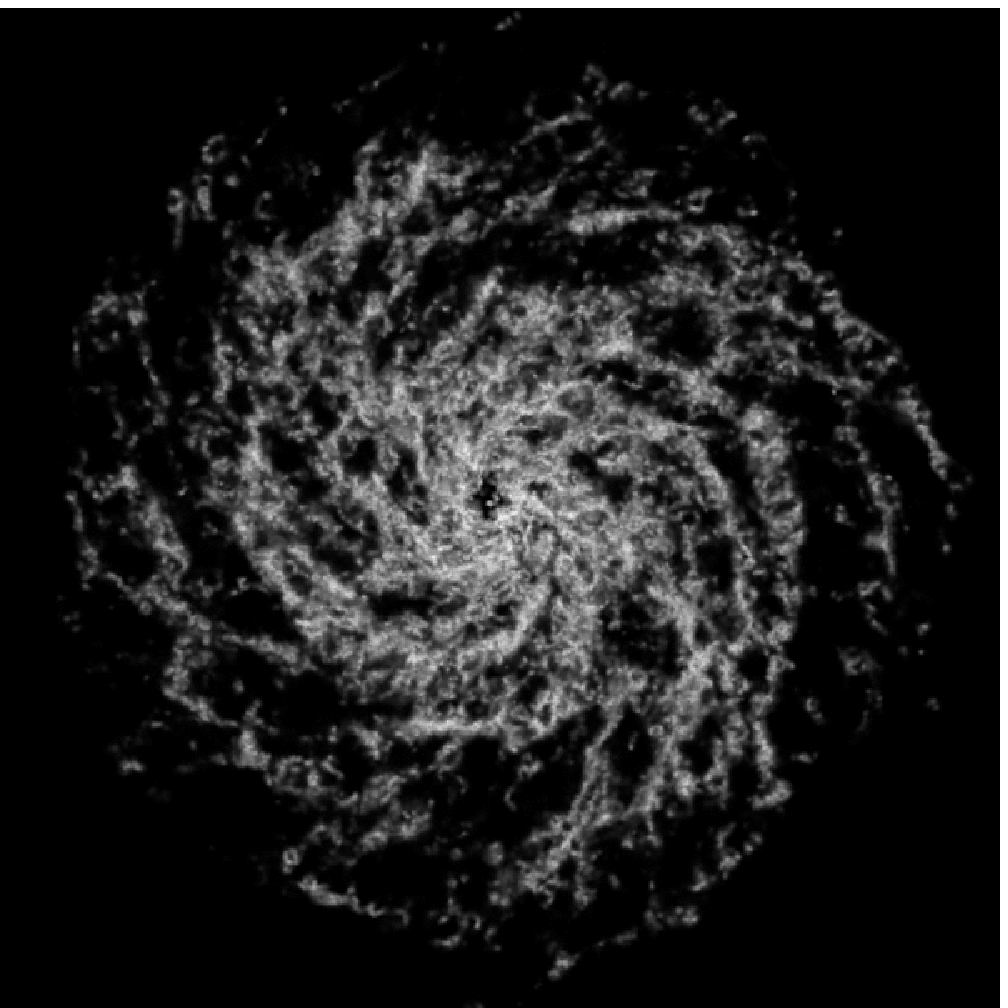}
 \plotone{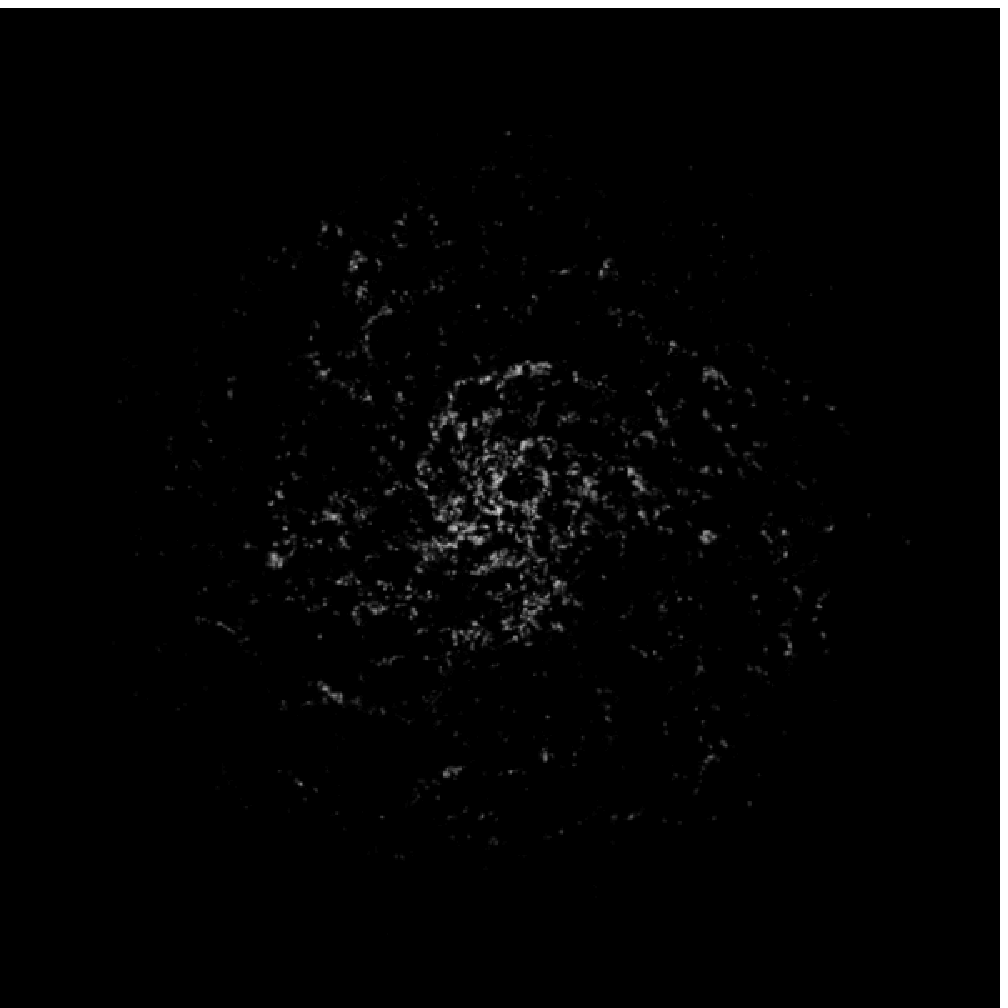}
 \epsscale{0.062}
 \plotone{figs/bar2.ps}
 \epsscale{0.16}

 \caption{Gas distribution  of simulations. Shown are (from  top to bottom)
 HI distribution, the $\rm H_2$ distribution and the CO map for runs (from
 left to right) A1-E1 after 1 Gyr of evolution and SD star formation. The
 A1 panels are 6 kpc across, the others 12 kpc.}
\label{fig:SD}
\end{figure*}

\begin{figure*}
 \centering
 \epsscale{0.16}
 \plotone{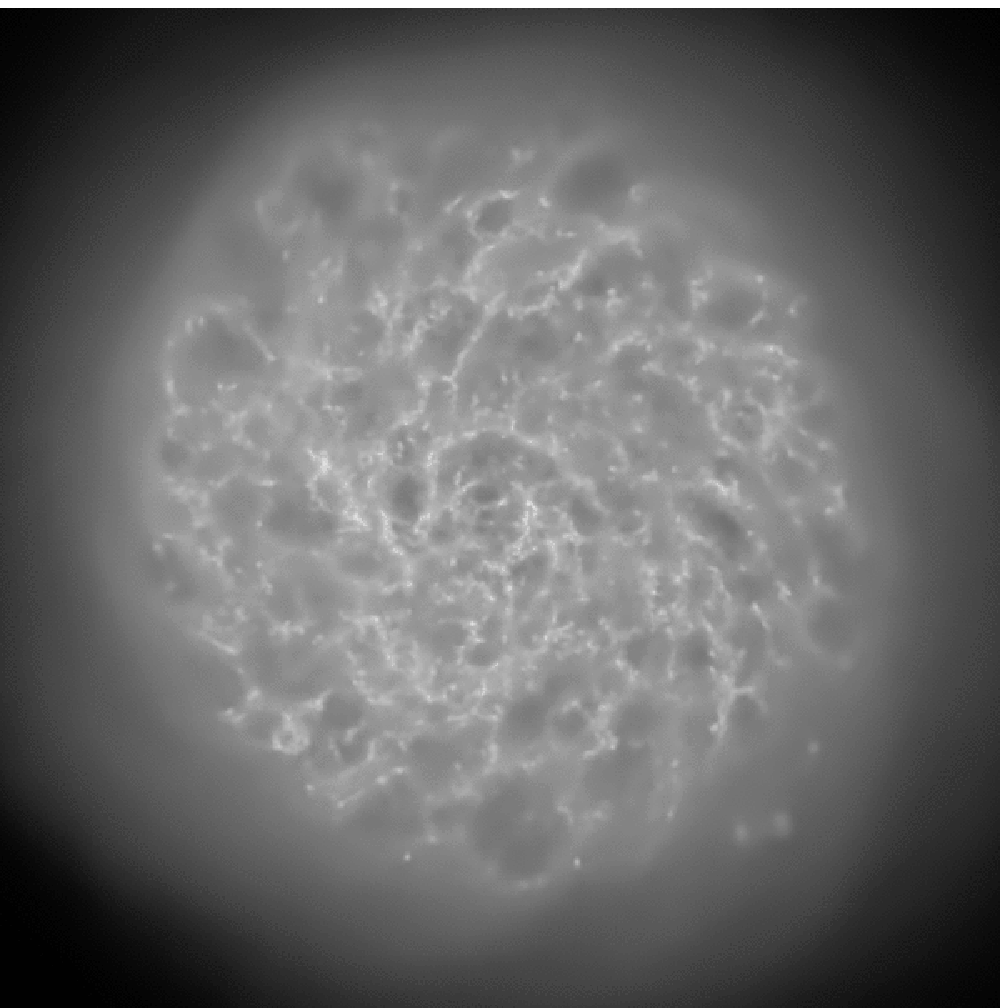}
 \plotone{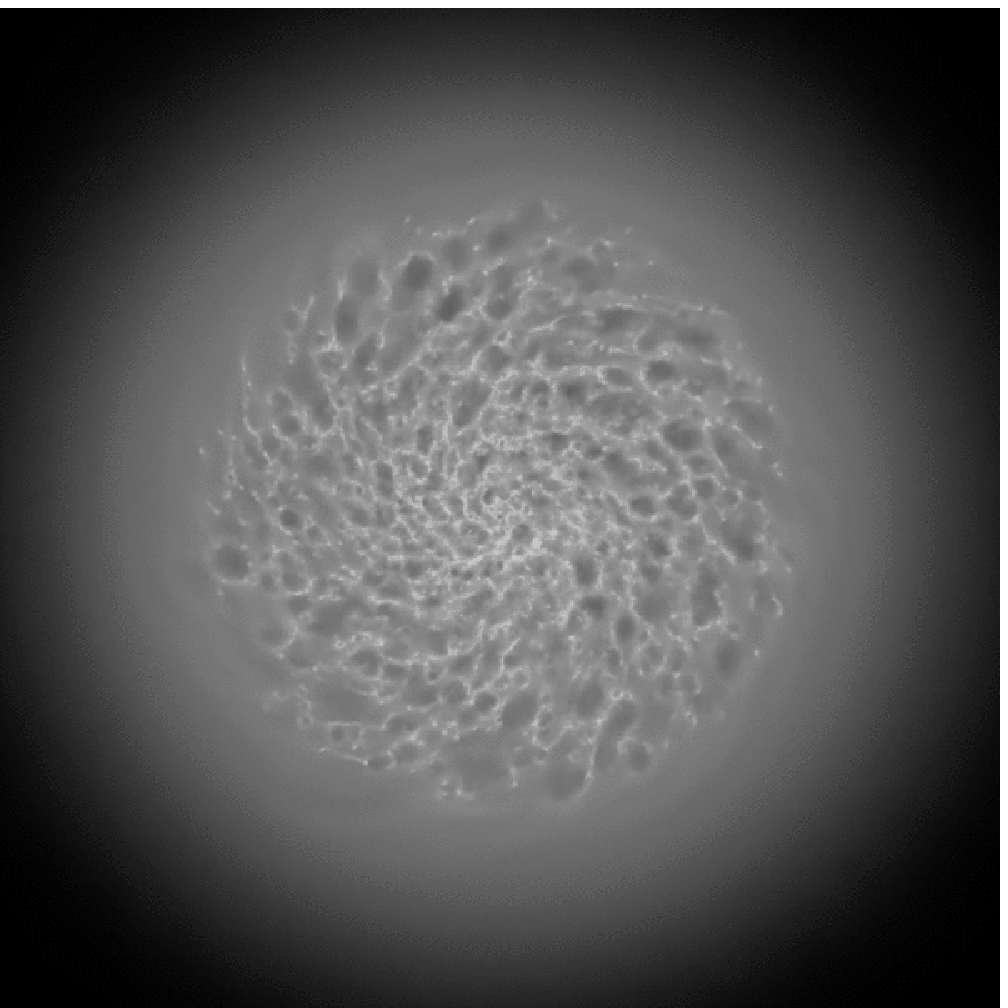}
 \plotone{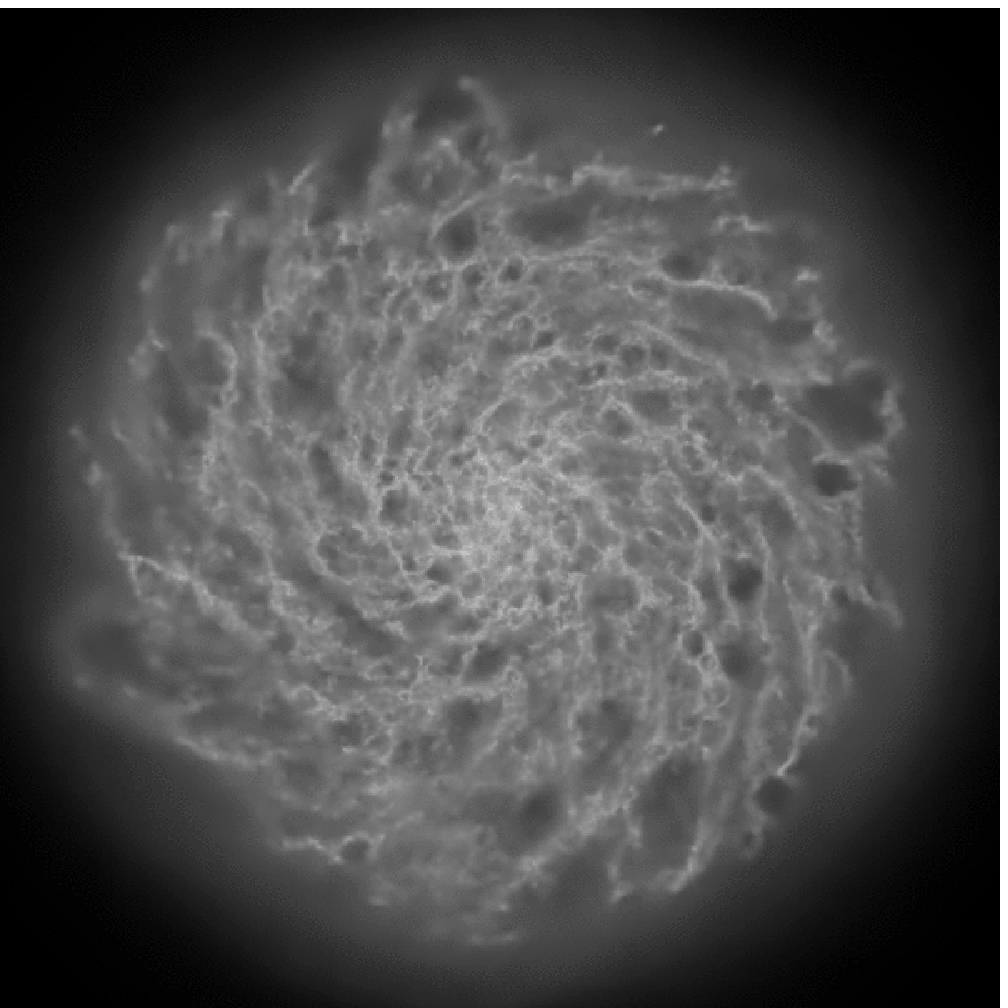}
 \plotone{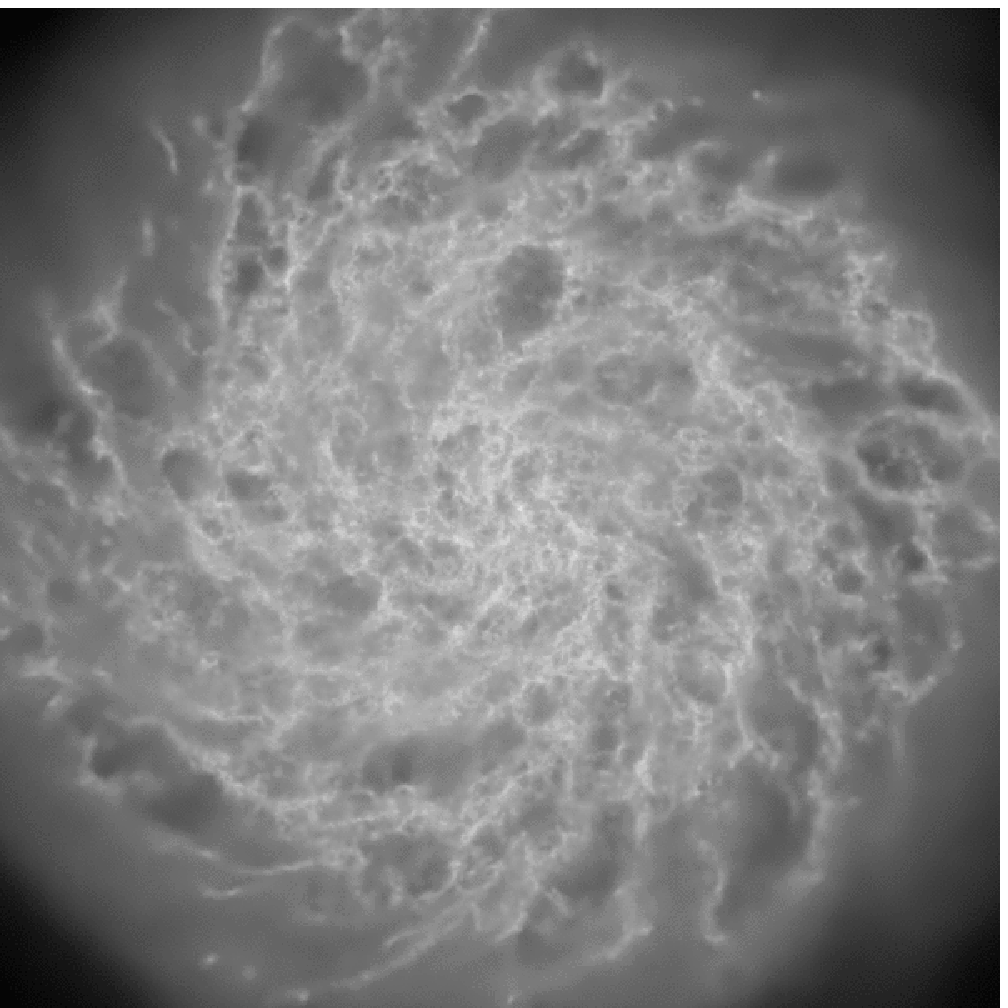}
 \plotone{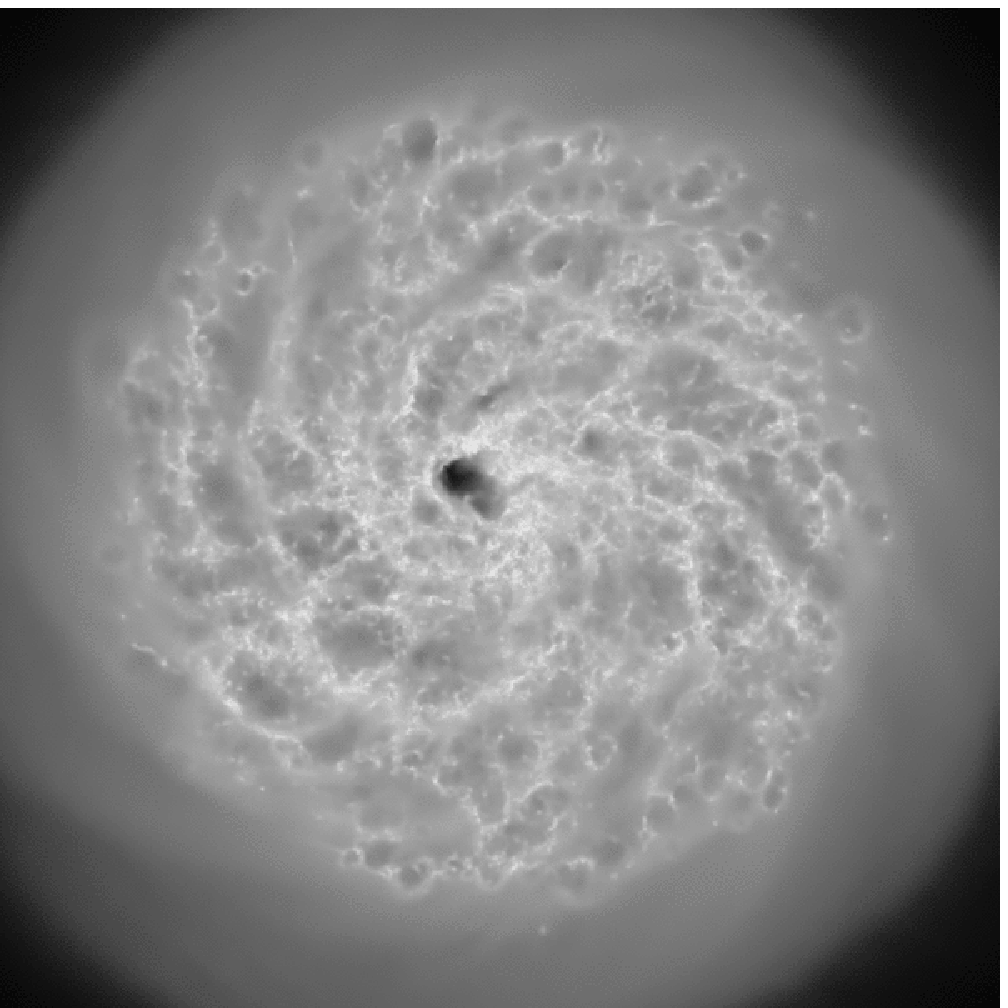}
 \epsscale{0.062}
 \plotone{figs/bar2.ps}
 \epsscale{0.16}

 \plotone{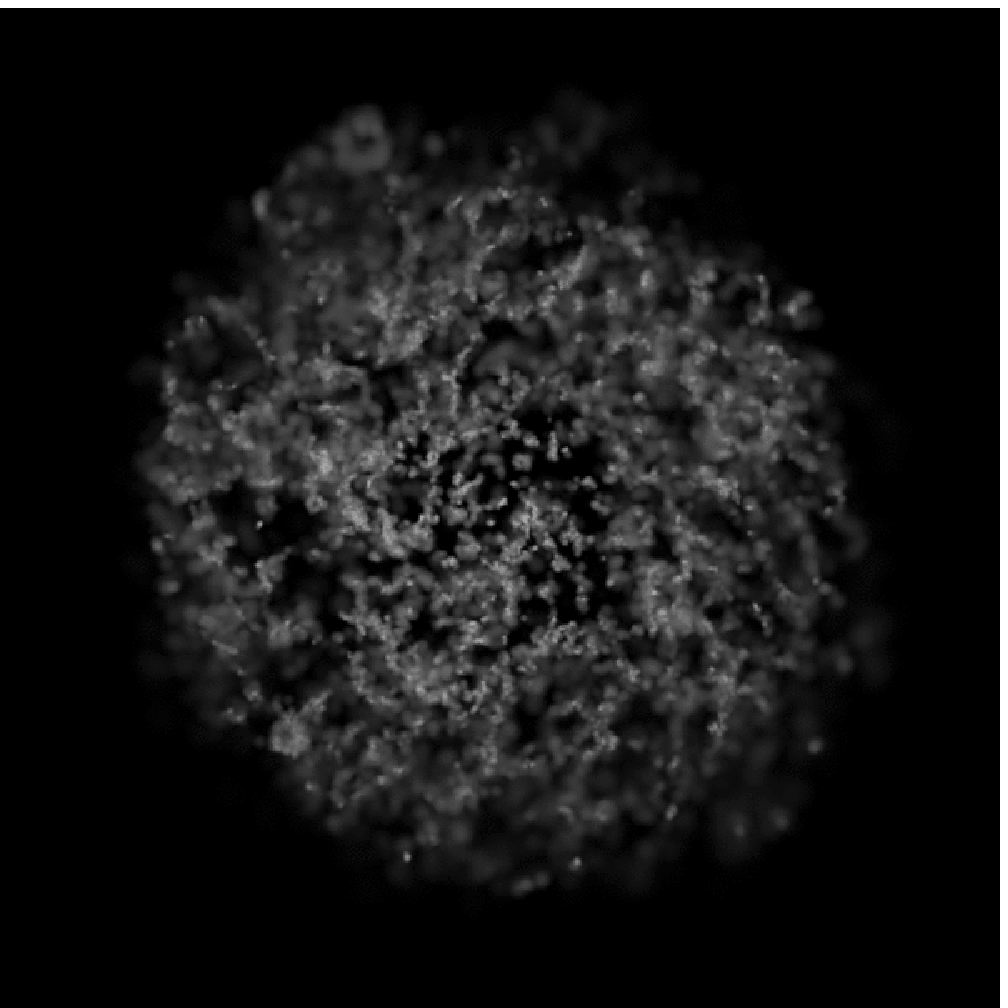}
 \plotone{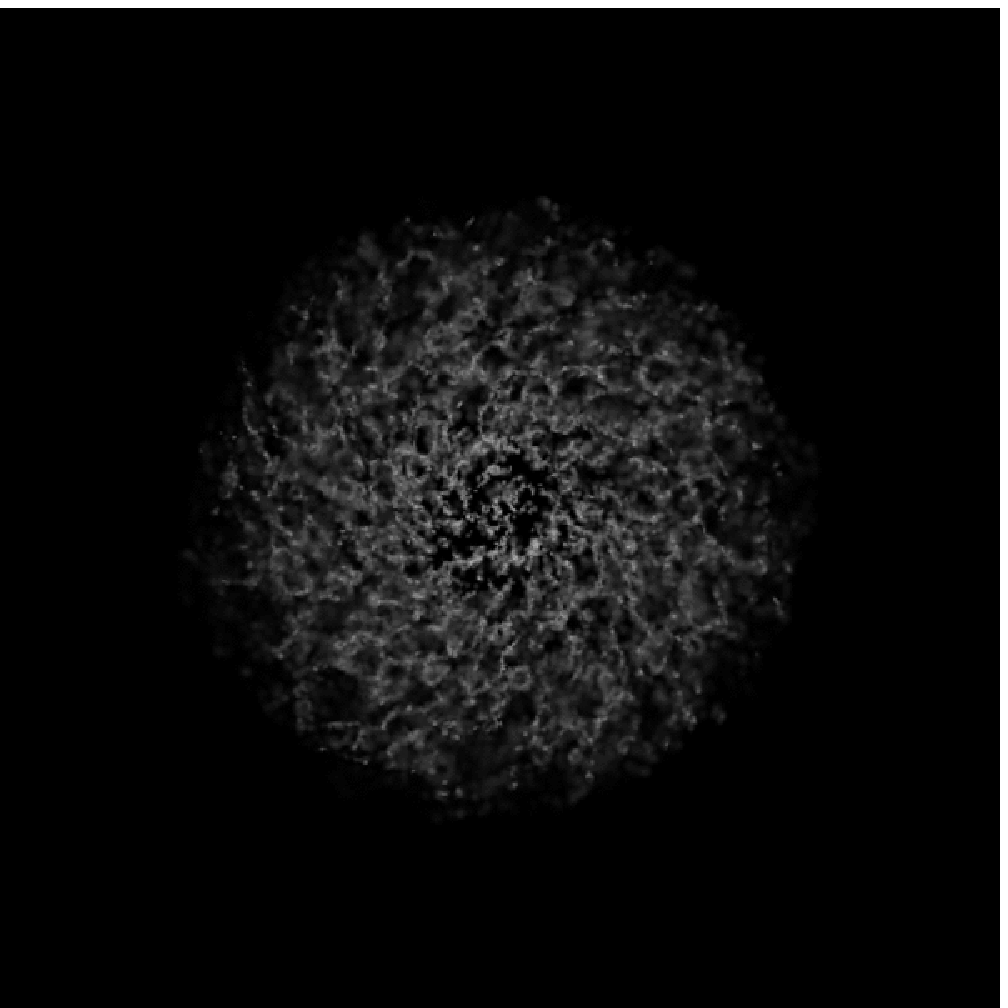}
 \plotone{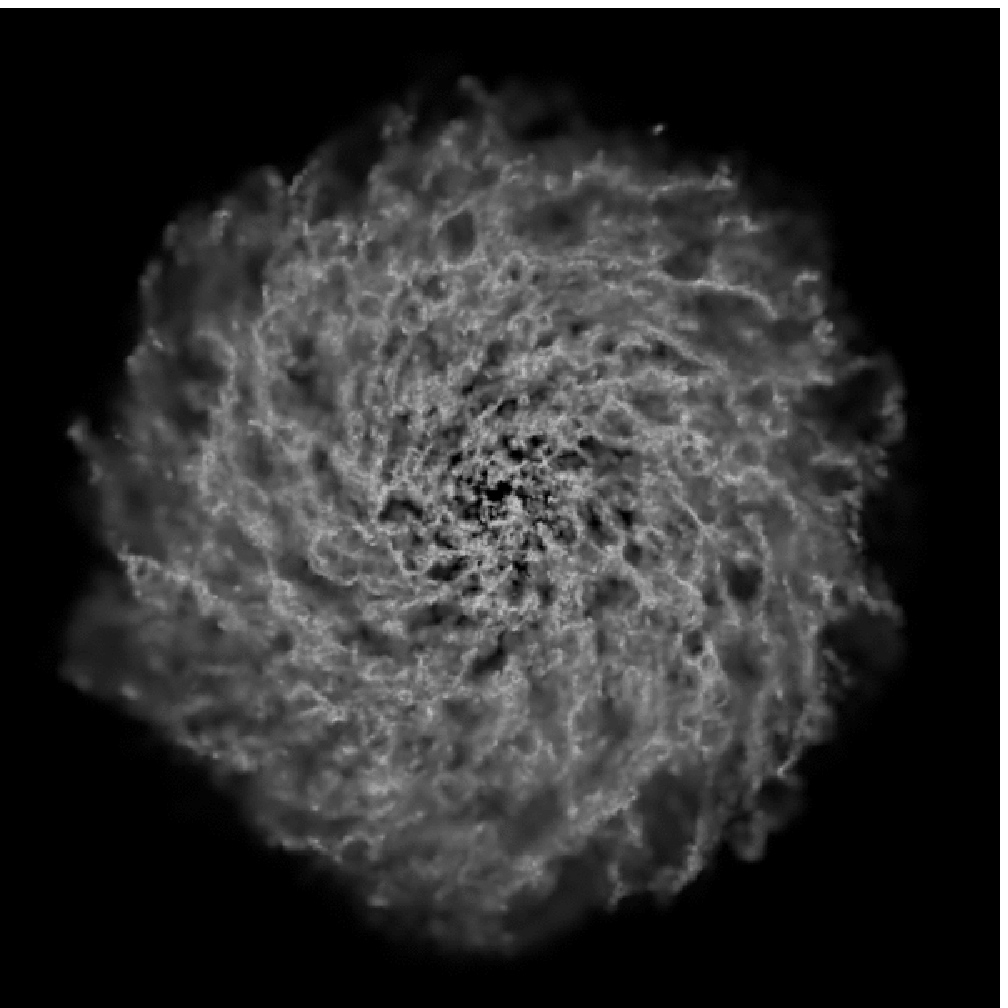}
 \plotone{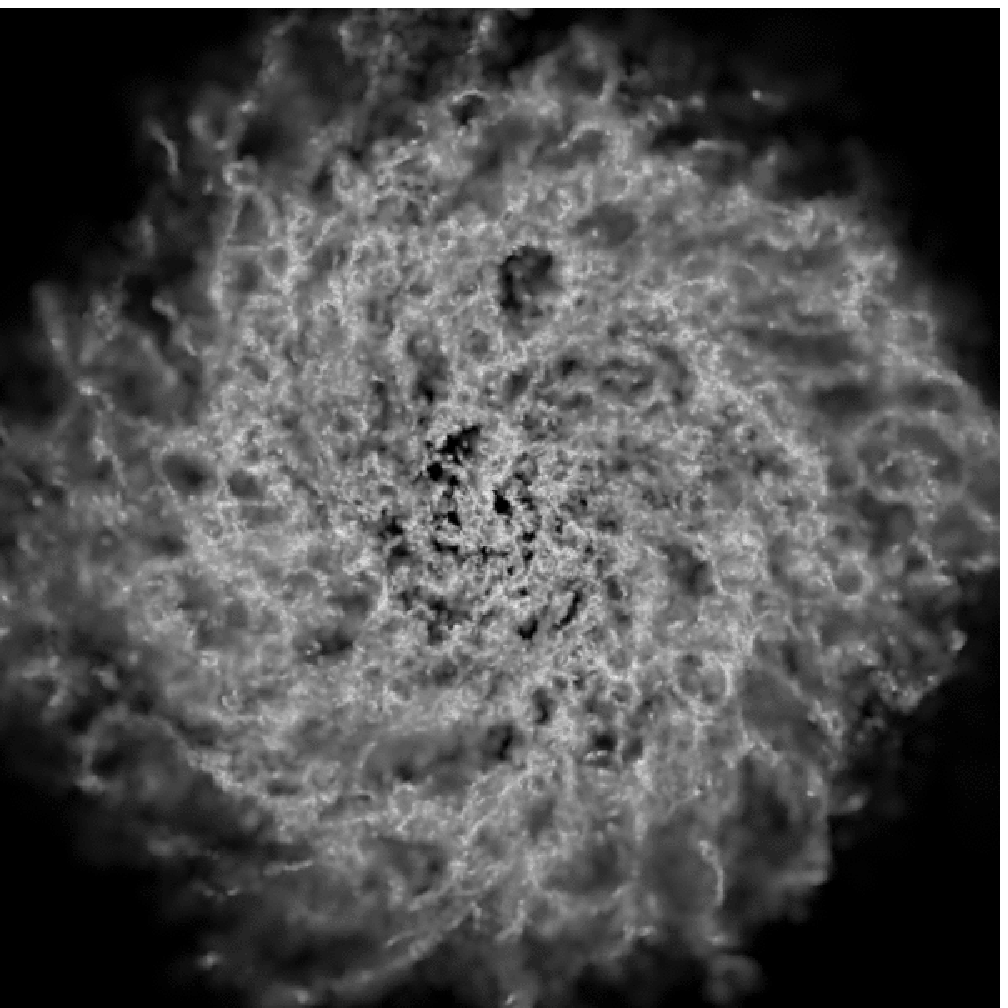}
 \plotone{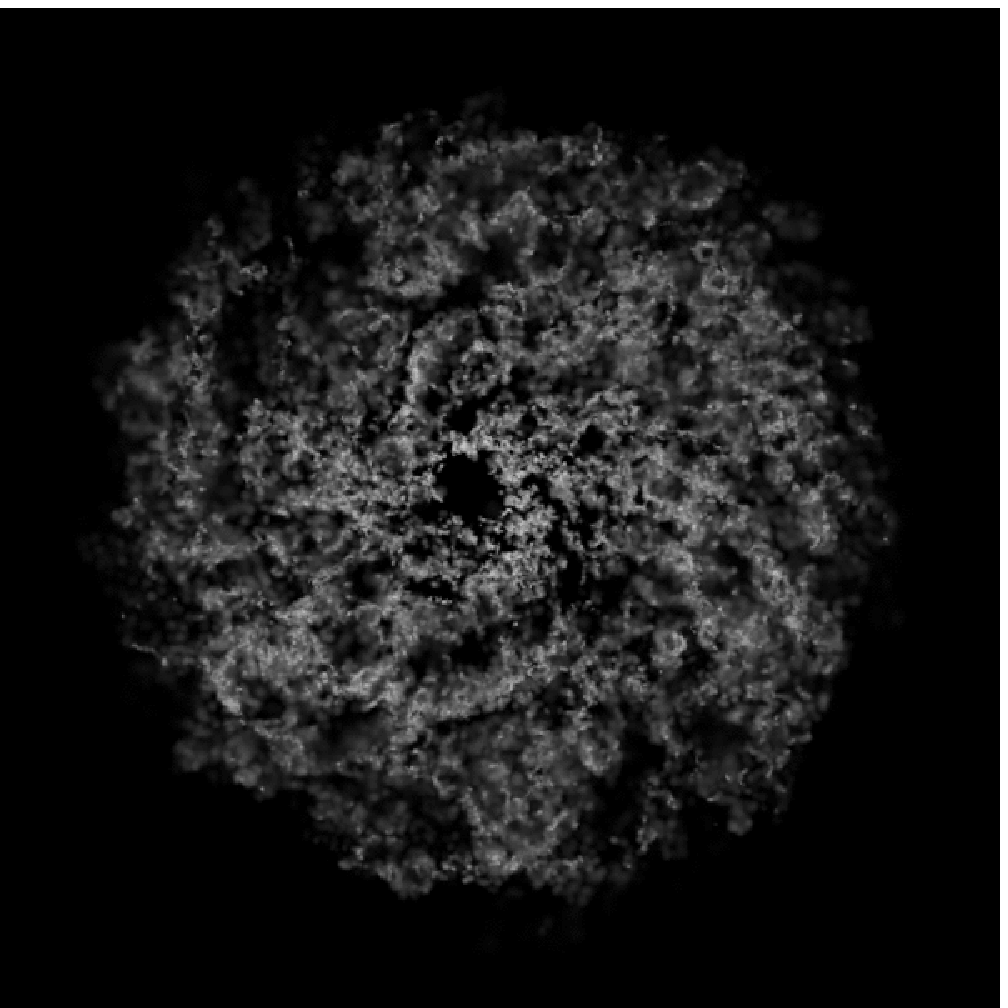}
 \epsscale{0.062}
 \plotone{figs/bar2.ps}
 \epsscale{0.16}

 \plotone{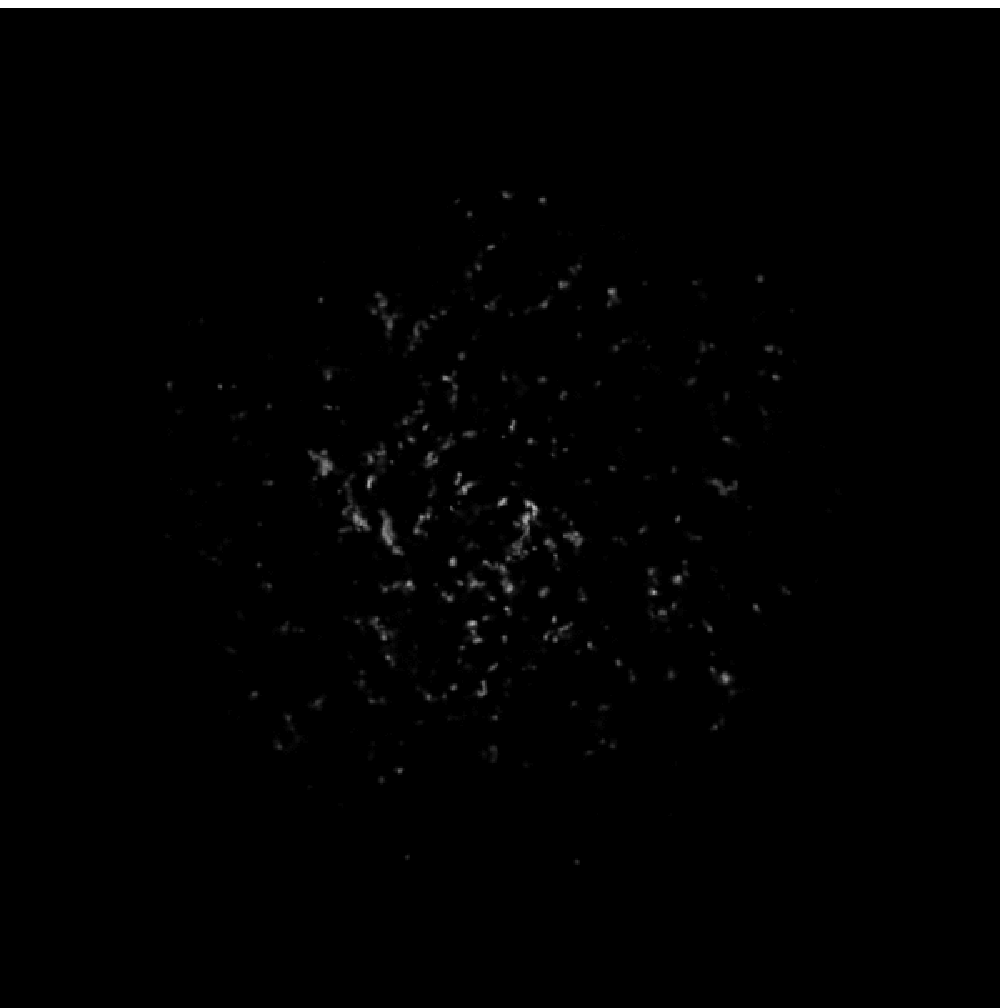}
 \plotone{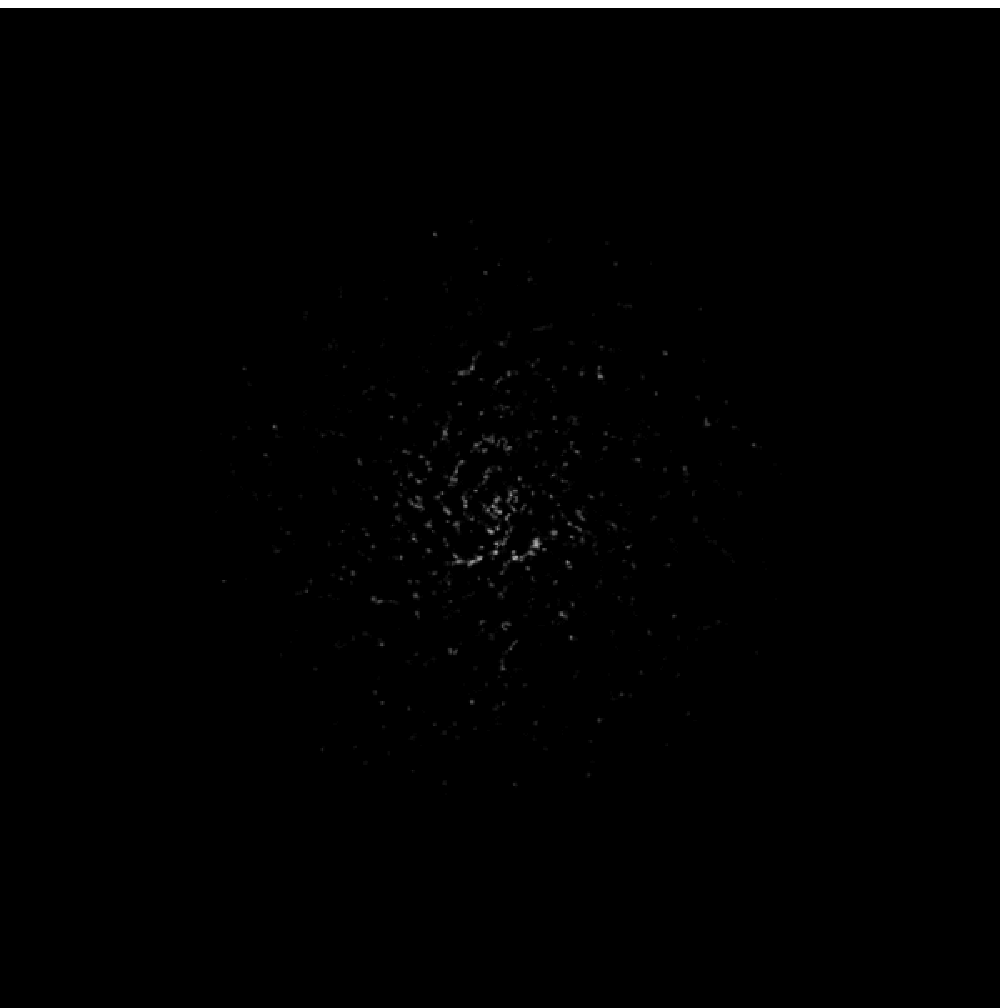}
 \plotone{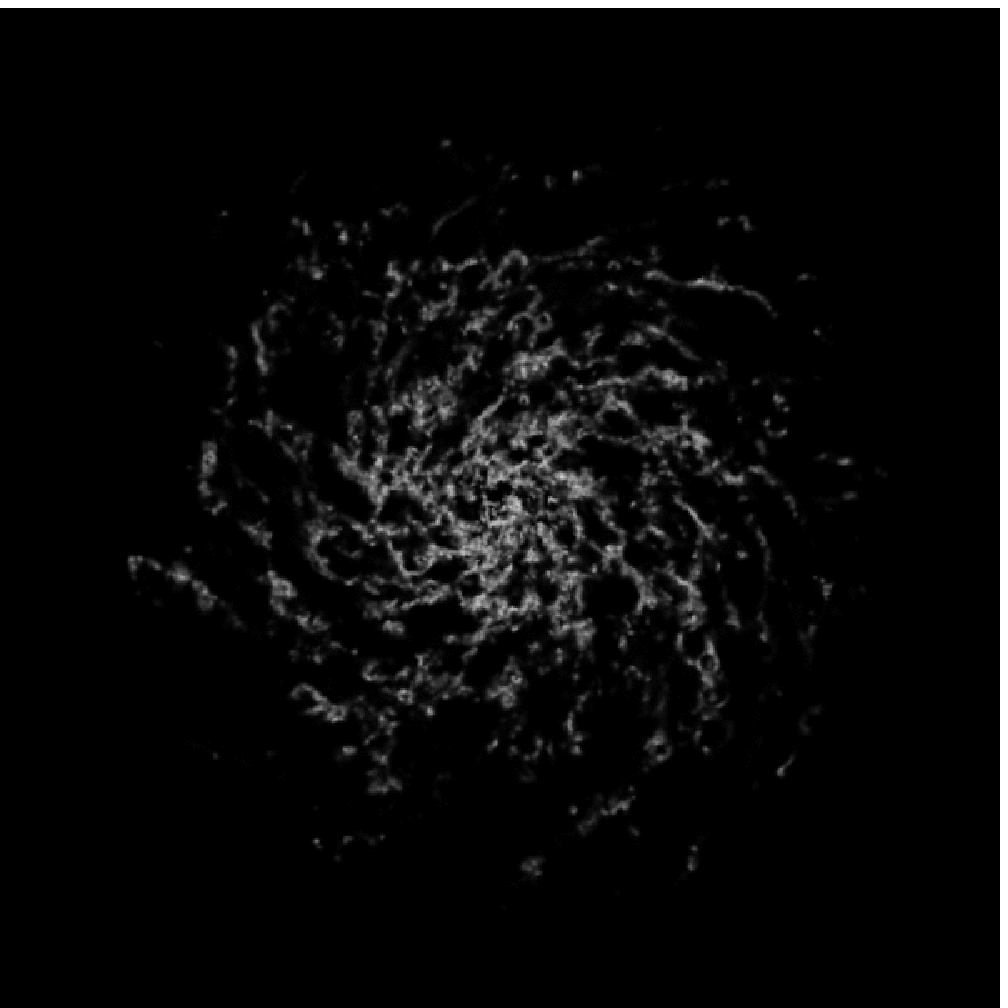}
 \plotone{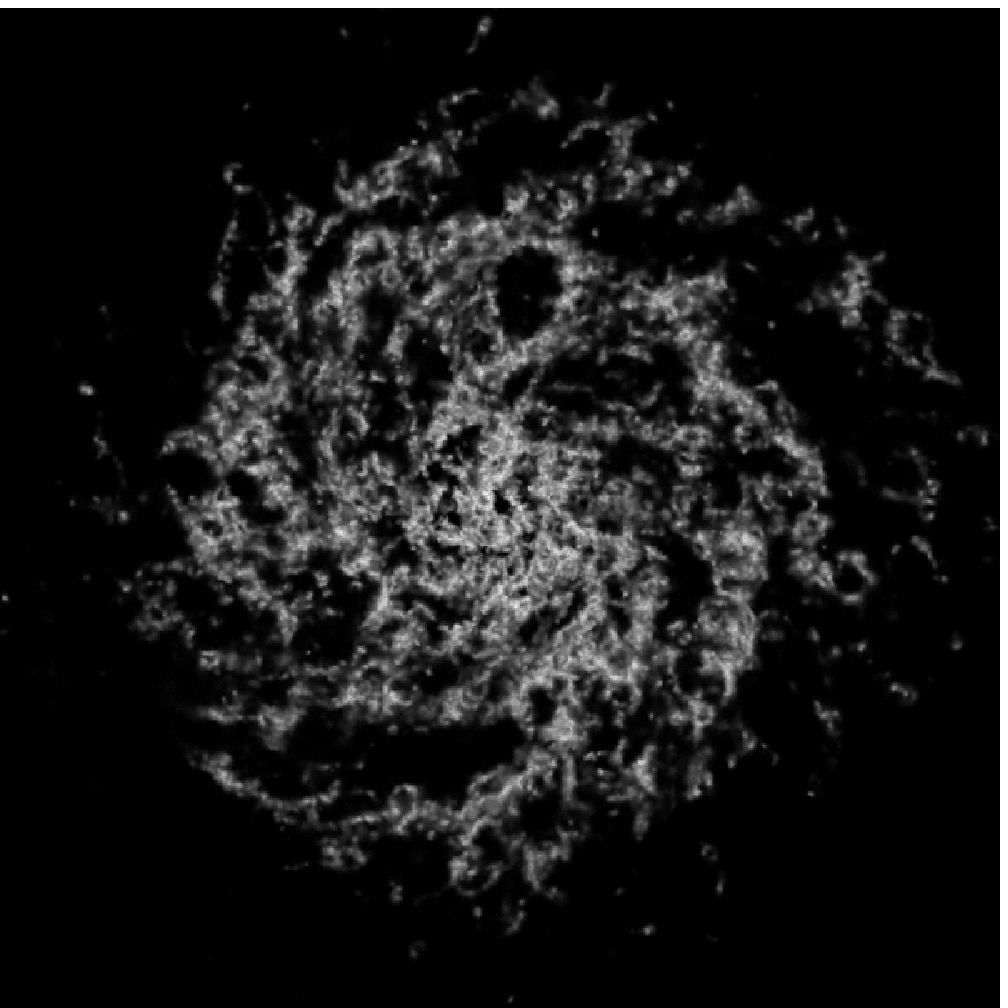}
 \plotone{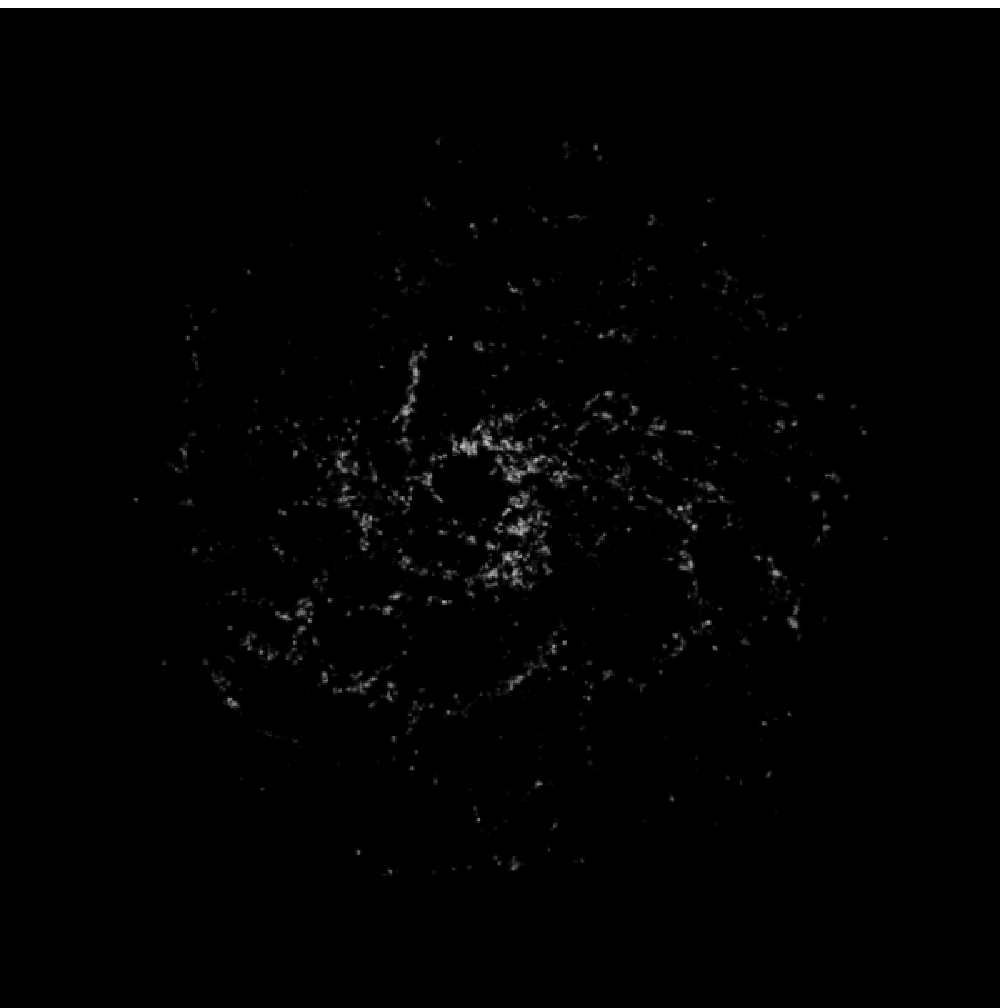}
 \epsscale{0.062}
 \plotone{figs/bar2.ps}
 \epsscale{0.16}

 \caption{Gas distribution  of simulations. Shown are (from  top to bottom)
 HI distribution, the $\rm H_2$ distribution and the CO map for runs (from
 left to right) A1-E1 after 1 Gyr of evolution and MR star formation.}
\label{fig:MR}
\end{figure*}

In Figures~\ref{fig:SD} and~\ref{fig:MR}  we show the gas distribution
from  the  simulation snapshots  of  models  A1-E1. The top  row  of
Figure~\ref{fig:SD} shows the HI gas maps obtained from projecting the 
neutral mass fraction of the particles for runs  using  the SD  star formation  model.
In the middel row the molecular gas distribution obtained from $f_m$ directly is shown, while
the bottom row shows the maps for  the   CO-rich    $\rm H_2$   distribution
 (determined    as   in Section~\ref{sec:CO}).
Figure~\ref{fig:MR}  shows  the  equivalent   maps  for  the  MR  star
formation  runs.   Some  features  of  these maps  are  common  across
different  models. For  example  the  panels for  the  C1 run  (spiral
galaxy/solar metallicity, middle panels) show the frothy appearance of
the  neutral  gas  distribution   typical  for  a  star  forming  ISM.
Comparing the HI  and $\rm H_2$ maps we see from  the enhanced contrast of
the molecular  map that the $\rm H_2$  tends to concentrate  in the higher
density regions. In the outer galaxy regions the $\rm H_2$ distribution cuts
off  before  the  HI  distribution  and  the  smoothness  of  the  gas
distribution shows  little feedback from  stars. The third  row panels
shows the  CO distribution concentrated towards  high column densities
in the  central regions  and dense clumps.   Much the same  pattern is
visible for  the equivalent low  metallicity run B1.   Low metallicity
means less $\rm H_2$  and CO formed, with CO restricted  to the very highest
density clumps.   Note also  that there is  a big smooth  region where
star formation  and $\rm H_2$ are absent,  a pattern repeated in  the A1 run
(note that the linear scale of the maps are different).
The  pure gas  models (two  right most panels  in 
figs.~\ref{fig:SD},~\ref{fig:MR})  are stable despite the high gas content,
stabilized by supernova feedback \citep{Springel2005}.

The  resulting  gas  distribution  for  the MR  star  formation  model
(fig.~\ref{fig:MR}) is  generally similar to the  SD model, especially
on large scales.  On small  scales the $\rm H_2$-regulated SF structures are
affected by feedback due to  a star formation more biased towards high
density regions,  which results in  a more bursty star  formation mode
with large ``bubbles''  (like that seen in the centre  of the panel of
E1 model) forming more~often.

\subsubsection{Non-equilibrium $\rm HI\leftrightarrow H_2$ mass exchange}
\label{sec:noneq}
 
For typical  CNM HI densities  the $\rm H_2$ formation timescales  are $\sim 5-50$
Myr (Eq.  4). The frothy gas disk  structures evident in Figures
~\ref{fig:SD}  and~\ref{fig:MR} indicates  the dynamic  nature  of the
processes.  In order to show this it is necessary to resolve the 
feedback effects  in enough detail.  Together with an assumed equilibrium
$f_m$  this will result in a more static picture for the ISM.

\begin{figure*}
 \centering
 \epsscale{0.49}
 \caption{Equilibrium  and   non-equilibrium   $f_m$  values versus
 gas density. The equilibrium $f_m$ of particles in the simulation 
 (diamonds) and the non-equilibrium $f_m$ (dots) as a function of
 density for $Z=\sun/5$ (left  panel) and $Z=\zsun$ (right panel). The
 equilibrium  $f_m$  is calculated  from  the instantaneous  particle
 properties.   The dashed line  marks a  time track  of a  typical gas
 particle  experiencing  a  cycle  of  collapse,  star  formation  and
 reheating, with square symbols  placed at 0.93 Myr intervals.
 }
 \plotone{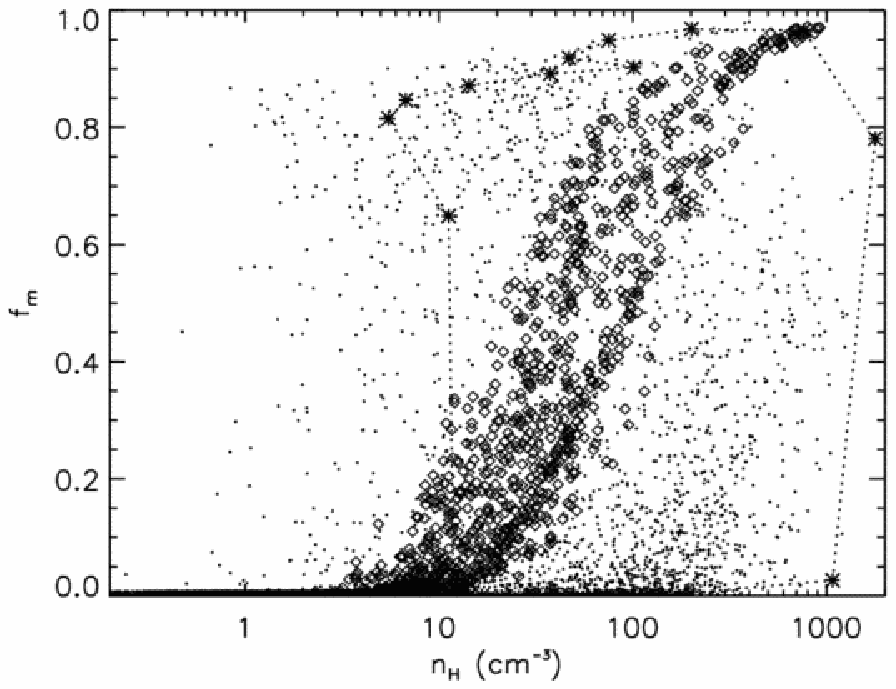}
 \plotone{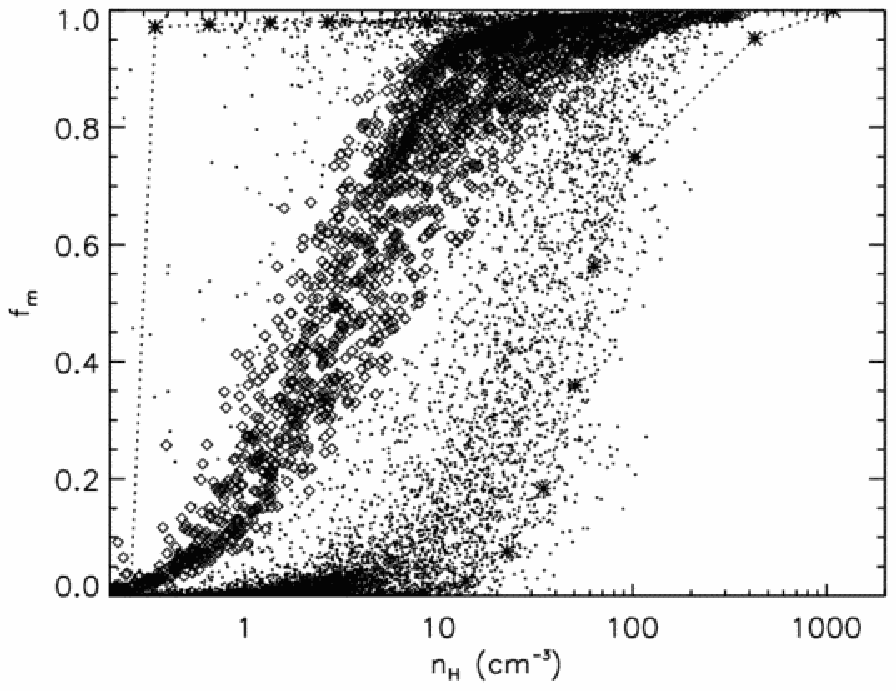}
 \label{fig:noneq}
\end{figure*}

In    figure~\ref{fig:noneq}     we    demonstrate    the   non-equilibrium
between the  HI and $\rm H_2$ gas phases  by plotting the equilibrium  $f_m$ 
from   section~\ref{sec:aneq}  as  a  function  of density, and  the actual 
non-equilibrium $f_m$ from  the simulations (for \Zsun and  $\zsun/5$).  As
we can see from  this snapshot the gas is out of HI/$\rm H_2$ equilibrium for a
large fraction of gas particles, with only  the highest densities 
converging to equilibrium.   The ``ergodic'' rather than  particle-ensemble
representation of  this is demonstrated by the evolutionary track of a 
single gas particle in the $\rm f_m-n$ plot (dashed line)  showing that it
spends most of  the time away from equilibrium areas.  This also means 
that both the collapse to higher densities and the destruction of $\rm H_2$  in
the diffuse gas phase is slow compared to the evolution of gas structures
and ambient ISM conditions (further justifying the use of a time-dependent
computation of $f_m$).

\subsubsection{The CO tracer molecule versus $\rm H_2$}

\begin{figure*}
 \centering
 \epsscale{0.49}
 \caption{ $\rm H_2$  versus CO-rich $\rm H_2$ mass  surface density.  Plotted
are the  pixel values  of the projections  of the gas  distribution for the 
 SD SF model for  low  (left upper panel)  and  solar  
(right upper panel) metallicities (models B1 and C1). Lower left and Lower
right panels show the corresponding results for the MR SF recipe.
The CO pixel values are normalized as  described in  the text.  
For reference  a long  dashed  line with linear slope ($n=1$) is plotted, 
as well as a short dashed line with a slope $n=2$.  } 
\plotone{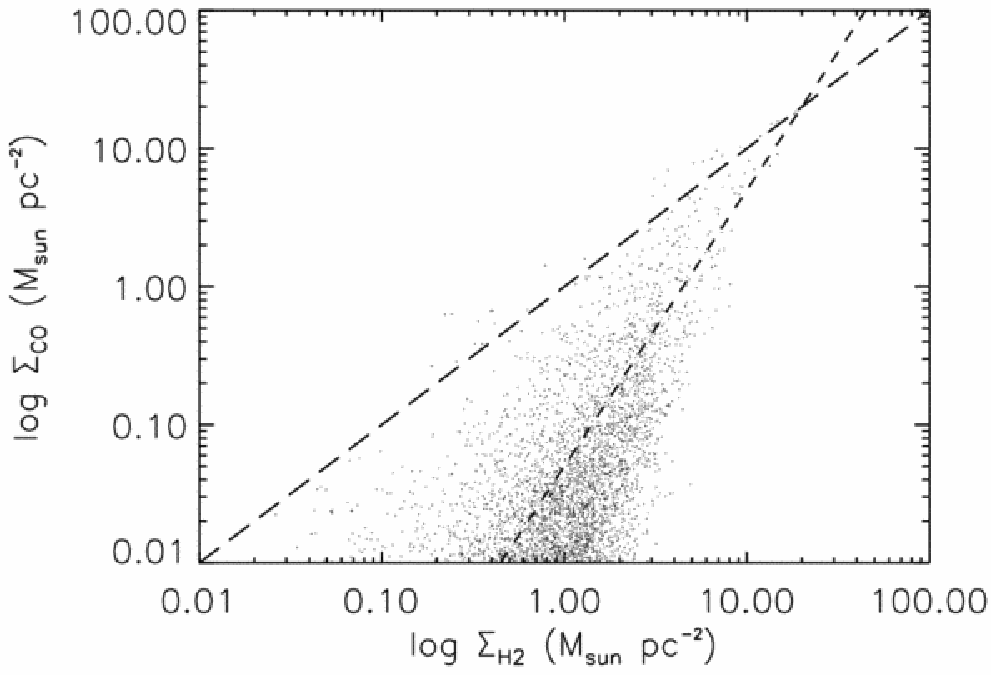} 
\plotone{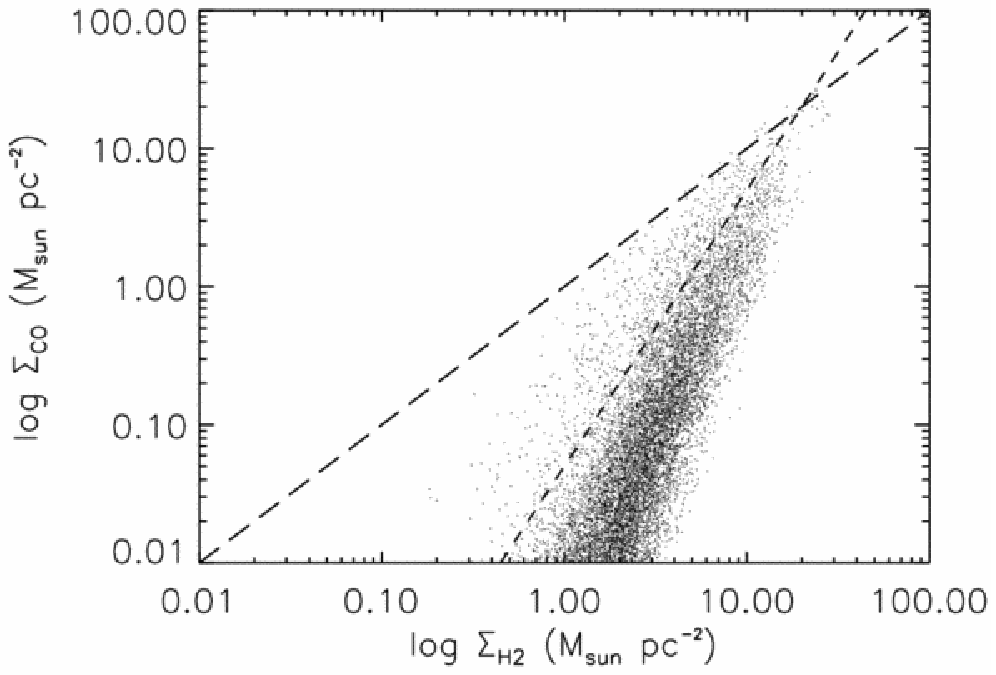}
\plotone{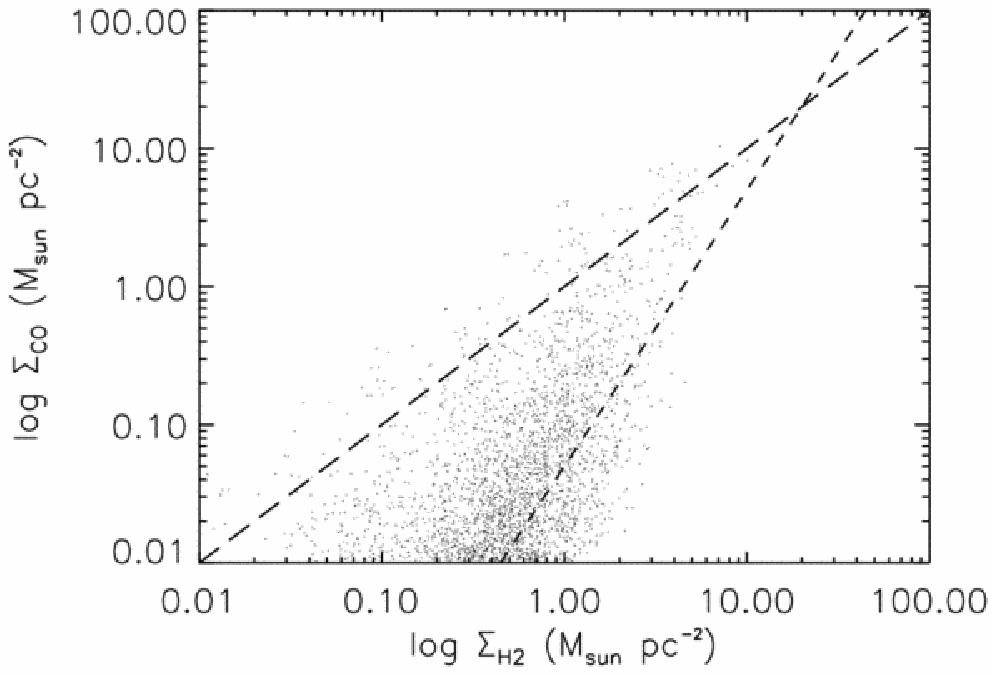} 
\plotone{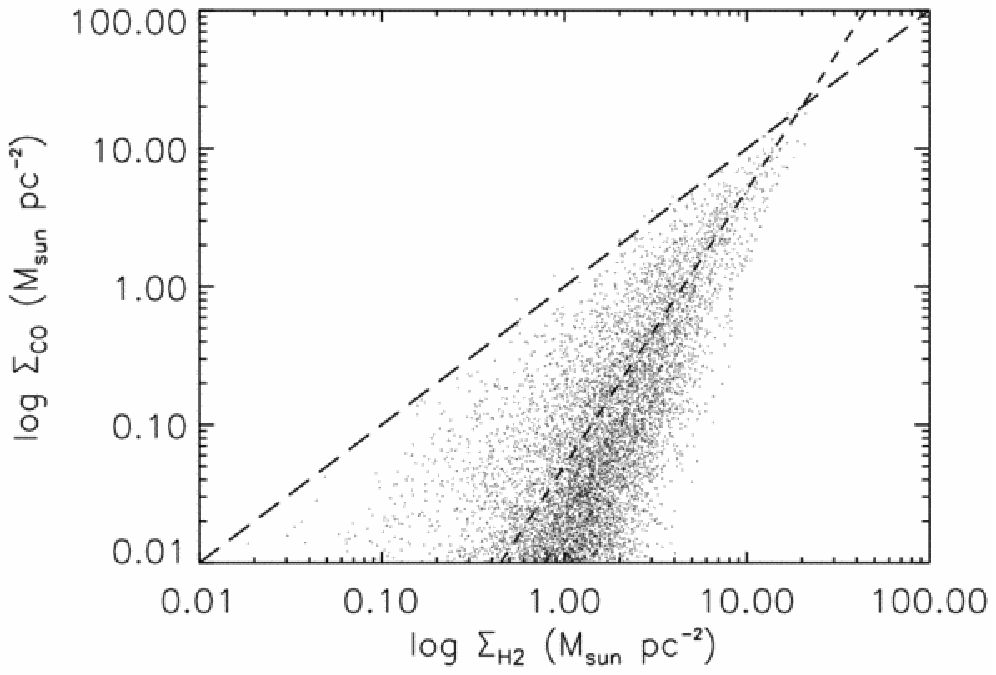}
 \label{fig:coh2}
\end{figure*}

The   CO-rich   $\rm H_2$   distribution   in   Figures~\ref{fig:SD}   and
 \ref{fig:MR} is markedly different  from the total $\rm H_2$ distribution
 demostrating  a  variable  CO-to-$\rm H_2$  mass  ratio,  especially  for
 metal-poor   systems.    This   can   be   seen   more   clearly   in
 Figure~\ref{fig:coh2}, where we plot  the pixel values drawn from the
 $\rm H_2$ distribution against  those drawn from the CO  map.  The values
 in figure~\ref{fig:coh2} are scaled  so that the maximum pixel values
 of  the  CO  and  the  $\rm H_2$  distributions  match,  which  could  be
 considered as an effective  calibration of the CO-to-$\rm H_2$ conversion
 factor  for our  simulations (observationally  also performed  at the
 bright end of  CO-luminous $\rm H_2$ clouds).  It  is clear that the
 CO-to-$\rm H_2$ factor  is not constant: the  CO-rich $\rm H_2$ distribution
 does show a  tight correlation to the total  $\rm H_2$ mass distribution,
 but  one that  is much  steeper than  linear (and  thus  difficult to
 calibrate   observationally    without   running   into   sensitivity
 limitations).  This has been suspected and argued widely in the
 literature  \citep[e.g.]{Maloney1988, Pak1998}, so it is interesting 
 that in our simple model this is  demonstrated  in  the context 
 of evolving  galaxy models.  Note that although a steeper CO-H$_2$
 relation seems to appear for metal-rich compared to the metal-poor 
 systems for the SD simulations \citep[contrary to what would be expected 
 for stationary cloud/radiation field models, e.g.][]{Maloney1988, Pak1998},
 this trend dissapears in the more realistic MR models.
 
 The non-linear relation between CO and $\rm H_2$ may also raise the 
 possibility  that (especially metal poor) galaxies  can be  ``CO-dark''  
 during certain epochs  of their evolution  (e.g. 
 immediately  after a burst of  star   formation  and   the  subsequent 
 enhancement   of  far-UV radiation),  while  $\rm H_2$  gas  is  still 
 there  continuing  forming stars. Such systems may appear as having much
 larger than usual star  formation efficiencies, i.e. forming stars out of
 seemingly very little CO-bright H2 gas. (we return to to this point in
 section \ref{sec:KS} and  \ref{sec:disc}).

Even  strongly  varying  CO-to-$\rm H_2$  relations can  be  difficult  to
discern  and calibrate  observationally. Direct  methods  would entail
independent  observations   of  CO  {\it  and}   $\rm H_2$  at  comparable
resolutions,  an  improbable   proposition  given  that  direct  $\rm H_2$
observations  are  difficult  and  thus  rare.   The  latter  are:  a)
observations of  its lowest excitation  S(0) line 
\citep[e.g.]{Valentijn1999}  which  can  be  excited for  CO-deficient  $\rm H_2$
\citep{Papadopoulos2002},  b)  $\rm H_2$  absorption  studies in  the
far-UV.   The  S(0)  line  emission  observations  at  28$\mu  $m  are
restricted  to Space,  where the  small apertures  deployed  until now
cannot match  the resolution  or gas-mass sensitivity  attainable with
ground-based  CO   observations  using  mm/sub-mm   telescopes.   Line
absorption  studies on  the other  hand  are restricted  by nature  to
single and special lines of  sight, making routine comparisons with CO
observations  difficult  especially  for  extragalactic  environments.
\cite{Krumholz2009}  find  considerably more  $\rm H_2$  detected by  far-UV
absorption  studies with  \emph{FUSE}  at low  column  densities than
their  CO,  $\rm H_2$  equilibrium   model  predicts,  indicating  that  a
CO-deficient diffuse  $\rm H_2$ phase  is indeed present  in the  Galaxy. An
indirect,  but nevertheless  powerful method  relies on  CO  and C$^+$
observations at 158$\mu $m  where any ``excess'' C$^+$ emission, after
correcting for contributions from H$^+$ and WNM, CNM HI gas phases, is
attributed to  $\rm H_2$.  Such observations have  indicated $\sim 10-100$
times more $\rm H_2$  gas than what CO emission  reveals in the metal-poor
and far-UV  intense environment of the dwarf  irregular IC\,10 (Madden
et  al. 1997). These  are indeed  the type  of environments  where the
largest disparities between $\rm H_2$  and CO-rich $\rm H_2$ distributions are
expected from  static \citep[e.g.][]{Bolatto1999} as well  as our own
dynamic models  (see Figures  2, 3).  Neverthless  C$^+$ observations
still suffer from similar limitations like those of the S(0) line with
the  aforementioned  example  being  one  of the  very  few  cases  of
meaningful  comparisons  with  CO  observations.   Finally,  with  the
CO-rich  regions restricted  deeper and  deeper into  $\rm H_2$  clouds as
metalicities  decrease (Figures  2,  3) a  rising ``overpressure''  on
these  regions is expected  by the  overlying CO-deficient  $\rm H_2$ gas.
Such an effect has been recently detected \citep{Bolatto2008}.

\subsubsection{The K-S empirical relation}
\label{sec:KS}

\begin{figure*}
 \centering
 \epsscale{0.99}
 \caption{ 
 Kennicut-Schmidt laws: The star formation density for the
 simulations A1-E1 (from top row to bottom row) using  the SD star formation
 recipe vs (panels from left to right)  the total, the neutral, $\rm H_2$ or CO-rich $\rm H_2$
 gas surface  density. Shown are mean relations (points are averaged in equal 
 logarithmic bins of surface density). The short dashed line shows the
 empirical \cite{Kennicutt1998} relation with a slope of $n=1.4$ as a
 reference, whereas the long dashed and dotted lines have of $n=2$ and $n=1$ 
 respectively.  Note that some outliers are probably affected by image 
 artefacts from the projection of the star particles.}
 \plotone{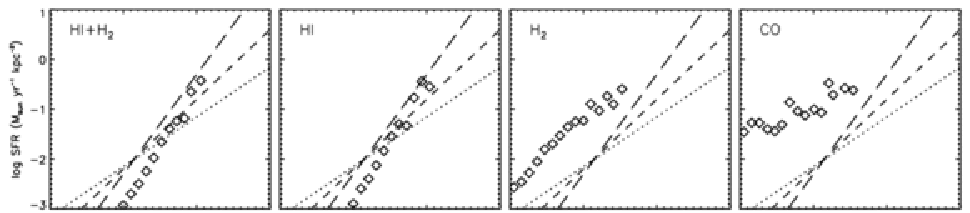}
 \plotone{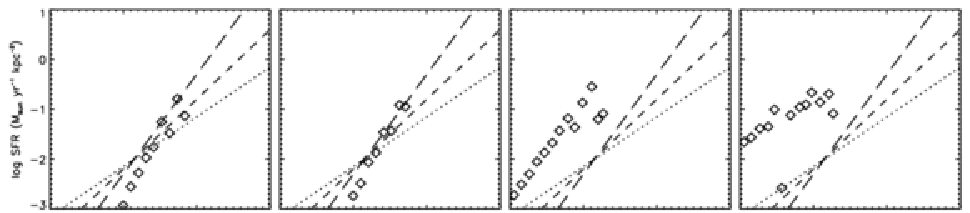}
 \plotone{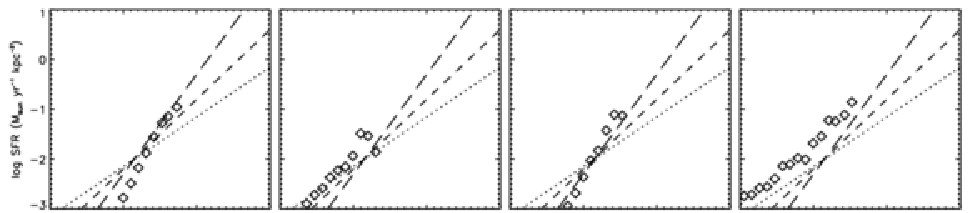}
 \plotone{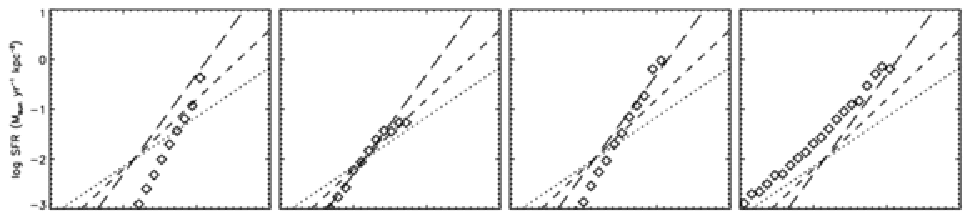}
 \plotone{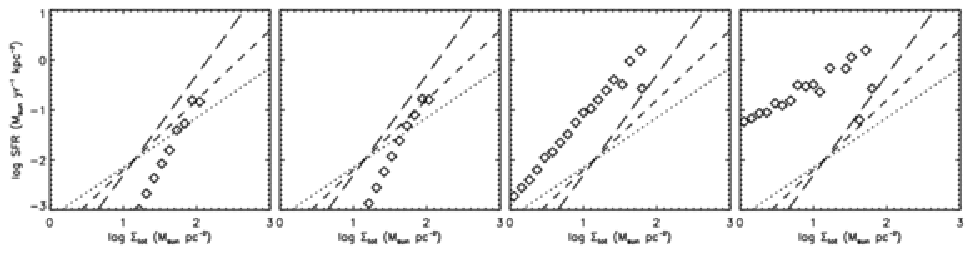}
 \label{fig:KS1}
\end{figure*}

\begin{figure*}
 \centering
 \epsscale{0.99}
 \caption{
  Kennicut Schmidt laws. Same as figure \ref{fig:KS1}, 
  but for the MR star formation recipe.
 }
 \plotone{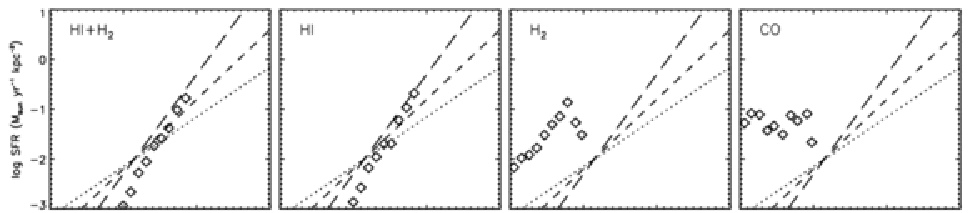}
 \plotone{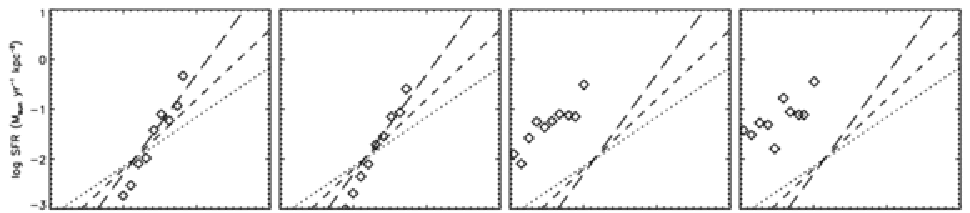}
 \plotone{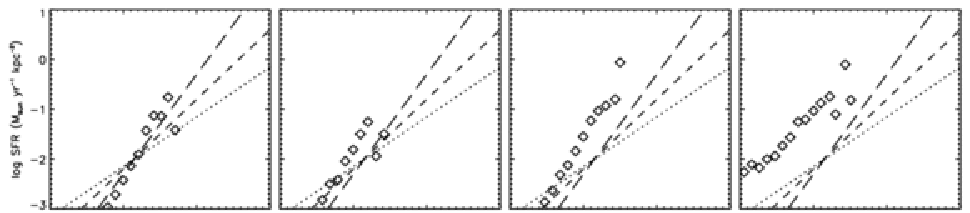}
 \plotone{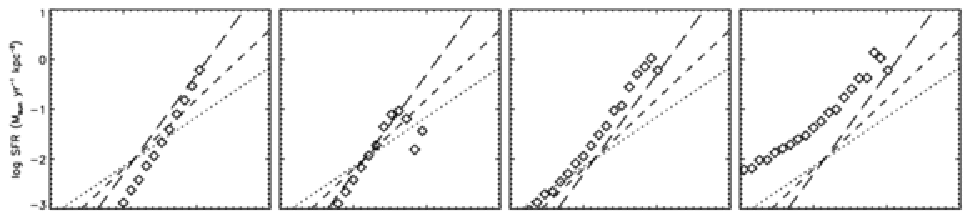}
 \plotone{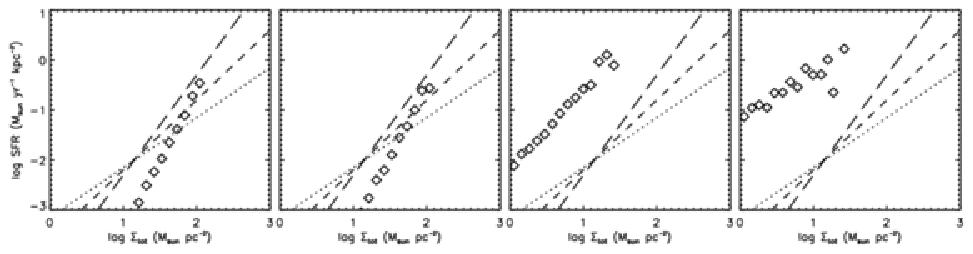}
 \label{fig:KS2}
\end{figure*}

Past investigations  of the K-S  empirical relations in  galaxies used
only stationary  models \citep[][]{Dopita1994, Robertson2008}. Testing 
for the  emergence  of  robust  K-S relations  using
dynamic  galaxy models  like  ours could thus be very interesting,
especially when the much more realistic $\rm H_2$-regulated star formation
recipe  is implemented.  The  relations between  the gas  mass surface
densities  (total, HI,  $\rm H_2$ and  CO-rich $\rm H_2$)  and the  local star
formation   density  in   our   simulations  are   shown  in   Figures
\ref{fig:KS1} (the SD model) and  \ref{fig:KS2} (the MR model) (the CO
surface density  is normalized as in  Figure~\ref{fig:coh2}).  In each
case  the K-S relation  as expressed  in \cite{Kennicutt1998}  is also
shown.

The  first notable  characteristic is  that the  K-S relation  for the
total gas or  the HI surface density is  closer to the observationally
derived  one for  all the  systems simulated  here, both  in  terms of
normalization and  slope.  The slopes are  generally between $n\approx
1.4$  and $n  \approx  2$.  In  addition,  the K-S  relations for  the
CO-rich $\rm H_2$ gas (tracing the  densest molecular phase present in our
simulations)  have a  more shallow  slope ($n  \approx  1-1.4$).  This
tendency  of  the K-S  relation  towards  more  linear slopes  as  one
progresses from HI  to CO-bright and then HCN-bright  $\rm H_2$ ($\rm n\ga
10^5\,cm^{-3}$) gas has also been noted observationally 
\citep{Wong2002, Gao2004}.  The  CO molecule will form in the highest
density peaks of the  $\rm H_2$ distribution where short dynamical scales
make them most intimately linked to the star formation.

There  is  some remaining  uncertainty  in  the  normalization of  our
derived empirical K-S  relations when it comes to  those involving the
CO-rich $\rm H_2$ gas (and thus directly comparable to observations). This
is because  far-UV absorption from the intervening  dusty ISM (besides
that in the natal clouds we  considered here) will affect CO much more
than the self-shielding $\rm H_2$  throught the galaxy, making the CO-rich
$\rm H_2$ distribution more extended than depicted in our models.  This is
expected to  ``shift'' all our  derived K-S relations involving  CO to
the  right, but  it  is  unlikely to  reduce  significantly the  large
deviations we  find for the  K-S relation in our  simulated metal-poor
systems ($Z=0.2Z_{\odot}$) where far-UV is much less absorbed in their
intervening ISM (i.e.  between molecular clouds).  In such systems the
corresponding K-S  relation is  shifted upwards by  a factor  of $\sim
5-10$ with respect to the metal-rich  ones (Fig 6, 7), yielding {\it a
much more efficient  star formation per CO-rich $\rm H_2$  mass.}  This is
because  in metal-poor  systems $\rm H_2$  and CO  can form  only  in much
denser gas (where star formation  is most efficient) deeper in the CNM
clouds, where  higher densities make up  for the loss  of dust surface
for $\rm H_2$ and eventually CO formation. A similar effect was noted 
in metal poor dwarf galaxy IC10 \citep{Leroy2006}.

It is  notable that the  high gas fraction  models (D1 and E1)  have a
somewhat \emph{lower}  SFR, i.e.  their star formation  rate per unit
surface at a given surface density in these systems is lower, although
the difference  is not large ($<50\%$).   This is likely due  to the 
fact that between two systems with similar total surface densities, 
the gas-rich one would necessarily have a smaller stellar component 
constraining the gas in the z-direction than the gas-poor one, resulting
to less star formation per total surface density. Such effects have been
described in the past, using K-S relations that involve the gaseous
as well as the stellar mass component \citep{Dopita1994}. Finally, 
the MR  models show a
slight upward  shift of all  the K-S relations, but  otherwise similar
behaviour, despite the  fact that the star formation  is formulated in
terms of the local molecular gas  fraction. A K-S like law is known to
arise  under a wide  set of  conditions when  the star  formation rate
scales with the reciprocal of the dynamical timescale $1/\sqrt{4 \pi G
\rho}$ \citep{Schaye2008}. However the  MR model inserts an additional
non-trivial criterion  (the $\rm H_2$ richness  of the star  forming gas),
and  thus there  is no  reason for  expecting so  similar  results (we
return to this point in the Discussion).

\subsubsection{The $\rm H_2$-pressure relation}
\label{sec:BS}

\begin{figure*}
 \centering
 \epsscale{0.45}

\plotone{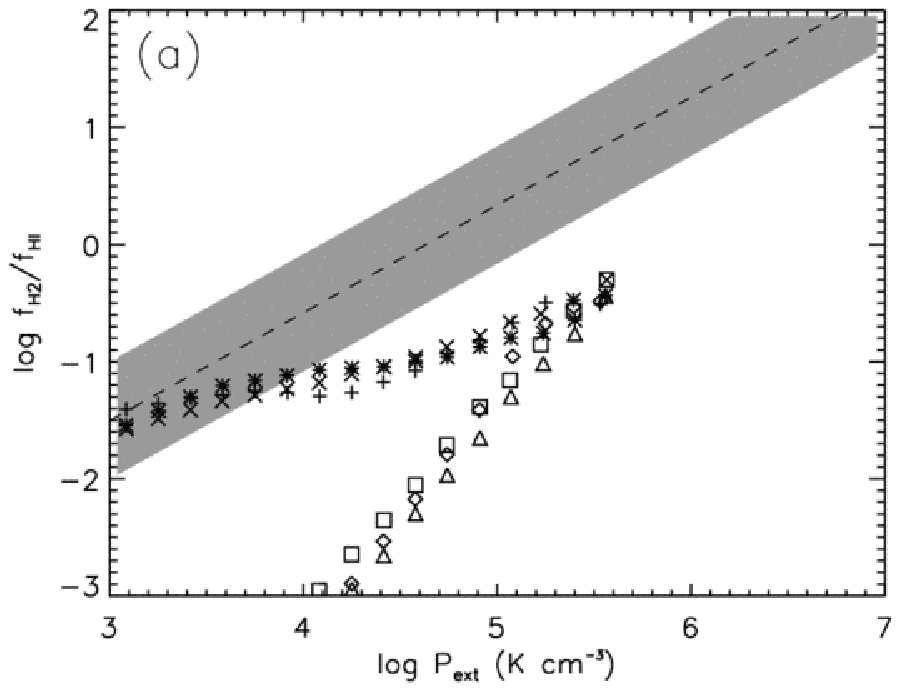}
\plotone{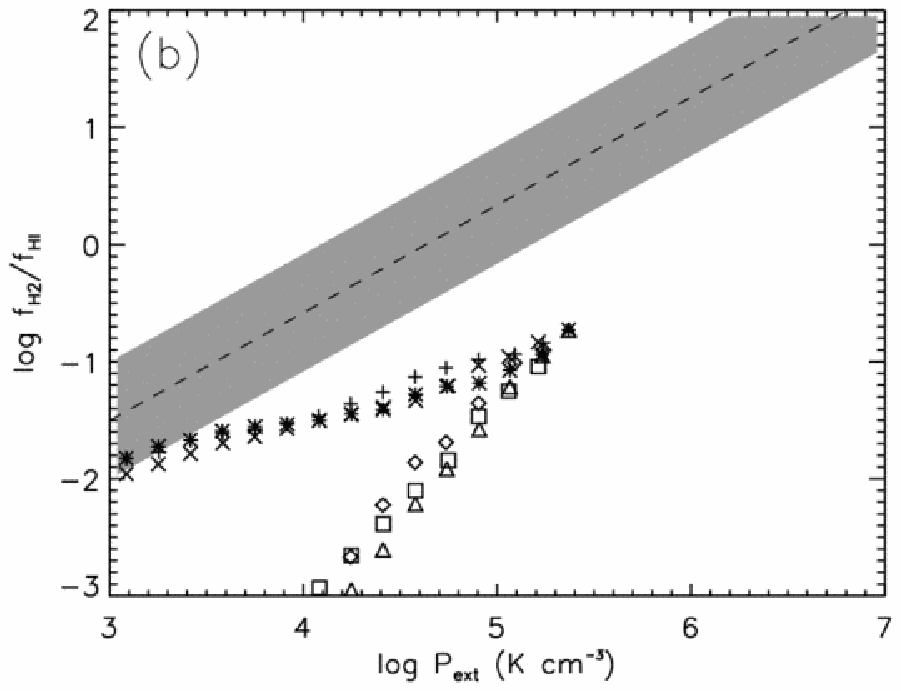}
\plotone{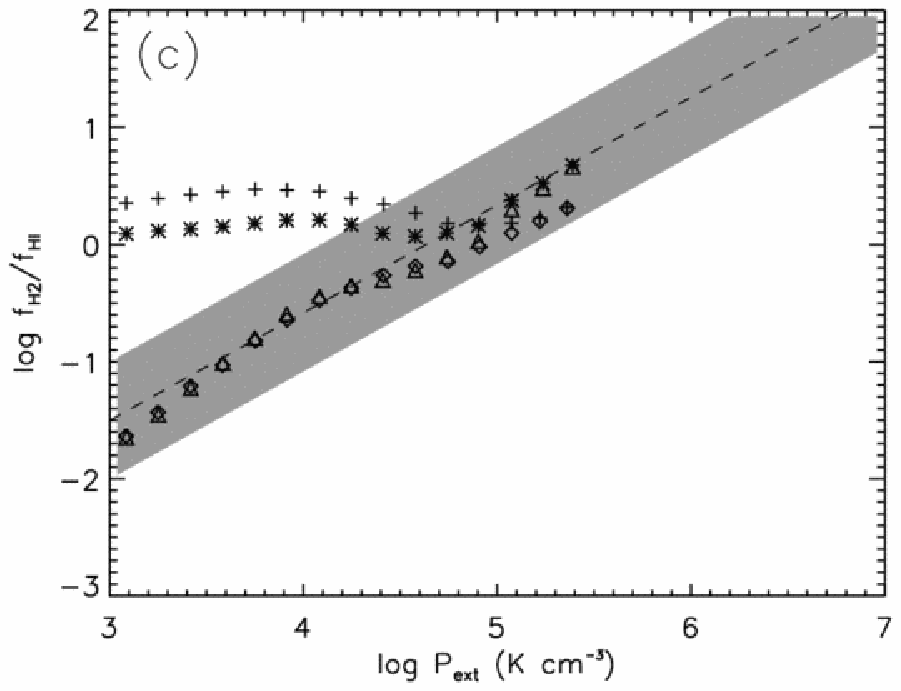}
\plotone{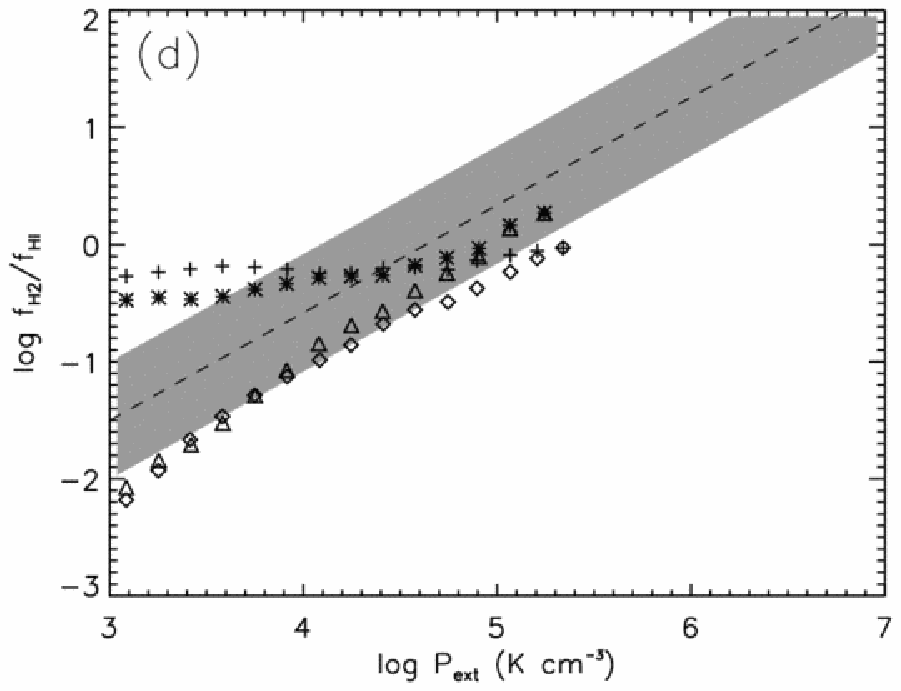}

 \caption{
 Simulation  results for  $R_m=f_{{\rm  H}_2}/f_{\rm  HI}$. Panels show
 $R-m$ as a function of midplane pressure for the low  metallicity models
 (A1,B1 and E1)  for the SD star formation model  (panel a) and the MR model
 (b) and for solar metallicity models (C1 and D1) with the SD model (c) and
 MR model (d). The star like symbols indicate that  $R_m$ is estimated from
 the $\rm H_2$ maps, while open symbols indicate that CO is used.  
 }
 \label{fig:pext_sim} 
\end{figure*}

In Figure~\ref{fig:pext_sim}  we show the  $\rm H_2$-pressure relation for
models  A1-E1, where  the  average $R_m$  is  plotted versus  midplane
pressure  $\rm P_{ext}$, as  estimated by \cite{Blitz2006}
(their equation 5),
\begin{eqnarray}
P_{ext} & = & 272 \ {\rm cm}^{-3} \ {\rm K} 
\left( \frac{\Sigma_g}{\Msun \ {\rm pc}^{-2}} \right)
\left( \frac{\Sigma_\star}{\Msun  \ {\rm pc}^{-2}} \right)^{0.5} \nonumber \\
& & \times \left( \frac{\sigma}{{\rm km \ s}^{-1}} \right)
\left( \frac{h_{\star}}{{\rm pc}} \right)^{-0.5}.
\label{eq:pblitz}
\end{eqnarray}
The  velocity  dispersion  $\sigma$   and  the  stellar  scale  height
$h_\star$ are taken to  be constants \citep[as in][]{Blitz2006}, while
the gas surface  density and stellar surface density  are derived from
the  projected gas  and stellar  distributions. The  $R_m$  plotted is
derived  either from  $\rm H_2$ or  the CO-rich  $\rm H_2$ surface  density.  As
discussed  previously  the  CO-rich   $\rm H_2$  surface  density  has  an
uncertainty in the absolute scaling, translating in some arbitrariness
in the vertical scaling of the corresponding $R_m$-$P_{ext}$ relation,
but the trend of the $R_m$  dependence on $P_{ext}$ is not affected by
this.   Compared   with  the  equilibrium   $\rm H_2$-pressure  relation  of
Figure~\ref{fig:pext_an}  the plots  here are  more comparable  to the
actual observed quantities,  given that the pressure used  is the same
indirect pressure estimate used in observations.

In  Figure~\ref{fig:pext_sim}a-d a  pressure  dependence much  flatter
 than  the  observed  $\rm H_2$-pressure   relation  is  found.   For  solar
 metallicity  the trend  with pressure  is  very flat  while at  lower
 metallicity  it is  steeper,  but  still short  of  a $n=0.92$  slope
 derived  by  \cite{Blitz2006}.   One  the other  hand,  the  relation
 derived for the CO-bright $\rm H_2$  mass fraction $R_{CO}$ has a steeper
 dependence and is very close to the  observed slope for \Zsun,
 while somewhat  steeper for the $\zsun/5$ simulations.   From what we
 have  seen in  figure \ref{fig:coh2},  this is  to be  expected: this
 difference  between  $\rm H_2$  and  CO-rich $\rm H_2$  in  the  $\rm H_2$-pressure
 relation derives from the bias of CO to form at higher gas densities,
 and thus pressures.   The steeper fall-off of the  CO-rich $\rm H_2$ mass
 at lower  densities compared to that  of $\rm H_2$ translates  to a steeper
 fall-off of $\rm R_{CO}$ at lower pressures.

Our results indicate that, like the K-S relation for the CO-rich $\rm H_2$
 phase,  the  CO-rich  $\rm H_2$-pressure  relation  should  show  a  strong
 dependence  on  metallicity, and  for  $\rm  Z=\zsun/5  $ it  is
 considerably  steeper than  the \cite{Blitz2006}  relation.  Although
 hampered  by the very  small number  of actual  CO detections  at low
 metallicities,  the  available data  suggest  no  strong trends  with
 metallicity  \citep{Blitz2006}.   Although  some shift  downwards  is
 expected  \citep{Krumholz2009},  the slope  should  be similar.   For
 example, the  IC\,10 data points shown by  ~\cite{Blitz2006} lie very
 close to the normal relation.   It will be interesting to see whether
 this  will be  borne out  by further  examination of  low metallicity
 dwarf galaxies.   In this context it  is worth pointing  out that the
 pressure  estimate,  based  on the  equation.~\ref{eq:pblitz} may not 
 be valid  for  such systems  due  to  their  low stellar  surface
 densities.
  
\section{Discussion}
\label{sec:disc}

\begin{figure*}
 \centering
 \epsscale{0.49}
 \caption{
  Time dependence of the SFR versus that expected
 from the K-S relation and the instantanous gas content. Shown are the
 actual total star formation (drawn black line) taking place in the models 
 A1, B1, C1 for SD or MR star formation recipe (indicated in the lower
 right of each panel). All other lines give the SFRs expected from
 various applications of the K-S relation (with slope n=1.4) using:
 total gas surface density (dashed), $\rm H_2$ gas surface density (dash-dotted), 
 and CO-bright $\rm H_2$ gas surface density (dotted). The KS
 star formation lines are normalized such that their average after 300 Myr
 matches the average star formation over the same period.
 }
 \plotone{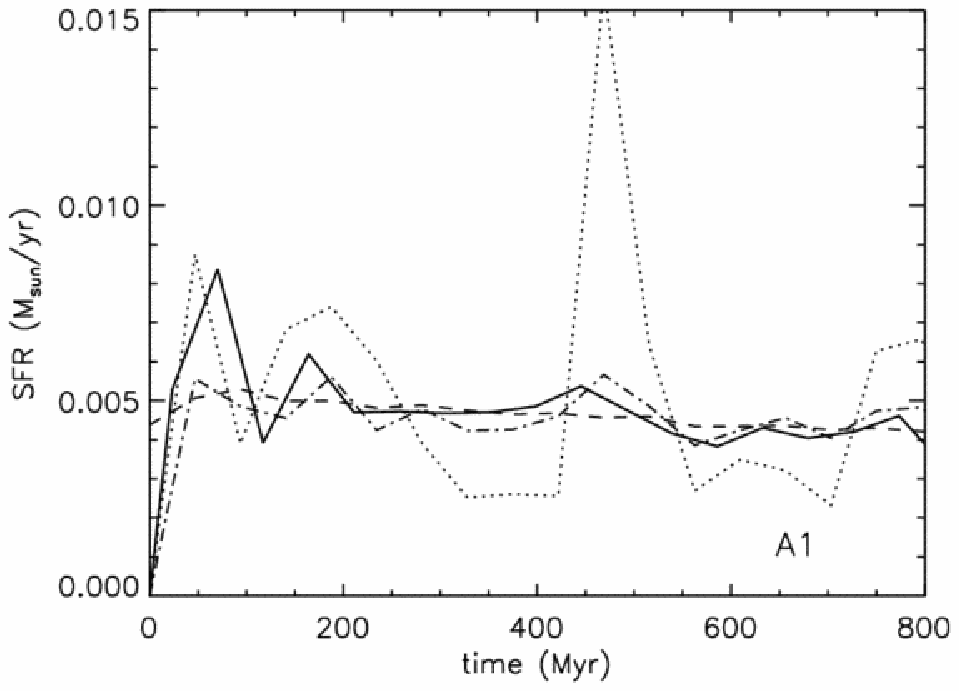}
 \plotone{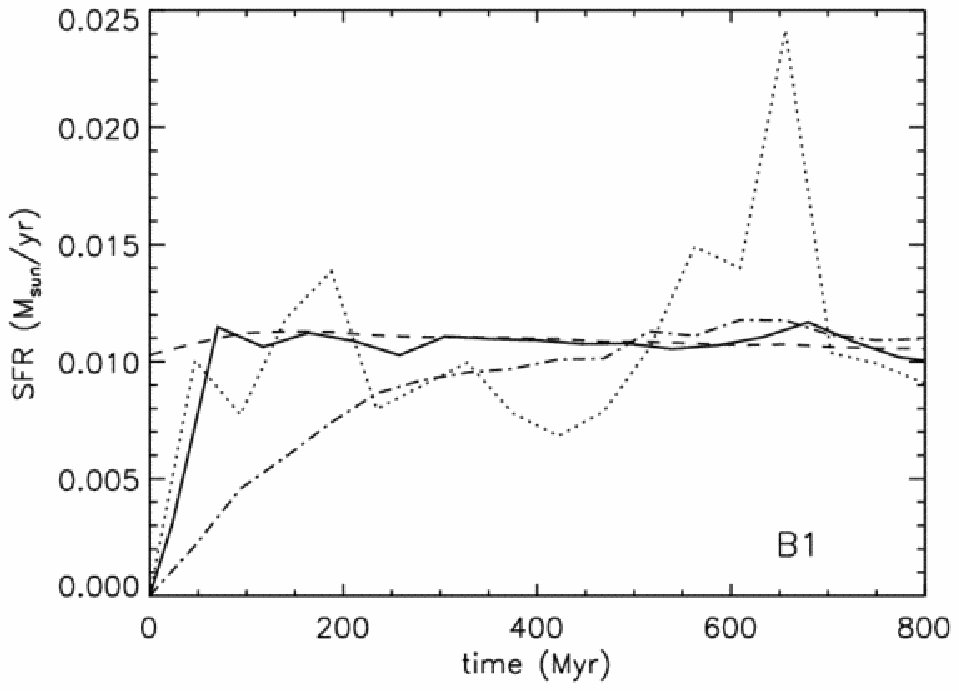}
 \plotone{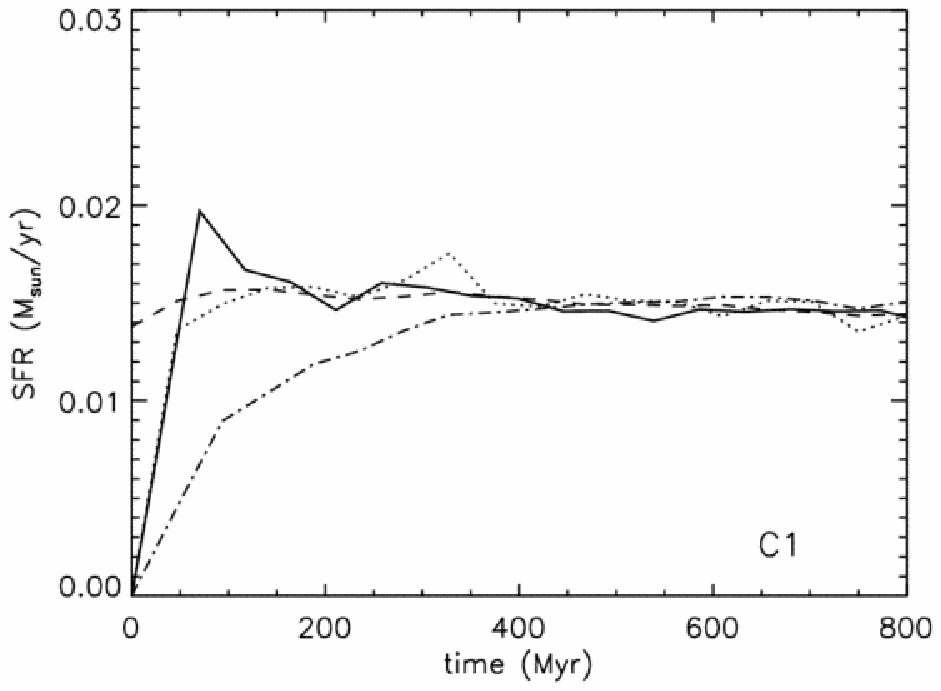}
 \plotone{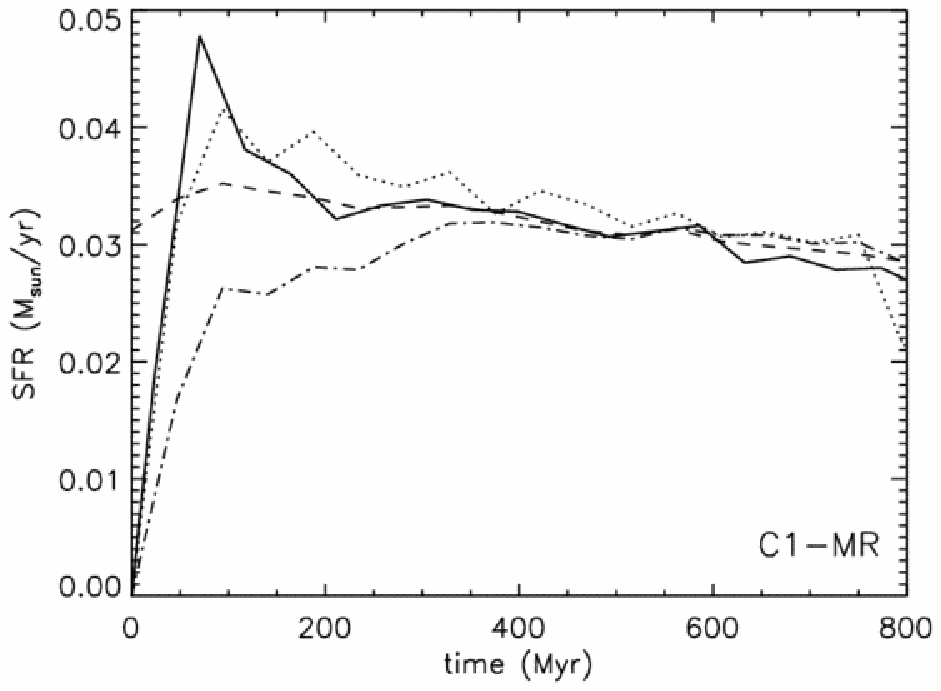}
 \plotone{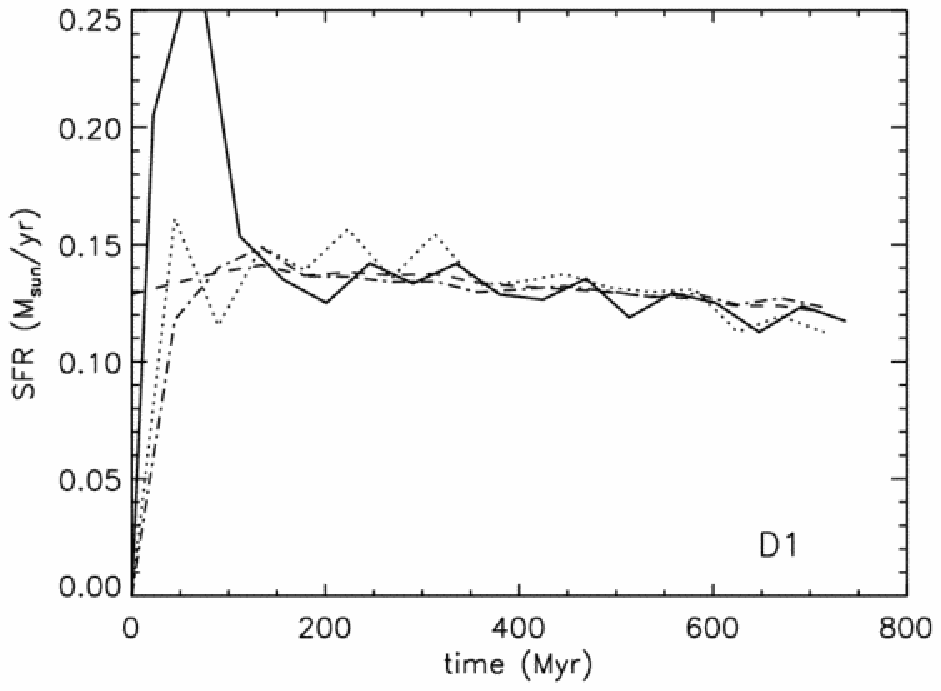}
 \plotone{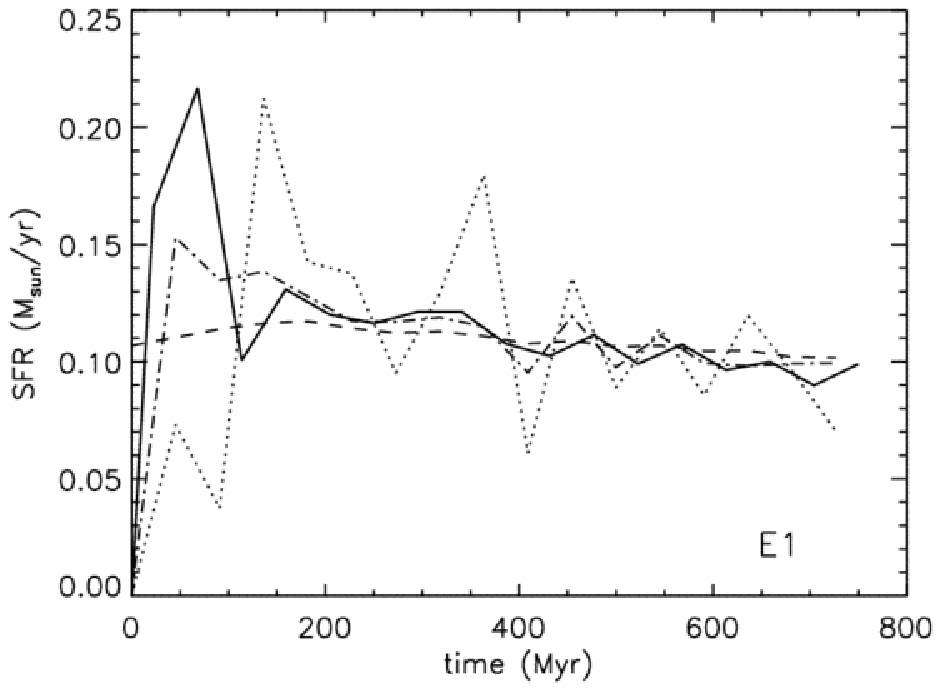}
 \label{fig:timeKS}
\end{figure*}

Our numerical models suggest that after a dynamic equilibrium sets in
  the differences  between  MR  and  SD star  formation
recipes shown by the simulations are small.  This is not only apparent
from  the star  formation rates  and/or a  cursory examination  of the
resulting  gas  morphologies -  which  are  determined mostly  through
feedback processes - but also borne out by the quantitative aspects of
the SF behavior  as revealed by the emergent K-S  and B-R relations in
our models.  For the star  formation process, within the range of SFRs
explored in  our simulations, the following picture  then emerges: for
diffuse  gas to  transform into  star-forming  gas it  must evolve  to
higher densities and the conversion into $\rm H_2$ proceeds passively at a
certain   high  density   threshold   (set  by   formation  rate   and
metallicity).  In  other words: star  formation is biased  towards the
higher densities  where $\rm H_2$  will form fast,  and this  becomes even
more pronounced  in low $Z$ cases.   However in the  MR star formation
model the phase  converted to molecular gas has  a star formation rate
sufficiently slow so that its conversion into molecular gas is not the
SF bottleneck (and  thus star formation will not  be affected, even in
the cases where $\rm H_2$ formation proceeds slowly).

We  have demonstrated that  robust phenomenological  relations linking
gas  content  to star  formation  (K-S  relation),  and molecular  gas
fraction (the SF fuel) to ambient ISM pressure (B-R relation), akin to
those found in the local  Universe,  emerge out of realistic
settings  of  dynamically evolving  galaxy-sized  systems  of gas  and
stars.  However, before considering the hereby presented investigation
as validating the  widespread practice of using K-S  type of relations
or SD star formation models  as sub-grid physics in large scale galaxy
evolution and cosmological models, we  must point out that a 
considerable fraction of the current stellar
mass has been asssembled in major ULIRG type mergers 
\citep{Smail1997, Hughes1998}. In  such systems destruction of $\rm H_2$ 
can have much  shorter timescales, induced  by faster  variations 
of the ambient far-UV field expected  in starbursts \citep{Parravano2003}
yet  also competing  against a  faster $\rm H_2$  formation in  their high
density ISM  gas.  In such  settings large deviations  from (K-S)-type
relations could occur,  given that the latter seem  to robustly emerge
only after  a dynamic  equilibrium between HI,  $\rm H_2$ gas  phases and
stars has been established. Finally, in most numerical models the sub-grid
K-S relation is set to use a much warmer gas phase
 ($\rm  \sim 10^3-10^4\,K$) thermodynamically 
far removed from the CNM HI and  the $\rm H_2$ gas that are directly linked to
star formation.

Modelling of ULIRG type systems is computationally more challenging as 
the high temperatures at high densities mean that timestepping of the
simulation will be slower and the implicit assumptions that we make about 
the transperancy of the interstellar medium break down as the star 
formation sites in these are obscured on galactic scales, so at the moment 
directly testing this is not possible. 
However, as an instructive first step to examine the validity of the K-S
relation during the early evolution of galaxies, we can examine our 
galaxy models during the early phases of the simulation, when the ISM phases
and star formation have yet to establish a dynamical equilibrium.
This is shown in in Figure~\ref{fig:timeKS},
from which it can readily be discerned 
that at early evolutionary timescales   significant deviations of 
the star formation from that expected from  the K-S relation do occur, 
and are especially pronounced for metal-poor systems  (A1, B1, and E1). In the
latter cases there  are periods when the CO-derived  K-S relation will
overestimate or underestimate the underlying star formation which, for
gas-rich {\it and}  metal-poor systems (E1), can last  well into later
evolution  times ($\rm  T\sim 0.2-0.5\,Gyrs$).   This seems  to  be an
effect  of   the  greater  sensitivity   of  CO  destruction   at  low
metallicities where this molecule survives  only in the densest of the
CNM gas, itself spawing star-forming  regions very fast, which in turn
destroy CO.  The fact that this behavior emerges for both a small (A1)
and a  10$\times $  larger metal-poor system  (B1) suggests  that this
``oscillating'' of the SFR with  respect to a K-S(CO) relation (Figure
9) is not  due to  large stochastic scatter  from a smaller  number of
star forming  sites.  On  the other hand  the CO-derived  K-S relation
seems to  remain a  good predictor of  the underlying  star formation,
even at early evolutionary times, for metal-rich systems with moderate
amounts of gas (C1, C1-MR), i.e. like those used for its establishment
in the local Universe.

The largest deviations between the  SFR predicted from the K-S relation
and  the  actual  one  occur  for  the  K-S(HI+$\rm H_2$)  and  K-S($\rm H_2$)
relations during early  evolutionary times ($\rm T\la 0.2-0.3\,Gyrs$),
and are particularly pronounced  for the very gas-rich sustems (Fig.9:
E1,   D1  models).   In   these  cases   even  the   K-S(CO)  relation
underpredicts the true SFR, even  for the metal-rich system (D1) where
CO tracks  $\rm H_2$ well, and thus cannot  be due to CO  failing to trace
$\rm H_2$ (the SF fuel) well.   During those early epochs gas-rich systems
can  appear as  undergoing periods  of very  efficient  star formation
(i.e.  little CO-bright $\rm H_2$ gas but lots of ongoing star formation),
where application  of the K-S(CO)  relation using their  observed SFRs
would  imply much  more molecular  gas  than there  is. Such  systems,
though more massive  that those modeled here, may  have been recently
observed at high redshifts \citep{Tacconi2008}.

The  failure of  the  standard K-S(HI+$\rm H_2$)  relation  to track  SFRs
  during  the  early evolution  of  very  gas-rich  systems is  rather
  expected given that this  relation is ``blind'' to the thermodynamic
  state  of the  gas, and  thus  can equally  well make  stars out  of
  $10^4\,K$ WNM HI  or $30\,K$ $\rm H_2$ gas. Only at  later times its SFR
  predictions become  valid, result of a  dynamic equilibrium among
  the various ISM phases  and the stellar component being established.
  This could have implications for modelling of the gas-rich  galaxies \
  found in the  distant Universe,  or systems  where major  gas  mass 
  accretion events ``reset''  their evolutionary  states back to  gas-rich ones.
  In such cases the non-equilibrium, non-linear, mass/energy exchange
  between the various ISM phases and the stellar component may come to
  dominate significant  periods of intense star  formation and stellar
  mass built-up during which  not even the most realistic, CO-derived,
  K-S relation seems applicable (Figure \ref{fig:timeKS}: 
  models E1, D1).

\section{Conclusions}
\label{sec:concl}

We use  our time-varying, galaxy-sized, numerical  models of gas+stars
 that track the ISM thermodynamics and the $\rm HI\leftrightarrow H_2$
 gas phase exchange, to investigate: a) the emergence of two prominent
 empirical relations  deduced for galaxies in the  local Universe: the
 Kennicutt-Schmidt (K-S) relation  and the $\rm H_2$-pressure relation, b)
 the  effects  of  a  more realistic  $\rm H_2$-regulated  star  formation
 recipe, and c) the evolution of very gas-rich systems. Our models now
 include  a separate treatment  for formation  and destruction  of the
 $\rm H_2$-tracing CO  molecule, which allows a direct  comparison of such
 models with observations, and  a new independent investigation of the
 CO-$\rm H_2$ concomitance in the  ISM of evolving galaxies. Our findings
 can be summarized as follows

\begin{itemize}

\item  For  ISM states  of  $\rm H_2$/HI  equilibrium, an  $\rm H_2$-pressure
      relation close to  the one observed robustly emerges  for a wide
      range  of  parameters,  with   a  strong  dependance  mostly  on
      metallicity.  For  the more realistic  non-equilibrium $\rm H_2$/HI
      states   {\it  only   the   CO-bright  $\rm H_2$   phase  shows   an
      $\rm H_2$-pressure relation similar to the one observed.}

\item The $\rm H_2$-regulated star  formation model successfully models
       star formation without  the adhoc  parameter of the local star 
       formation efficiency
       adopted   by   most   galaxy-sized  numerical   models,   while
       incorporating  a  fundamental  aspect  of  the  star  formation
       process.

\item  A   comparison  between   numerical  models  using   the  usual
      simple-delay  (SD)  and the  new  molecular-regulated (MR)  star
      formation  recipes reveals  very  few differences.  It shows  a
      factor of $\sim  3-4$ more efficient star formation  per CNM gas
      mass than the case of MR star formation. 

\item  We  find  little   sensitivity  of  the  global  SF  efficiency
       M(HI+$\rm H_2$)/SFR   to  the  SF   recipe  chosen,   once  dynamic
       equilibrium  between  ISM  phases  and stars  is  established,
       yielding confidence  to (K-S)-type  of relations emerging  as a
       general characteristic of galaxies.

\item A  non-equilibrium $\rm HI\leftrightarrow H_2$ gas  mass exchange is
      revealed  taking   place    under   typical   ISM   conditions,
      demonstrating the need for a full dynamic rather than stationary
      treatment of these ISM  phases.

\item The  CO molecule can be  a poor, non-linear, tracer  of the true
      underlying  $\rm H_2$  gas  distribution, especially  in  metal-poor
      systems, and  even in  those with very  high gas  mass fractions
      (more typically found at high redshifts).

\item A K-S relation  robustly emerges from our time-dependent models,
      irrespective  of   the  SF   recipe  used,  after   a  dynamical
      equilibrium is established  ($\rm T\ga $1\,Gyr). The CO-derived
      K-S relation has a more shallow slope than the one involving the
      total gas mass, and as  in the $\rm H_2$-pressure relation, a strong
      dependance on metallicity is found.

\item  At  early evolutionary  timescales  ($\rm  T\la 0.4\,Gyr$)  our
  models show  {\it significant and systematic deviations  of the true
  star formation from that expected from the K-S relation,} which seem
  especially pronounced  and prolonged for  metal-poor systems.  These
  deviations  occur even  for the  CO-derived K-S  relation  (the more
  realistic one since CO is directly observable and traces the densest
  $\rm H_2$ gas  which ``fuels'' star formation), and  even for metal-rich
  systems where CO tracks the $\rm H_2$ gas well.

\item  The largest  deviations  from  the K-S  relation  occur at  the
       earliest evolutionary stages of  the systems modeled here ($\rm
       T\la 0.2 Gyr$) and for the most gas-rich ones. During this time
       significantly higher  star formation rates  per CO-bright $\rm H_2$
       gas mass  occur, and such  star-forming galaxies may  have been
       already observed at high redshifts.

\end{itemize}

Finally  we must  note that  when it  comes to  the  gas-rich galaxies
  accessible to current  observational capabilities at high redshifts,
  our   results,  drawn   for  much   less  massive   systems,  remain
  provisional.   Nevertheless for  more massive  gas-rich  systems the
  larger amplitudes of  ISM equilibrium-perturbing agents (e.g.  SNs,
  far-UV  radiation  fields), and  the  shorter  timescales that  will
  characterize their variations are more likely than not to exaggerate
  the deviations  of true star  formation versus the one  derived from
  (K-S)-type  phenomenological relations.   A  dedicated observational
  effort  to  study  such  galaxies  at high  redshifts  (soon  to  be
  dramatically  enhanced  by  ALMA),  as well  as  extending  detailed
  numerical  modeling   of  gas  and  stars  to   larger  systems  (as
  computational  capabilities  improve),  can help  establish  whether
  (K-S)-type relations  remain valid during  most of the  stellar mass
  built-up in galaxies, or  only emerge after dynamic equilibrium has
  been reached during much latter evolutionary stages.

\bibliographystyle{astron}

\end{document}